\documentclass[pdftex,twocolumn,epjc3]{svjour3}        
\journalname{Eur. Phys. J. C}
\def\makeheadbox{\fbox{Submitted to arXiv \hspace{14.3cm}~}}
\usepackage{here}

\usepackage{graphicx}				%
\usepackage{amsmath,bm}			%
\usepackage{caption}
\usepackage{amssymb}
\usepackage{url}
\usepackage{multirow}
\usepackage{array,booktabs}	
\usepackage{lineno}
\usepackage{physics}
\usepackage{float}
\usepackage{pifont}
\usepackage{comment}
\usepackage{rotate}
\usepackage[overlay]{textpos}
\usepackage{braket}
\usepackage{tabularx}
\usepackage[latin1]{inputenc}

\usepackage{times}
\usepackage{tipa}
\usepackage{xspace}

\newcommand{\ce}[1]{Eq.~(\ref{#1})}
\newcommand{\cf}[1]{{Fig.~\ref{#1}}}

\DeclareMathAlphabet{\pazocal}{OMS}{zplm}{m}{n}
\newcommand{\Q}{\pazocal{Q}}

\newcommand{\med}[1]{\text{med}(#1)}
\usepackage{color}
\usepackage{ulem}
\usepackage{cancel}
\usepackage{txfonts}
\newcommand{\old}[1]{{\color{blue}\sout{#1}}}

\newcommand{\pT}{\ensuremath{P_T}\xspace}
\newcommand{\jpsi}{\ensuremath{J/\psi}\xspace}
\newcommand{\psip}{\ensuremath{\psi(2S)}\xspace}

\usepackage{float} %
\usepackage[caption=false]{subfig} %

\usepackage{tablefootnote}

\usepackage{scalerel}
\usepackage{tikz}
\usetikzlibrary{svg.path}

\definecolor{orcidlogocol}{HTML}{A6CE39}
\tikzset{
  orcidlogo/.pic={
    \fill[orcidlogocol] svg{M256,128c0,70.7-57.3,128-128,128C57.3,256,0,198.7,0,128C0,57.3,57.3,0,128,0C198.7,0,256,57.3,256,128z};
    \fill[white] svg{M86.3,186.2H70.9V79.1h15.4v48.4V186.2z}
                 svg{M108.9,79.1h41.6c39.6,0,57,28.3,57,53.6c0,27.5-21.5,53.6-56.8,53.6h-41.8V79.1z M124.3,172.4h24.5c34.9,0,42.9-26.5,42.9-39.7c0-21.5-13.7-39.7-43.7-39.7h-23.7V172.4z}
                 svg{M88.7,56.8c0,5.5-4.5,10.1-10.1,10.1c-5.6,0-10.1-4.6-10.1-10.1c0-5.6,4.5-10.1,10.1-10.1C84.2,46.7,88.7,51.3,88.7,56.8z};
  }
}

\newcommand\orcidicon[1]{\href{https://orcid.org/#1}{\mbox{\scalerel*{
\begin{tikzpicture}[yscale=-1,transform shape]
\pic{orcidlogo};
\end{tikzpicture}
}{|}}}}

\usepackage[colorlinks,citecolor=blue,linktoc=all,linkcolor=blue]{hyperref}

\author{
Jean-Philippe~Lansberg\thanksref{addr1}\protect\orcidicon{0000-0003-2746-5986}
\and 
Kate~Lynch\thanksref{addr1,addr2,corr}\protect\orcidicon{0000-0002-7053-4951}
        \and
        Charlotte~Van~Hulse\thanksref{addr3} \protect\orcidicon{0000-0002-5397-6782}
        \and
        Ronan~McNulty\thanksref{addr1,addr2}\protect\orcidicon{0000-0001-7144-0175}
}

\thankstext{corr}{Corresponding author}

\institute{Universit\'e Paris-Saclay, CNRS, IJCLab, 91405 Orsay, France \label{addr1}
\and
School of Physics, University College Dublin,
Dublin 4, Ireland \label{addr2}
           \and
           University of Alcal\'{a}, Alcal\'{a} de Henares (Madrid), Spain \label{addr3}
           \\~ \\
\email{Jean-Philippe.Lansberg@in2p3.fr, kate.lynch1@ucdconnect.ie, charlotte.barbara.van.hulse@cern.ch \& ronan.mcnulty@ucd.ie}
}

\date{\today}
\begin{document}

\title{Inclusive photoproduction of vector quarkonium in ultra-peripheral collisions at the LHC}

\maketitle
\begin{abstract}
We explore the possibility of using ultra-peripheral proton-lead collisions at the LHC to study inclusive vector-quarkonium photoproduction,  
that occurs when a quasi-real photon emitted by a fully stripped lead ion breaks a proton to produce a vector quarkonium. Owing to the extremely large energies of the colliding hadrons circulating in the LHC, the range of accessible photon-nucleon centre-of-mass energies, $W_{\gamma p}$, largely exceeds what has been and will be studied at lepton-hadron colliders, HERA and the EIC. We perform a tune to HERA photoproduction data, use this tune to predict the yields of photoproduced $J/\psi$, and estimate the corresponding transverse-momentum reach at LHC experiments. We also model the hadroproduction background and demonstrate that inclusive photoproduction can be isolated at the LHC from such background by imposing constraints on the hadronic activity in the lead-going direction at mid, forward, or far-forward rapidities depending on the capability of the detector under consideration. We find that the resulting cross sections are large enough to be measured by ALICE, ATLAS, CMS, and LHCb. We estimate the background-to-signal ratio after isolation to be of the order of 0.001 and 0.1 in the low and large transverse-momentum regions, respectively. In addition, we propose and assess the Jacquet-Blondel method to reconstruct the photon-nucleon centre-of-mass energy and the fractional energy of the quarkonium with respect to the photon.
\end{abstract}

 \section{Introduction}\label{intro}

Quarkonia (hereafter denoted $\Q$) offer a unique platform to probe the interplay between the perturbative and non-pertubative domain{s} of the strong interaction. Since the discovery of the first quarkonium, named $J/\psi$, almost half a century ago, they have been the object of extensive study (see \cite{Lansberg:2006dh,QuarkoniumWorkingGroup:2004kpm,Brambilla:2010cs,Andronic:2015wma,Lansberg:2019adr,Kramer:2001hh} for reviews). Yet, currently, there is no model of quarkonium production that can encompass all of the existing experimental data. In particular, no description can reconcile the data from hadroproduction (from hadron-hadron collisions), photoproduction (from real-photon--hadron collisions), leptoproduction (from off-shell photon-hadron collisions), and from lepton--anti-lepton annihilation. 

\begin{figure}[htbp!]
         \centering
             \subfloat[]{\label{fig:CSM-photo}\includegraphics[width=0.24\textwidth]{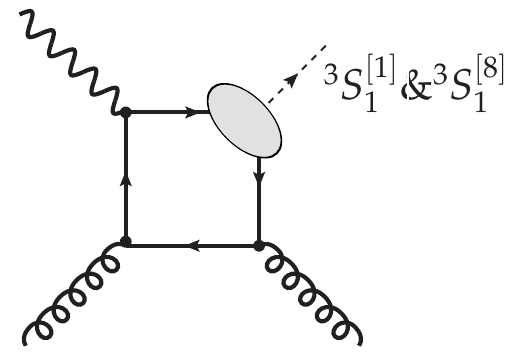}}
             \subfloat[]{\includegraphics[width=0.24\textwidth]{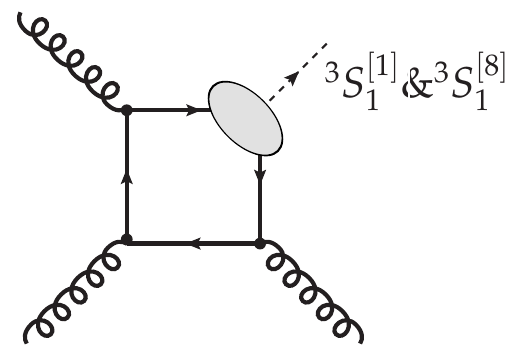}\label{fig:CSM-hadro}}
\caption{Representative diagrams contributing to quarkonium (a) photoproduction and (b) hadroproduction at finite \pT via, respectively, photon-gluon and gluon-gluon fusion. The quantum numbers of the heavy-quark--anti-quark pair  {which will form the quarkonium} are indicated using the usual spectroscopic notation (see  Section~\ref{sec:models}).}
\label{photprod}
\end{figure}

High-energy inclusive photoproduction, produced mainly via photon-gluon fusion (Fig. \ref{fig:CSM-photo}), is {simpler}\footnote{Here we refer to photoproduction by a \textit{direct} or point-like photon. A photon, due to its coupling to quarks, has a hadronic component. Photoproduction may proceed through coupling to a \textit{resolved} photon or a hadronic component of the photon. This resolved photon has a non-perturbative element and as such renders the simplicity over hadroproduction void. The resolved-photon contribution increases with increasing photon-proton centre-of-mass energy, $W_{\gamma p}$.} to describe and thus, in principle, is computable with smaller uncertainties than inclusive hadroproduction, which is produced mainly via gluon-gluon fusion (Fig. \ref{fig:CSM-hadro}). This follows from the Abelian character of the electromagnetic interaction. As a result, the limited photoproduction data from HERA are more constraining than the very precise hadroproduction data from the LHC~\cite{Lansberg:2019adr}. Inclusive quarkonium photoproduction data was collected at HERA but since its shut down, no more data has been recorded at any other facility. Additional inclusive photoproduction data is therefore welcome, preferably before the advent of the US EIC ten years from now. 

The cross sections for photoproduction are smaller than those for hadroproduction and thus measurements require larger luminosities. In addition, they are constrained to a restricted phase space compared to hadroproduction. Inclusive \jpsi photoproduction has been studied at HERA \cite{H1:1996kyo,H1:2002voc,H1:2010udv,ZEUS:1997wrc,ZEUS:2002src,ZEUS:2009qug,ZEUS:2012qog}, with limited statistical samples, and consequently, limited reach of the \jpsi transverse momentum, \pT, up to 10 GeV. This is ten times smaller than what is now routinely achieved for hadroproduction at the LHC~\cite{Chapon:2020heu}. %
{Furthermore}, there is practically no photoproduction data of \psip nor bottomonium to be compared to hadroproduction data. 

In this context, we stress that inclusive quarkonium photoproduction measurements at the LHC are not only possible but, as we demonstrate in the present study, extend to higher centre-of-mass energies, to higher \pT, and to more quarkonium species with respect to existing HERA measurements. Moreover, contrary to the measurements at HERA, all four experiments at the LHC can disentangle
prompt-quarkonium production (where the quarkonium originates from the primary interaction vertex) from non-prompt where the quarkonium comes from $b$-hadron decays. This is important since it has been recently noted~\cite{Lansberg:2019adr,Flore:2020jau} that the yield of non-prompt photoproduced $J/\psi$ at large $P_T$ at HERA might be as large or larger than that of \old{the} prompt $J/\psi$.

Although inclusive quarkonium production at the LHC, in nucleus-nucleus ($AA$), proton-nucleus ($pA$), or proton-proton ($pp$) collisions, is dominated by hadroproduction, we show that inclusive photoproduction cross sections off the proton {for $J/\psi$, $\psip$ and $\Upsilon(nS)$} are large enough to be observed at the LHC {in $p$Pb collisions} with  {existing or expected data} and may be selected by a characterisation of the final state. We show that this characterisation can be based on the level of {hadronic} activity in the central, forward {and far-forward} regions. To date, this has only been performed in PbPb collisions{,} to study azimuthal correlations and %
{identify inclusive di-jet photoproduction}~\cite{ATLAS:2021jhn,ATLAS:2022cbd}.

In exclusive quarkonium photoproduction at the LHC, the detection of the quarkonium is sufficient to reconstruct the photon energy and to determine\footnote{Up to {small} kinematical corrections.} the photon-nucleon centre-of-mass energy, $W_{\gamma N}$, without measuring the momentum of the photon emitter. In inclusive quarkonium photoproduction, the detection of the quarkonium {alone} is not sufficient. Nevertheless, we show that the existing LHC detectors allow for a satisfactory reconstruction of the initial photon energy via the measurement of hadrons {that are nearly} collinear to the photons, in a method similar
to the determination of the kinematics of charged-current DIS with an unobserved final-state neutrino~\cite{Amaldi:1979qp,Pawlak:1999ph}.

In the current study, we focus on $p$Pb collisions at the LHC as the Pb ion is the most probable photon emitter, while in $pp$ and PbPb collisions, one must address the ambiguity in the identity of the photon source. LHC Run 3 will feature a limited run of $p$O collisions, however, we do not anticipate sufficient yields to measure cross sections. In addition, in $pp$ collisions, photoproduction cross sections are much smaller than in $p$Pb collisions due to the reduced size of the proton photon flux. Such a reduction is partly 
compensated by the much higher luminosity recorded in LHC $pp$ collisions. This, however, is at the cost of a much higher pile-up, which may prevent an efficient characterisation of the final state in order to select photoproduced events. 

To demonstrate the feasibility of tagging inclusive photoproduction of quarkonium at the LHC,
we focus on the easiest quarkonium to study, namely the $J/\psi$ meson. For $\psip$ and $\Upsilon(nS)$, we restrict {the discussion} to 
{quoting} expected rates.

The structure of the manuscript is as follows. In Section \ref{sec:photon}, we discuss photon-induced reactions in hadron-hadron collisions and introduce the concept of ultra-peripheral collisions (UPCs). Section~\ref{sec:IQPLHC} gives a theoretical overview of quarkonium photoproduction and additionally discusses LHC-specific background contributions. Section~\ref{sec:simulation} introduces our simulation set-up, which we use to assess the feasibility of an inclusive quarkonium photoproduction measurement at the LHC. Section~\ref{sec:expselection} outlines our proposed photoproduction selection strategy, gives the main results, and features a brief discussion of the resolved photon contribution. Section~\ref{sec:reco} assess the reconstruction capability of kinematic variables at the LHC, namely $W_{\gamma N}$ and the elasticity, $z$, via the Jacquet-Blondel method. Finally, Section \ref{outlook} presents our outlook and conclusions.

\section{Photon-induced reactions at the LHC}\label{sec:photon}

This section deals with photon-induced interactions at the LHC. Section~\ref{sec:photon1} discusses theoretical generalities related to photon-induced reactions in hadron-hadron collisions as well as the kinematic region accessible at the LHC. Section~\ref{sec:photon2} mentions experimental studies of photon-induced interactions perfomed using LHC data and the associated event-selection strategies involved. 

\subsection{Photon-induced interactions in high-energy hadron-hadron collisions}
\label{sec:photon1}
At ultra-relativistic velocities, such as those reached at the LHC, the electromagnetic-field generated by a moving %
electric charge distribution is highly boosted, resulting in a high density of electromagnetic field lines transverse {to the motion of the charge}. The semi-classical Weisz\"acker-Williams approximation of equivalent photons \cite{vonWeizsacker:1934nji,Williams:1934ad} relates this electromagnetic field to a number of equivalent photons of energy $E_\gamma$ at some transverse distance from the charge {and moving in the same direction. When these photons are the result of the coherent action of the moving charge distribution, they do not resolve its internal structure. 
Thus, the corresponding photon flux is proportional to the square of its charge, $Z^2$, and the wavelength of the emitted photon is larger than the charge-distribution radius, $R$, in its rest frame.  This places a constraint on the virtuality, $Q^2$, of such coherently emitted photons
\begin{equation}
\begin{split}
Q^2 \lesssim \frac{1}{ R^2 },
\label{HUP}
 \end{split}
\end{equation}
with $Q^2 \equiv -P_\gamma^2$, where $P_\gamma$ is the photon four momentum. When the 
moving charge distribution is that of a hadron or nucleus, the virtuality of these coherent photons {can be neglected\footnote{The proton charge radius is $R_p \approx 0.7$~fm and the charge radius of a nucleus with mass number $A>16$ is $R_{A} \approx1.2 A^{1/3}$~fm, which, following \ce{HUP}, corresponds to $Q^{2}_\text{max}\approx 0.1$~GeV$^2$ and $Q^2_\text{max}\approx 0.03A^{-2/3}$~GeV$^{2}$, respectively. For a Pb ion, $Q^2_\text{max}\approx0.001$~GeV$^2$. This scale, $Q^2$, is always smaller than scales involved in a hard process, and therefore it is neglected.}. They are then considered as quasi-real and particle interactions involving these photons are termed {\it photoproduction} as opposed to {\it leptoproduction} when a highly virtual photon is emitted by a (point-like) lepton.}

Anticipating the advent of the first heavy-ion colliders, RHIC and the LHC, it was proposed~\cite{Bertulani:1987tz}, already 35 years ago, that photoproduction could be studied at hadron-hadron colliders.  The first observation of such photon-induced reactions was indeed made in AuAu collisions at RHIC~\cite{STAR:2002caw} {via the photoproduction of $\rho^0$ mesons}. The isolation of photoproduction in such hadron-hadron collisions was then -- and systematically so far is -- performed through {the selection of}  so-called  ultra-peripheral collisions (UPCs). These are defined as interactions mediated over distances $b$ larger than the sum of the radii of the colliding hadrons, as depicted in Fig. \ref{fig:imp}. At such impact parameters, there is no hadronic overlap, strong interactions are suppressed, and photon-induced interactions become dominant. 

 \begin{figure}[h!] %
    \centering
   \subfloat[]{\includegraphics[width=0.5\textwidth]{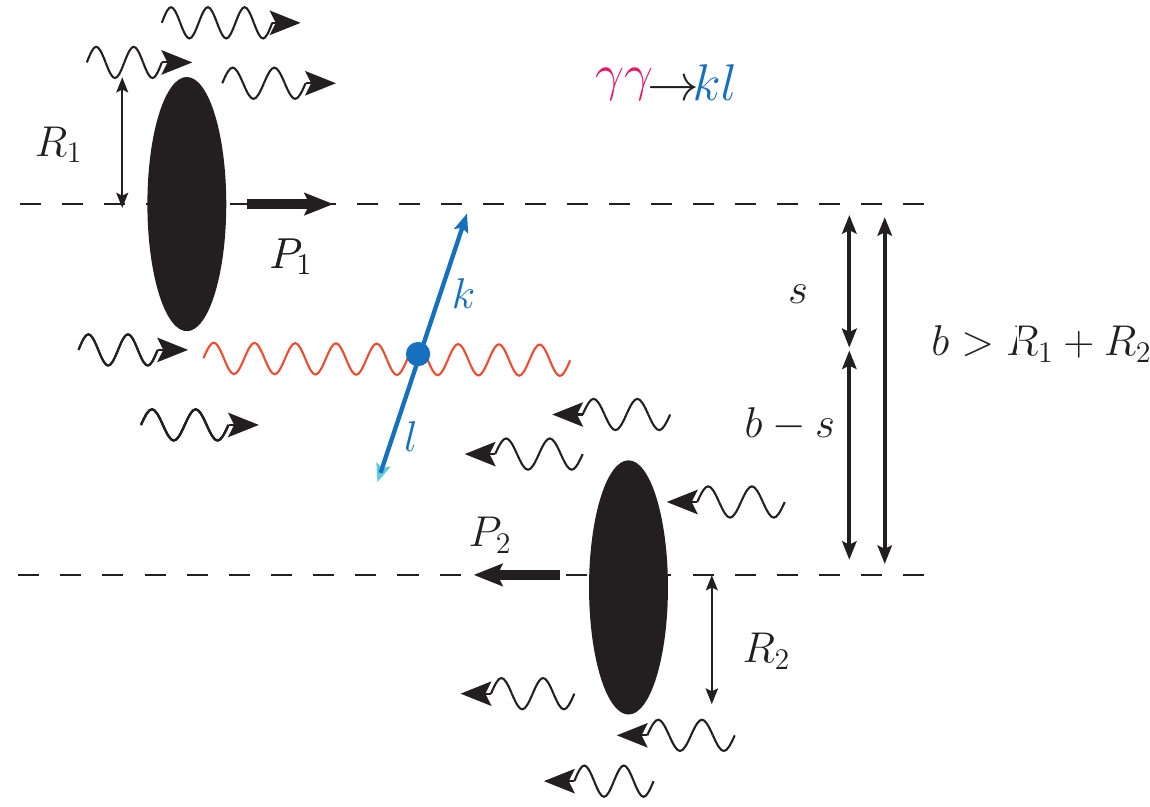}\label{fig:gammagamma}}\\
    \subfloat[]{\includegraphics[width=0.5\textwidth]{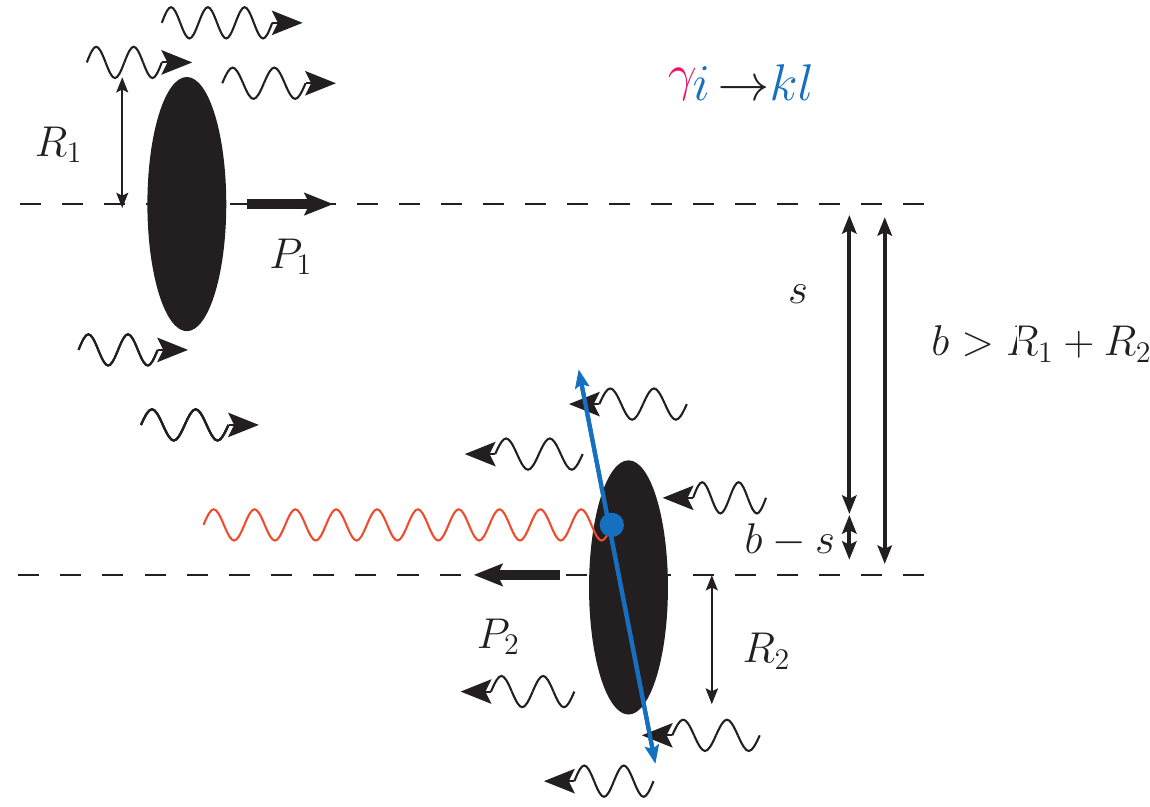}\label{fig:gammaH}}
      \caption{A UPC is mediated over distances, $b$, larger than the sum of colliding radii{, $R_1$ and $R_2$,} and can result in (a) photon-photon {($\gamma \gamma \to k l$)} or (b) photon-hadron {($\gamma i \to k l$)} interactions{, where the curly lines are photons resulting from the electromagnetic fields of the colliding nuclei, $i,k,$ and $l$ are partons, $P_{1}$ and $P_{2}$ are the momenta of colliding nucleons, and $s$ ($b-s$) is the distance between the centre of nucleon 1 (2) and the interaction point.} 
      }
    \label{fig:imp}
 \end{figure}

One distinguishes two classes of photon-induced reactions in hadron-hadron collisions: those induced by two photons emitted by each hadron (photon-photon interactions, Fig. \ref{fig:gammagamma}) and those induced by only one photon (photon-hadron interactions, Fig. \ref{fig:gammaH}). In the latter case, the {\it source} hadron is defined as that that emits the photon and the other hadron is defined as the \textit{target} hadron. Since the electromagnetic interaction is long range, the emitted photon can interact with the partons in the  target hadron even when the colliding hadrons themselves do not overlap. It is important to note that such photon-hadron reactions necessarily takes place within the cylindrical volume traced by 
the trajectory of the target hadron (see \cf{fig:gammaH}).

\begin{table*}[h!]
\caption{The energy{, $E$,} per nucleon of the (a,b) beam particles, where $E= Z/A E_p$ 
for an ion with charge $Z$ and mass number $A${, and $E_p$ is the corresponding LHC energy for a proton beam},  (c) the centre-of-mass energy per nucleon, (d) the Lorentz boost between nucleon rest frames, (e) the minimum impact parameter, (f) the maximum photon energy in the rest frame of the emitting particle, (g) the maximum energy of the photon in the rest frame of the target particle, (h) the maximum photoproduction centre-of-mass energy, and (i) the maximum energy fraction taken by the photon from the beam{. The energies and boosts are computed for different colliding and fixed-target systems for the LHC setting corresponding to $E_p=6500$~GeV.}  } 
\resizebox{2\columnwidth}{!}{        \centering
\begin{tabular}{|c|c|c|c|c|c|c|c|c|c|}
\hline
System  & (a) $E_{1}$&(b) $E_{2}$&(c)  $\sqrt{s_{NN}}$ &(d)  $\gamma_L$& (e) $b_\text{min}$& (f)$E^\text{max}_{\gamma}$ &(g) $E^{\prime \text{ max}}_{\gamma}$& (h) $W^\text{max}_{\gamma N}$&(i) $x_{\gamma}^\text{max}$ \\
 &&&  & $=s_{NN} /(2m_N^2)$&$=R_1+R_2$ & $=1 /b_\text{min}$ & $=\gamma_L \times E_\gamma^\text{max}$ & $=\sqrt{2 m_N E_\gamma^{\prime \text{ max}}}$ & $=\lambda_{C_N}/b_\text{min}$\\ 
\hline 
    \textbf{Collider} & & &=$\sqrt{4 E_1 E_2}$ & & & &&& \\
$pp$ & 6500 GeV & 6500 GeV &  13.0 TeV & 9.6 $\times 10^7$& 1.4 fm& 141 MeV & 13.5 PeV & 5.0 TeV & 0.15 \\
$p$Pb & 6500 GeV & 2562 GeV &  8.16 TeV & 3.8 $\times 10^7$& 7.8 fm& 25 MeV & 1.0 PeV & 1.3 TeV & 0.03 \\
$p$O & 6500 GeV & 3250 GeV &  9.19 TeV & 4.8 $\times 10^7$& 3.7 fm& 53 MeV & 2.5 PeV & 2.2 TeV & 0.06 \\
PbPb & 2562 GeV & 2562 GeV &  5.13 TeV & 1.5 $\times 10^7$& 14.2 fm& 14 MeV & 0.2 PeV & 0.6 TeV & 0.01 \\

\textbf{Fixed target} & & & =$\sqrt{2 E_1 m_N}$ & & &&&& \\
$p$Ar & 6500 GeV & $m_N$ & 110 GeV & 6.9 $\times 10^3$& 4.8 fm& 41 MeV & 0.3 TeV & 23 GeV & 0.04 \\
$p$He & 6500 GeV & $m_N$ & 110 GeV & 6.9 $\times 10^3$& 2.4 fm& 83 MeV & 0.6 TeV & 33 GeV & 0.09 \\
$p$Ne & 6500 GeV & $m_N$ & 110 GeV & 6.9 $\times 10^3$& 4.0 fm& 50 MeV & 0.3 TeV & 25 GeV & 0.05 \\

\hline
\end{tabular}}

       \label{photons}
\end{table*}

In photon-hadron collisions, it is natural to work in the rest frame of the hadron. In this frame{,} soft photons coherently emitted by the source are highly boosted. This boost factor, $\gamma_L \approx  {s_{NN}}/({2m_N^2})$, where $m_N$ is the nucleon mass and $\sqrt{s_{NN}}$ is the centre-of-mass energy per nucleon of the source-target collision system, can be as large as $10^8$ at the LHC. Together with the distance between target and source, $b$, it determines $W_{\gamma N}$. In photon-photon collisions, it is most instructive to work in the centre-of-mass frame where the boost is $\gamma'_L \approx  {\sqrt{s_{NN}}}/({2m_N})$.

Table \ref{photons} lists the associated boost factor $\gamma_L$, the maximum photon energy in the rest frame of the source, $E_\gamma^\text{max}$, and of the target, $E_\gamma^{\prime \text{ max}}$, the maximum photon-nucleon centre-of-mass energy, $W^\text{max}_{\gamma N}$, for a selection of collider and fixed-target configurations at the LHC with beam set-up equivalent to 6.5~TeV protons. It also lists the maximum energy fraction taken by the photon from the source, $x^\text{max}_\gamma$, which is equivalently the ratio of the Compton wavelength of a nucleon, $\lambda_{C_N} = m_N^{-1}$, to the radius of the nucleus. It can be seen from the table that $x^{\text{ max}}_\gamma(pp)> x^{\text{ max}}_\gamma(\text{Pb}p)> x^{\text{ max}}_\gamma\text{(PbPb)}$, which is a consequence of the fact that interactions between more compact charges can be mediated over shorter distances. Table \ref{photons} also shows that the $W_{\gamma N}$ range achievable at the LHC in collider mode is two orders of magnitude larger than what can be achieved in fixed-target mode and is an order of magnitude larger than the operating energy of HERA and the future EIC.

\subsection{Photon{-induced} studies at the LHC}\label{sec:photon2}

{Photon-induced studies at the LHC in $p$Pb and PbPb UPCs as well as in $pp$ collisions %
until recently have focused on exclusive reactions, i.e., processes resulting in a fully determined final state, where it is implied that both beam particles remain intact\footnote{Here, we leave aside potential neutron emissions from the Pb ion.}. They range from the production of light vector mesons \cite{ALICE:2023kgv,ALICE:2020ugp,ALICE:2021jnv,CMS:2019awk,ALICE:2015nbw}, vector quarkonia \cite{ALICE:2021tyx,ALICE:2018oyo,ALICE:2012yye,CMS:2018bbk,LHCb:2021bfl,LHCb:2021hoq,ALICE:2014eof,LHCb:2022ahs,ALICE:2013wjo,ALICE:2019tqa,ALICE:2021gpt,CMS:2016itn}, dijets \cite{CMS:2022lbi}, and dileptons \cite{ALICE:2013wjo,ATLAS:2020epq}, to light-by-light scattering \cite{ATLAS:2017fur,CMS:2018erd} in $p$Pb and PbPb UPCs, while for $pp$ collisions the focus was on vector quarkonia and light-by-light scattering \cite{LHCb:2013nqs,LHCb:2014acg,LHCb:2018rcm,LHCb:2015wlx,ATLAS:2017sfe,ATLAS:2015wnx,CMS:2011vma}.} 

In  these LHC measurements, the event selection is typically based on a combination of several criteria: 
\begin{enumerate}
    \item the reconstruction of the newly produced particles of interest and the absence of additional particles in  {\it central} detectors\footnote{The reconstruction of particles at the LHC experiments 
    is performed with {\it central} detectors within $-5\lesssim \eta \lesssim 5$, thus far from the beam rapidities, which in collider mode is $y_\text{b} = \pm\ln(\sqrt{s_{NN}}/m_N)$ equal to $\pm9.5$ for $pp$ at $\sqrt{s}=13$~TeV and $\pm9.1$ for $p$Pb at $\sqrt{s_{NN}}=8.16$~TeV.};
    \item the presence of rapidity gaps;
    \item (a)~the absence of signal in the far-forward and far-backward detectors, (b)~an explicit tagging of the intact beam particles, and/or (c)~the characterisation of the \pT of the produced system to ensure that the beam particles are intact.
\end{enumerate}

{In $p$Pb and PbPb collisions, photon-induced reactions via UPCs are usually selected by imposing criteria 1--3. Such processes proceed via photon-photon or via photon-hadron reactions depending on the quantum number of the produced system, e.g. photon-photon for a scalar, pseudoscalar, or tensor meson and photon-hadron for a vector meson. Note that even if criterion 3 is not satisfied and if the target (or the source) dissociates, photon-induced reactions would still likely dominate over any kind of hadronic exchanges as the typical impact parameter of the collision would still be much larger than the hadronic radius of the colliding objects with coherent-photon emissions. }

In $pp$ collisions, (double) diffractive interactions like pomeron-pomeron or pomeron-odderon reactions, i.e., non photon-induced reactions, would also satisfy criteria 1 and 2. Let us cite the case of $H^0$, $\chi_c$ or di-$\jpsi$ production by pomeron-pomeron fusion~\cite{Kaidalov:2003fw,Khoze:2004yb,Harland-Lang:2011scf,LHCb:2014zwa} or $J/\psi$ and $\Upsilon$ production by pomeron-odderon fusion~\cite{Bzdak:2007cz}. In this case,  double diffractive hadronic interactions from pomeron-odderon fusion can compete with photon-pomeron reactions depending on the kinematics (criterion 3c) of the centrally produced system: the photon-induced reactions usually occur at slightly smaller transverse momenta.

Photon-induced production of $J/\psi$ has also been isolated in peripheral, as opposed to ultra-peripheral, PbPb collisions~\cite{ALICE:2022zso,ALICE:2015mzu,LHCb:2021hoq} by searching for an excess of low-\pT \jpsi with respect to the expected hadroproduction yield. In such a case, as the $J/\psi$ is accompanied by many other particles, the only signature of photoproduction is to be found in the kinematics of the measured \jpsi (criterion 3c). {Such kinematic information can in principle also be used in exclusive \jpsi photoproduction in PbPb UPCs to tell if the photon has interacted with a single nucleon or coherently with the entire nucleus. In the latter case, the \pT spectrum is steeper, characteristic of a larger interaction zone.}

{Recently, \jpsi production by photon-Pomeron fusion has been accessed~\cite{ALICE:2023mfc} in UPCs with target dissociation by imposing criteria 1 and 2 but none of 3.}
One can also look for exclusive \jpsi production accompanied by additional photonuclear excitation, resulting in neutron emissions, which has been motivated theoretically in \cite{Baltz:2002pp} to gain insight into the impact parameter of the collisions. Along the same lines, the accompanying neutron emissions in the measurement of exclusive $J/\psi$ in PbPb collisions have been used to disentangle the identity of the photon emitter \cite{CMS:2023snh,ALICE:2023jgu}.

However, the bulk of the photoproduction yield is not contained within any of the above UPC studies. It misses the most probable configuration where 
the photon breaks the proton by interacting with a single parton,
producing many more hadrons along with the particles of interest. 
To retain the corresponding events, still selecting UPCs, it is sufficient to impose criteria 2 and 3(a or b) on the side of the photon emitter. Doing so, we gain access to the \textit{inclusive} photoproduction yield. 
It is our purpose to show that it is feasible for quarkonium at the LHC. 
\section{Inclusive quarkonium photoproduction at the LHC}\label{sec:IQPLHC}

This section gives a theoretical overview of inclusive quarkonium photoproduction at the LHC. 
In Section~\ref{sec:models}, a general theoretical description of quarkonium production is presented in terms of the three most commonly {used} 
models.  
Inclusive photoproduction of quarkonium is discussed in
Section \ref{sec:theory-photoprod}, while Section \ref{sec:theory-photoprod-lhc} focuses on the LHC. The major expected background contributions to the experimental isolation of inclusive quarkonium photoproduction in $p$Pb collisions at the LHC are discussed  in Section \ref{sec:theory-background-lhc}.

\subsection{Inclusive quarkonium production}\label{sec:models}

{

Several mechanisms  have been proposed to explain the hadronisation of quarkonia but none of them is fully satisfactory in describing the variety of quarkonium-production data in inclusive reactions~\cite{Lansberg:2019adr}.
The three most popular approaches are the Colour Singlet Model (CSM)~\cite{Chang:1979nn,Berger:1980ni,Baier:1981uk}, the Colour Octet Mechanism (COM) within non-relativistic QCD (NRQCD)~\cite{Bodwin:1994jh,Cho:1995vh,Cho:1995ce} and the Colour Evaporation Model (CEM)~\cite{Halzen:1977rs,Fritzsch:1977ay}. They differ in their treatment of hadronisation, and 
in particular, in the evolution of the quantum numbers of the heavy-quark ($Q\Bar{Q}$) pair during their transition to the final $\Q$:
\begin{itemize}
    \item The CSM requires that the $Q\bar{Q}$ pair is produced by the hard process with the same quantum numbers as the final $\Q$ and hence, as the name implies, the $Q\bar{Q}$ pair must be produced in a colour-singlet state. In addition, it must be produced on-shell with zero relative momentum in the $\Q$ rest frame.
    \item The COM extends the treatment of hadronisation by considering states with colour and angular momentum different from the final $\Q$. Indeed, the angular momentum and colour can be changed by the emission of soft gluons. These are treated within NRQCD, which involves an expansion in $v$ (the relative velocity between the $Q$ and $\Bar{Q}$ in the rest frame of the pair) in addition to the expansion in $\alpha_s$ of perturbative QCD (pQCD).  Each quantum state of the $Q\bar{Q}$ pair, typically denoted using spectroscopic notation $^{2S+1}L^{[c_f]}_J$, where $S$ is the spin, $L$ the angular momentum, $J$ the total angular momentum, and $[c_f]$ the colour state of the pair (1=singlet and 8=octet), has a different probability of hadronisation into a particular quarkonium, which are given by non-perturbative, Long-Distance Matrix Elements (LDMEs).  In addition, the $Q\bar{Q}$ pair is produced by the hard process with different kinematic distributions and different (soft and hard) radiation patterns\footnote{This is in direct analogy with the fact that the $\chi_{c2}(^3P_2)$ and $\jpsi(^3S_1)$ mesons minimally couple to two and three gluons, respectively, resulting in different accompanying radiation patterns.}.
    \item The CEM places no constraint on the quantum numbers of the $Q\bar{Q}$ system, and during the transition from $Q\bar{Q}$ pair to $\Q$, the $Q\bar{Q}$ pair radiates soft gluons, which decorrelate the initial- from the final-state quantum numbers. {The CEM, however, requires that the invariant mass of the $Q\bar{Q}$ pair is below the threshold for open heavy-flavour production}.
\end{itemize}  

For a detailed and up-to-date discussion of the successes and failures of these approaches, we guide the reader to a recent review~\cite{Lansberg:2019adr}. We stress that HERA photoproduction data, despite their limited precision, better discriminate between the models above than the very precise LHC hadroproduction data. The reason for such a discriminating power is connected to the different hard-radiation patterns in hadro- and photoproduction, which result in different expectations for the energy and transverse-momentum spectra as discussed in the next section. More photoproduction data should thus be collected wherever possible and the LHC can be used to do so. 

\subsection{Inclusive $\jpsi$ and $\Upsilon$ photoproduction off protons}\label{sec:theory-photoprod}

Contrary to far off-shell photons, quasi on-shell photons can either directly interact with the proton content or  undergo a hadronic fluctuation,\footnote{Such a description effectively amounts to resum collinearly enhanced contributions that appear at higher order in $\alpha_s$ into a non-perturbative distribution of the partonic content of the photon.} which then interacts with the proton content. When discussing photoproduction, one distinguishes between \textit{direct} and \textit{resolved} photons. Resolved photoproduction is a high-energy phenomenon taking place when a small fraction of the photon energy is sufficient to produce the quarkonium, the rest being converted into additional particles. As it is similar to hadroproduction and more complex to describe, kinematic constraints are usually placed to suppress it. At HERA, this was performed using the elasticity of the photon-proton reaction, $z$, which is defined as
\begin{equation}
    z = \frac{P_p \cdot P_\Q}{P_p \cdot P_\gamma},
\end{equation}
where $P_i$ corresponds to the four momentum of $i=p, \Q$, and $\gamma$.
In the proton rest frame, $z$ is the fractional energy of the photon taken by the quarkonium, $E_\Q/E_\gamma$. In exclusive photoproduction in the limit $W_{\gamma p}\gg m_p, m_\pazocal{Q}, \pT$,  $z=  1-P_T^2/W_{\gamma p}^2+{\cal O}\left(p_T^2 m_p^2W_{\gamma p}^{-4}, p_T^2 m_\Q^2W_{\gamma p}^{-4}\right)$, and so, for large $W_{\gamma p}$, $z\simeq 1$. As $z$ deviates from unity, some of the photon energy is used to create additional particles and the relevant partonic scatterings are $2 \to n$, with $n\geq2$. 
Resolved-photon contributions are expected to be maximal at small $z$, and can thus be minimised by requiring $z>z_\text{min}$ for a chosen value $z_\text{min}$. The amount of suppression depends on the assumed quarkonium-production model~\cite{Kramer:2001hh}.

Similar to deep-inelastic--scattering (DIS) processes, the partonic content of the proton probed by inelastic quarkonium-production reactions is described via a parton distribution function (PDF). However, quarkonium production at $z\neq1$ can also feature a rapidity gap, similar to diffractive DIS events~\cite{H1:2006zyl,H1:2007oqt}, and the proton content should then be described via a diffractive PDF (DPDF). To the best of our knowledge, such diffractive photoproduction has never been experimentally studied at HERA. 

Within the CEM, 
the energy dependence of the cross section was computed at NLO (${\cal O}(\alpha \alpha^2_s)$) in 1996~\cite{Amundson:1996qr}. It was found to be compatible with existing fixed-target data. However, the $z$ dependence~\cite{Eboli:2003fr} was later shown to require non-perturbative effects beyond the hadronisation probabilities of the CEM: these effects were then tuned to describe the H1~\cite{H1:1996kyo} and ZEUS~\cite{ZEUS:1997wrc} data. {This hinders global CEM analyses including hadroproduction and photoproduction. For this reason we do not discuss the CEM further.}

Within the CSM, computations of inclusive $\jpsi$ (and $\Upsilon$) photoproduction off protons from direct photons were performed as early as 1981~\cite{Berger:1980ni,Berger:1982fh} at Leading Order (LO) in $\alpha_s$, corresponding to ${\cal O}(\alpha \alpha^2_s)$ (see \cf{fig:CSM-photo}). In 1995, they were extended to Next-to-Leading Order (NLO)~\cite{Kramer:1995nb} (${\cal O}(\alpha \alpha^3_s)$, see \cf{fig:ag3S11gg-FSR-HQ}-\ref{fig:ag3S11g-loop}) and it was found that a class of real-emission NLO QCD corrections is enhanced by a kinematic factor proportional to $\pT^2$, (Figs. \ref{fig:ag3S11gg-ISR}-\ref{fig:aq3S11qg}). The NLO \pT-differential spectrum is thus considerably harder, in agreement with the HERA data~\cite{H1:1996kyo,H1:2002voc,H1:2010udv}. In 2009, these NLO computations were further extended to polarisation observables~\cite{Artoisenet:2009xh,Chang:2009uj} and the cross sections recomputed with different theoretical inputs. The agreement with the HERA data was found to be worse and the CSM was then claimed to fail to describe photoproduction data. The NLO CSM cross sections at HERA were reevaluated along with predictions for the EIC with yet different  theoretical inputs~\cite{Flore:2021rlc}. They were found to describe the latest and most precise H1 data~\cite{H1:2010udv}, having taken into account significant $b$-hadron feed-down contributions at large \pT, which are not part of the prompt CSM computations and therefore were not included in previous calculations.

\begin{figure}[hbt!]
\centering 

\subfloat[]{\includegraphics[scale=.351]{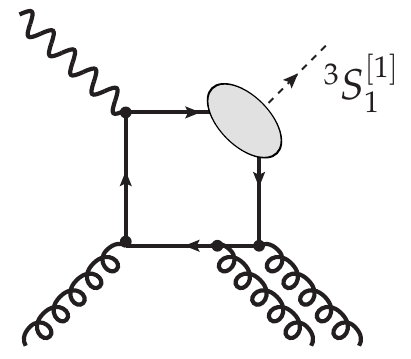}\label{fig:ag3S11gg-FSR-HQ}}
\subfloat[]{\includegraphics[scale=.351]{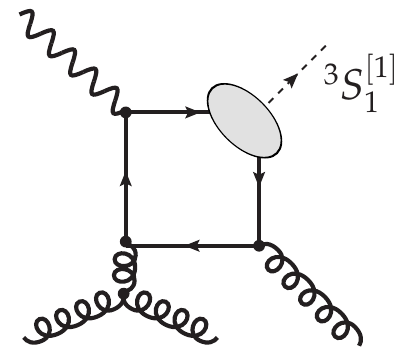}\label{fig:ag3S11gg-ISR}}\\
\subfloat[]{\includegraphics[scale=.351]{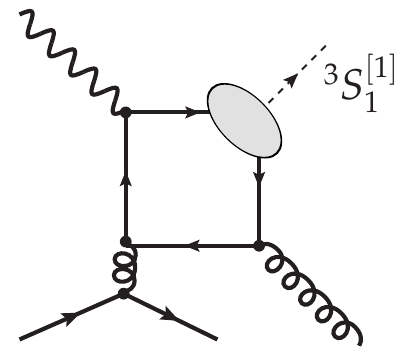}\label{fig:aq3S11qg}}%
\subfloat[]{\includegraphics[scale=.351]{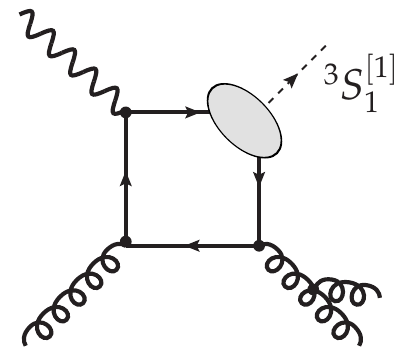}\label{fig:ag3S11gg-FSR-gl-splitting}}
\subfloat[]{\includegraphics[scale=.351]{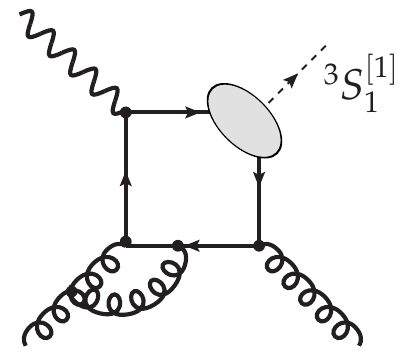}\label{fig:ag3S11g-loop-vertex}}\\
\subfloat[]{\includegraphics[scale=.351]{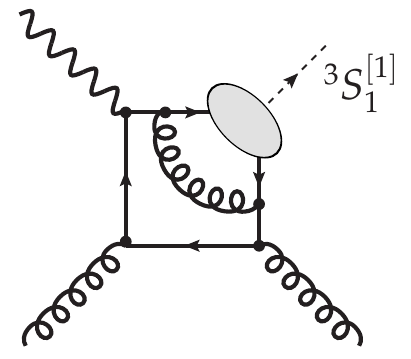}\label{fig:ag3S11g-loop-Coulomb}}
\subfloat[]{\includegraphics[scale=.351]{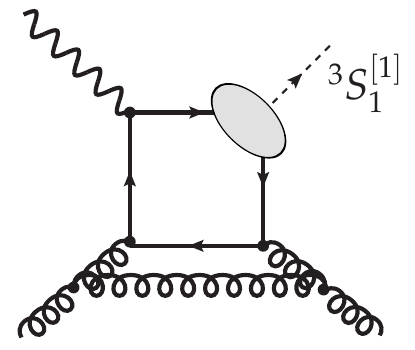}\label{fig:ag3S11g-loop}}\vspace*{-0.15cm}
\caption{Representative Feynman diagrams for inelastic $\Q$ photoproduction contributing via $^3S_1^{[1]}$ CS channels at order $\alpha \alpha_s^3$.}
\label{diagrams-CSM-photoproduction}
\end{figure}

Within the COM, LO cross sections for  inelastic reactions (${\cal O}(\alpha \alpha^2_s)$, see Figs. \ref{fig:CSM-photo}, \ref{gammag-oniumg-LO-inel1}, and \ref{gammag-oniumg-LO-inel2}) were first computed in 1996~\cite{Cacciari:1996dg}. 
With the use of LO NRQCD LDME values close to those compatible with hadroproduction data~\cite{Cho:1995ce}, a peak at $z\to 1$,  typical of $2\to 1$ partonic subprocesses (see  \cf{gammag-oniumg-LO}), was predicted but not seen in HERA data. However, with the inclusion in 2009~\cite{Butenschoen:2009zy} of NLO COM computations (${\cal O}(\alpha \alpha^3_s)$, see Figs.~\ref{COM-NLO1}--\ref{COM-NLO4}), and thanks to a fine tuning of the NRQCD LDME values, it was possible to dampen this peak, bringing the predictions closer to data. The cost of this fine tuning is twofold: first, at large \pT the hadroproduced \jpsi yield is transversely polarised~\cite{Butenschoen:2012px}, which is at odds with Tevatron and LHC hadroproduction data~\cite{Andronic:2015wma,Chapon:2020heu}, and second, there is an overestimate of the $\eta_c$ hadroproduction yield~\cite{Butenschoen:2014dra}. 
Photoproduction is indeed sensitive to different combinations of NRQCD LDMEs and 
comparisons with data bring to light strong tensions that are otherwise only faintly visible in hadroproduction~\cite{Lansberg:2019adr}.

\begin{figure}[hbt!]
\centering
\subfloat[]{\includegraphics[scale=.351,draft=false]{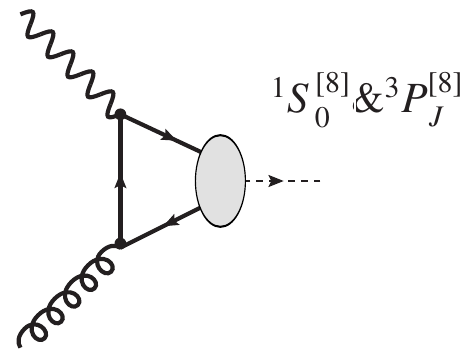}\label{gammag-oniumg-LO}}
\subfloat[]{\includegraphics[scale=.351,draft=false]{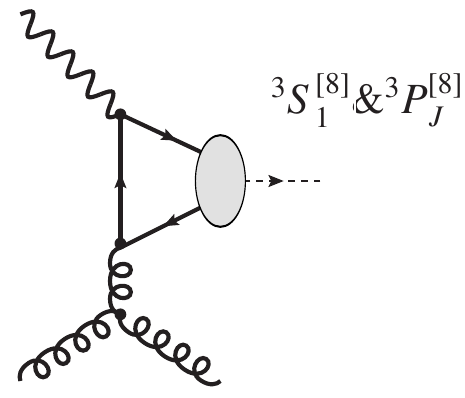}\label{gammag-oniumg-LO-inel1}}\\
\subfloat[]{\includegraphics[scale=.351,draft=false]{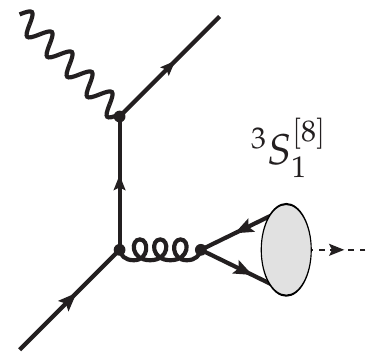}\label{gammag-oniumg-LO-inel2}}
\subfloat[]{\includegraphics[scale=.351,draft=false]{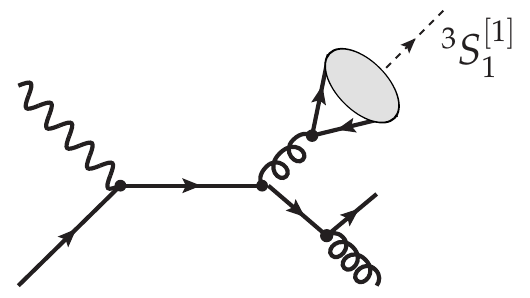}\label{COM-NLO4}}
\subfloat[]{\includegraphics[scale=.351,draft=false]{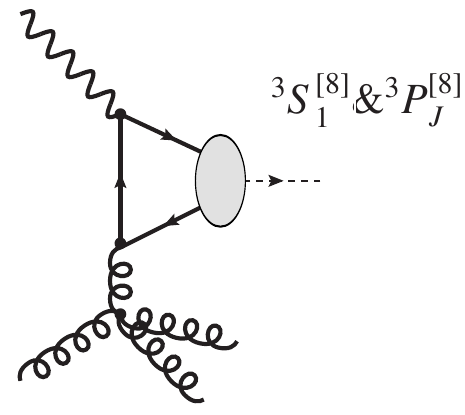}}\\
\subfloat[]{\includegraphics[scale=.351,draft=false]{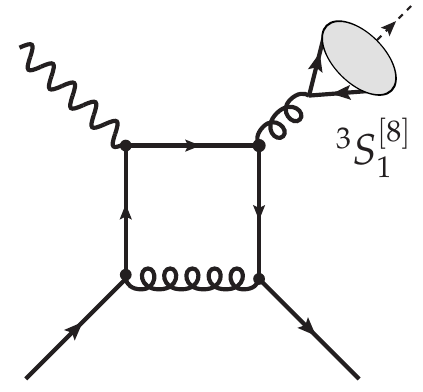}}
\subfloat[]{\includegraphics[scale=.351]{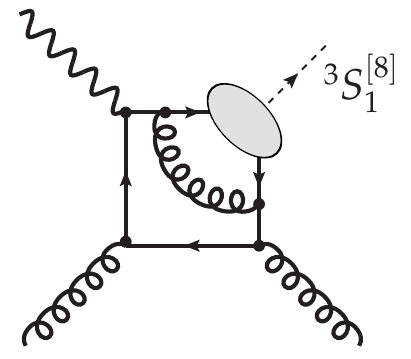}\label{COM-NLO1}}

\caption{Representative Feynman diagrams for $\Q$ photoproduction contributing via $^3S_1^{[8]}$, $^1S_0^{[8]}$, and $^3P_J^{[8]}$ CO channels at orders (a) $\alpha \alpha_s$, (b,c) $\alpha \alpha_s^2$, and (d--g) $\alpha \alpha_s^3$. }
\label{diagrams-COM-photoproduction}
\end{figure}

It is important to note that the cross sections for the $\mathcal{O}(\alpha \alpha_s)$ processes (Fig. \ref{gammag-oniumg-LO}) are non-zero. They are responsible for the peak at $z\to 1$ as well as for a divergence at $\pT\to 0$. This is why when the $\pT$ dependence is studied at Fixed Order (FO) in $\alpha_s$ within NRQCD, as discussed above, a $\pT$ cut is applied to avoid the region where resummation is necessary. Along the same lines, the low-$\pT$ spectrum of a FO computation would be strongly altered by the parton shower and fits of NRQCD LDMEs to data based on a FO computation, would not describe the same data after the inclusion of parton-shower effects~\cite{Cano-Coloma:1997dvl}. 

From the observations described above, it is clear that photoproduction data have discriminanting power and they are complementary to hadroproduction data.
In addition, NLO QCD corrections play a crucial role in the description of important features of the data, in particular its $\pT$ dependence, and should systematically be accounted for. However, the current implementation of quarkonium production in event-generator codes is limited to LO\footnote{However, these can still be used for partial NLO computations in the region where real-emission corrections are dominant~\cite{Flore:2021rlc}, i.e., at large $\pT$.}, which required us to tune our Monte-Carlo (MC) sample to data, as discussed in Section~\ref{sec:signal}. %

\subsection{Inclusive photoproduction of $J/\psi$ and $\Upsilon$ at the LHC}
\label{sec:theory-photoprod-lhc}

Various {predictions}, specifically for inclusive quarkonium photoproduction at the LHC have been performed. 
First, using LO CSM, cross sections were found to be large enough to result in measurable yields of $J/\psi$ and $\Upsilon(1S)$ in $pp$, $p$Pb, and PbPb collision systems at the LHC~\cite{Goncalves:2013ixa}. It was further found that, in $p$Pb collisions at $\sqrt{s_{NN}}=5$~TeV, $10\%$ and $2\%$ of this yield comes from non-dissociative diffraction for $J/\psi$ and $\Upsilon(1S)$, respectively~\cite{Goncalves:2017bmo}.

Another study~\cite{Tichouk:2019lxk}, using NRQCD at LO, investigated the prospect of using forward proton spectrometers, such as TOTEM at CMS~\cite{CMS:2022hly} and AFP at ATLAS \cite{Adamczyk:2015cjy}, to tag the intact photon emitter. However, contrary to what is suggested by this study, such detectors are incapable of tagging lead ions, which are negligibly deflected by the photon emission. Consequently, this proposal is only relevant for $pp$ collisions, which is experimentally hindered by pile-up at high luminosity interactions such as expected by CMS and ATLAS during Run 3 and 4. The use of proton spectrometers to tag inclusive photoproduction events may require special run conditions with reduced pile-up. It was found, in~\cite{Tichouk:2019lxk}, using a nominal acceptance of $0.0015<x_\gamma<0.5$ (resp. $0.015<x_\gamma<0.15$) for the TOTEM (resp. AFP) detector, that $50\%$ (resp. $5\%$) of the inclusive $pp$ \jpsi yield can be tagged. We note that for $x_\gamma\gtrsim 0.1$ the impact parameter of the collision is close to 2~fm{ and}
{the photoproduction and hadronic cross sections could become similar.}
{Thus,} this large $x_\gamma$ region should be vetoed to obtain a clean photoproduction sample. Such a veto requires a good photon-energy resolution for $x_\gamma>0.1$, which appears possible according to simulations of these detectors~\cite{CMS:2022hly,Adamczyk:2015cjy}.

\subsection{Specific backgrounds to inclusive photoproduction in $p$Pb collisions}\label{sec:theory-background-lhc}
To realistically evaluate the feasibility of measuring inclusive photoproduction, background processes need to be considered. We discuss here hadroproduction, which is by far the dominant background, diffractive production, and the feed-down contribution from excited states.

The cross section for quarkonium hadroproduction, which proceeds through the exchange of quarks and gluons between colliding hadrons, as shown in \cf{fig:CSM-hadro}, is orders of magnitude larger than that for quarkonium photoproduction. 
The final-state particles resulting from the partonic interaction for both processes are identical; the difference lies in the emission of a photon versus a parton from one of the colliding hadrons, the identification of which is the key to  measuring inclusive photoproduction.

At LO, the $2\to2$ partonic process for photo- and hadroproduction occurs at $\alpha \alpha_s^2$ and $\alpha_s^3$, respectively. %
Hence, photoproduction is suppressed by a factor $\alpha/\alpha_s$ with respect to hadroproduction. 
Additionally, at high energies, photoproduction is further suppressed with respect to hadroproduction at large \pT.
This is because, at LO, hadroproduction exhibits gluon-induced fragmentation\footnote{We recall that, at large \pT, fragmentation contributions are enhanced up to $\pT^4$ compared to other LO contributions.} to $^3S_1^{[8]}$ states
(\cf{diagram-COM-PT-c}), which is favoured due to the large gluon PDF at low $x$, whereas photoproduction is restricted to quark-induced fragmentation (\cf{gammag-oniumg-LO-inel2}).

\begin{figure}[bt!]
\centering
\subfloat[]{\includegraphics[scale=.35,draft=false]{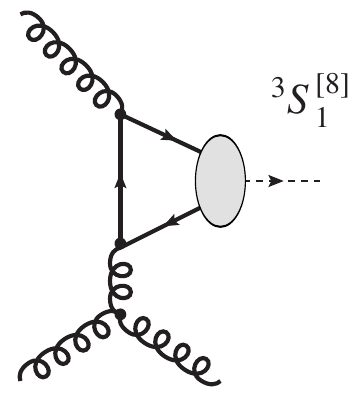}\label{diagram-COM-PT-b}}%
\subfloat[]{\includegraphics[scale=.35,draft=false]{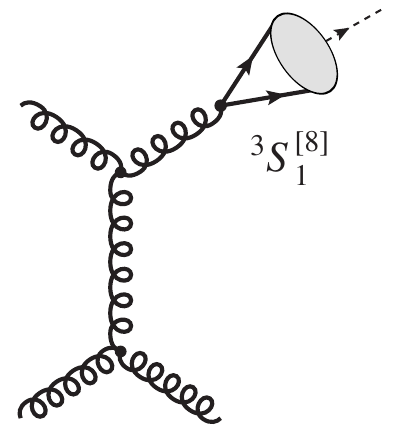}\label{diagram-COM-PT-c}}%
\caption{Representative Feynman diagrams for $\Q$ hadroproduction contributing via the $^3S_1^{[8]}$ CO channel at order $\alpha^3_s$.} 
  \label{fig:COM-hadro}
\end{figure}

Vector-quarkonium production in $p$Pb collisions can also proceed through pomeron-odderon exchange~\cite{Bzdak:2007cz}. This type of diffractive background could in principle leave one hadron intact but owing to its hadronic nature would occur at small impact parameters and is unlikely to leave the lead ion intact.

The feed-down contribution from the decay of heavier particles, both prompt and non-prompt, is a further source of background. 
The non-prompt component, from the decay of $b$ hadrons, can be identified using lifetime information from vertex detectors at the four LHC experiments {,} whereas the feed-down decay from quarkonium-excited states is usually inferred from the available cross-section measurements of these excited states and from their branching to the lower lying state.
Prompt decays, depending on \pT, are estimated to account for 20--40\% and 30--60\% of the \jpsi and $\Upsilon(1S)$ yields, respectively~\cite{Lansberg:2019adr}. In photoproduction the only expected feed-down contribution is from the decays of radial excitations, $n'S$.

Like for photoproduction, hadroproduction is sensitive to higher-order QCD corrections~\cite{Lansberg:2019adr} and parton-shower effects~\cite{Cano-Coloma:1997dvl}. In addition, there are no MC tools available at NLO accuracy for hadroproduction. 
This necessitates tuning our MC sample to data, which will be  discussed in Section~\ref{sec:background}.

\begin{table*}[hbt!]
 \centering
  \caption{ Kinematic coverage in ALICE \cite{ALICE:2020eji,ALICE:2021qlw}, ATLAS \cite{ATLAS:2016ydt,ATLAS:2017prf}, CMS \cite{CMS:2017exb,CMS:2010nis,CMS:2023lfu}, and LHCb \cite{LHCb:2018rcm,LHCb:2018yzj} for $J/\psi$ and $\Upsilon$ reconstruction via dimuon decay (and dielectron decay for $J/\psi$ in ALICE \cite{Lofnes:2019jns}) as well as requirements placed on the momenta of particles reconstructed in the different pseudorapidity regions for ALICE \cite{ALICE:2014sbx}, ATLAS \cite{ATLAS:2012djz}, CMS \cite{CMS:2023lfr}, and LHCb \cite{LHCb:2014set}, where the particles are reconstructed using calorimeters (cal) or are charged (ch) and reconstructed using tracking detectors. The transverse momentum \pT, the total momentum, $|p|$, $y$, and $\eta$ are given in the laboratory frame.}
\resizebox{\textwidth}{!}{\begin{tabular}{cccccc}

 \multicolumn{2}{c}{\textbf{ALICE} } & \textbf{ATLAS}&  \multicolumn{2}{c}{\textbf{CMS} }   & \textbf{LHCb} \\
  \hline
  \multicolumn{6}{c}{Kinematic constraints for $J/\psi$ reconstruction} \\   \hline
$|y^{J/\psi }|<0.9$ & $2.5<y^{J/\psi }<4.0$& $|y^{J/\psi }|<2.1$& $|y^{J/\psi }|<2.1$& $\pT^{ J/\psi } > 6.5 $~GeV for $|y^{J/\psi }|<1.2$& $2.0<y^{J/\psi }<4.5$ \\
&&$P^{J/\psi }_T > 8.5$~GeV &$\pT^{ J/\psi } > 6.5 $~GeV&$P^{ J/\psi }_T > 2$~GeV for $1.2 < |y^{  J/\psi }| < 1.6$  \\
  &&&  &$ \pT^{ J/\psi }> 0$~GeV for $1.6< |y^{J/\psi}| < 2.4$\\
  \hline \multicolumn{6}{c}{Kinematic constraints for $\Upsilon$ reconstruction} \\   \hline
  \multicolumn{2}{c}{$2.5<y^{\Upsilon }<4.0$}&$|y^{\Upsilon }|<2.0$& \multicolumn{2}{c}{$|y^{\Upsilon }|<2.4$} & $2.0<y^{\Upsilon }<4.5$ \\
    \hline   \multicolumn{6}{c}{Kinematic constraints on particle reconstruction} \\   \hline
 \multicolumn{2}{c}{$|\eta_\text{ch}|<0.8$ \quad $P_{T\,\text{ch}}>$0.2~GeV} &$|\eta_\text{ch}|<2.5$\quad $P_{T\,\text{ch}}>$0.2~GeV&  \multicolumn{2}{c}{ $|\eta_\text{ch}|<2.5$\quad $P_{T\,\text{ch}}>$0.2~GeV} &$2<\eta_\text{ch}<5$ \quad $|p_\text{ch}|>5$~GeV\\  
 & &$2.5<|\eta_\text{cal}|<4.9$\quad $P_{T\,\text{cal}}>$0.2~GeV  &  \multicolumn{2}{c}{ $2.5<|\eta_\text{cal}|<5.2$\quad $P_{T\,\text{cal}}>$0.2~GeV} & \\  
    \hline
 \end{tabular}}

       \label{Acc}
\end{table*}

\section{Simulation set-up}
\label{sec:simulation}
In order to assess the feasibility of measuring inclusive photoproduction at the LHC, it must be shown that large hadronic backgrounds can be significantly suppressed, while limiting signal reduction, in a model-independent way. We have built %
MC samples for the description of 
both the photoproduction signal and hadroproduction background, including the hadronic particle activity. These are discussed in Sections \ref{sec:signal} and \ref{sec:background}.

\begin{table*}[hbt!]
        \centering
 \caption{Proton-lead luminosity for Run 2 data recorded by ALICE \cite{ALICE:2020tsj}, ATLAS \cite{ATLAS:twiki}, CMS \cite{CMS:twiki}, and LHCb \cite{LHCbTwiki} as well as Run 3 and Run 4 targets \cite{LPCreport}, where $p$Pb and Pb$p$ correspond to the different beam configurations. }

         \begin{tabular}{c|cccccc}

       &\multicolumn{2}{c}{\textbf{ALICE}} &\textbf{ATLAS} &\textbf{CMS} &\multicolumn{2}{c}{\textbf{LHCb}} \\ 
       &$p$Pb & Pb$p$ &$p$Pb + Pb$p$ &$p$Pb + Pb$p$& $p$Pb & Pb$p$\\ \hline
        Run 2 (nb$^{-1}$)& 8.4&12.8&180&180 &12.5 &17.4 \\
        Runs 3 \& 4 (pb$^{-1}$)& \multicolumn{2}{c}{0.5 } & 1  & 1  &\multicolumn{2}{c}{0.2 } \\

 \end{tabular}
 \label{lumi}
\end{table*}
We focus on vector quarkonia, \jpsi, $\psi(2S)$, $\Upsilon(1S)$, $\Upsilon(2S)$, and $\Upsilon(3S)$, decaying to dimuons,
which offer a clean experimental reconstruction. 
Photoproduction yields are presented in Section \ref{sec:signal}.
The kinematic acceptance for \jpsi and $\Upsilon$ reconstructed within the four main LHC detectors and the existing and forecast data-taking luminosities are presented in Tables \ref{Acc} and \ref{lumi}. %
A positive rapidity is assigned to the proton direction. Because of the forward muon acceptance of the ALICE and LHCb detectors, both beam configurations are considered separately: one where the proton flies into the detector ($p$Pb) and the other where the lead ion flies into the detector (Pb$p$)\footnote{{For CMS and ATLAS, we only consider the $p$Pb beam configuration. Small differences are expected between the $p$Pb and Pb$p$ configurations due to the shift in rapidity between the centre-of-mass and laboratory frame ($\Delta y = 1/2\ln(Z/A)$), but we neglect them.}}.
In the latter case the acceptance given in Table~\ref{Acc} is quoted with opposite rapidity.
For CMS, two different acceptances are considered: one, in which all \jpsi mesons are collected above a \pT threshold; and the other, where this threshold decreases with increasing $|y^{\jpsi}|$, reaching 0 for $1.6<|y^{\jpsi}|<2.4$.

\subsection{Monte-Carlo simulation of inclusive quarkonium photoproduction at the LHC}\label{sec:signal}

\begin{figure*}[hbt!]
    \centering
    \subfloat[]{\label{fig:photoTune}\includegraphics[scale=0.47]{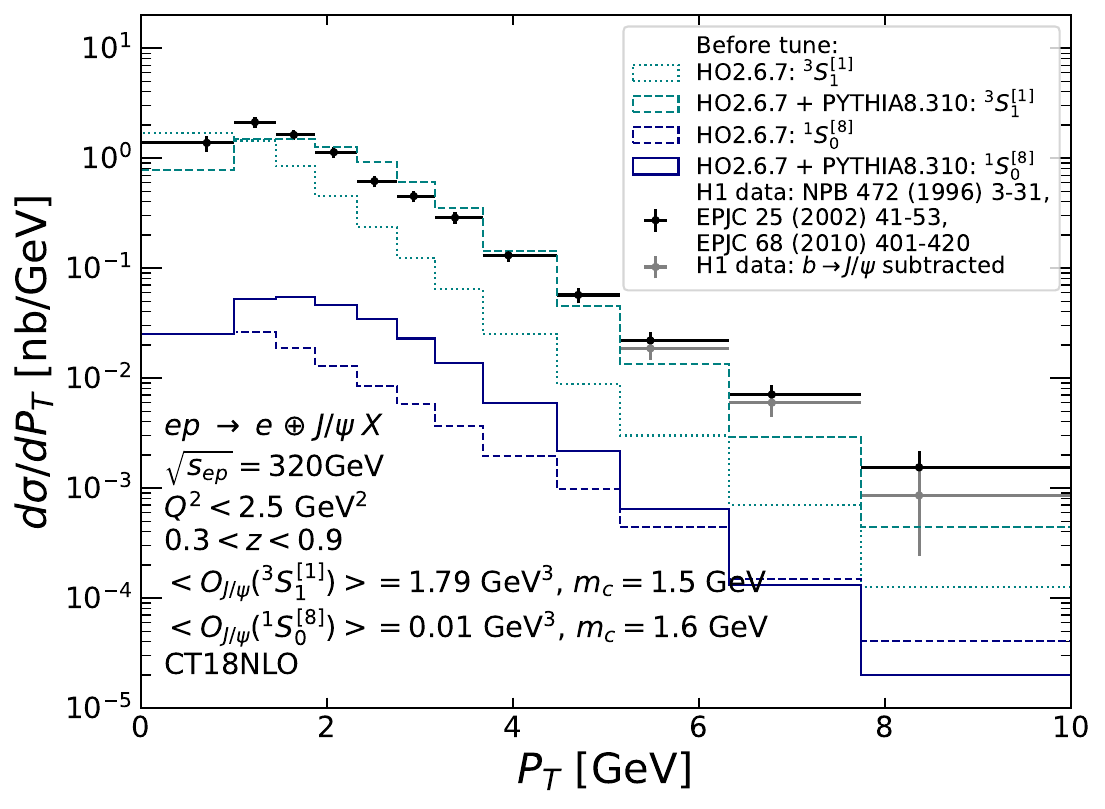}}
    \subfloat[]{\label{fig:photoVal}\includegraphics[scale=0.47]{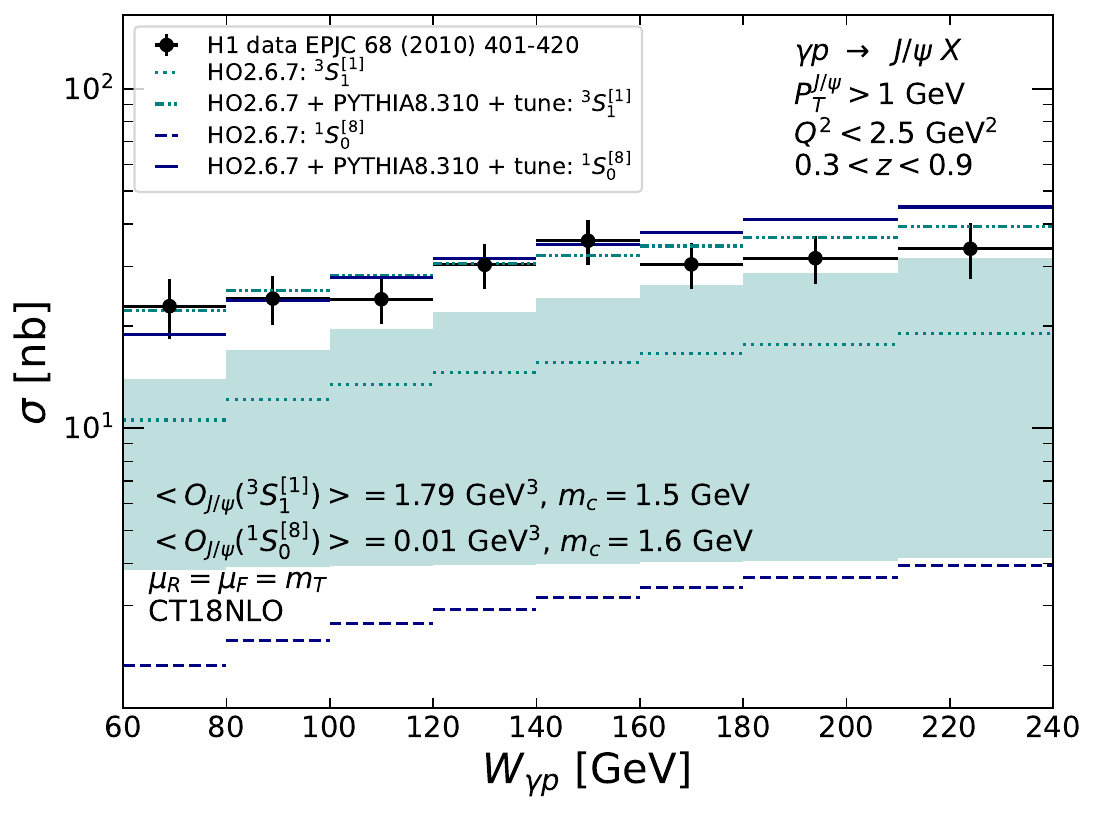}}
    \caption{
    {(a) \pT-differential cross section for \jpsi photoproduction computed at LO for the $^3S_1^{[1]}$ (teal) and $^1S_0^{[8]}$ (navy blue) states with the quoted LDME values with (\texttt{HO2.6.7 + PYTHIA8.310}) and without (\texttt{HO2.6.7}) parton shower compared to H1 data \cite{H1:1996kyo,H1:2002voc,H1:2010udv} (black) and with $b$ subtraction (grey). (b) \pT-integrated \jpsi photoproduction cross section as a function of $W_{\gamma p}$ computed at LO for the $^3S_1^{[1]}$ and $^1S_0^{[8]}$ states without parton shower (teal, dotted and navy-blue, dashed lines), and with parton shower and the tune to H1 data \cite{H1:2010udv} ((teal, dot-dashed and navy-blue, solid lines) compared to the H1 data. The teal band is the scale uncertainty of the LO $^3S_1^{[1]}$ cross section before the tune and without parton shower (see text).}}
    \label{fig:photoprod-validation}
\end{figure*}

The only public code to generate MC events for inclusively photoproduced quarkonium is \texttt{HELAC-Onia} (\texttt{HO})~\cite{Shao:2012iz,Shao:2015vga}, which is a parton-level matrix-element generator capable of performing LO computations within the NRQCD framework in colliding systems of electrons, protons, and their antiparticles. It can be interfaced to \texttt{PYTHIA}~\cite{Bierlich:2022pfr,Sjostrand:2014zea} for parton-shower and hadronisation effects via the LHE format~\cite{Alwall:2006yp}.
For our simulation we use \texttt{HO} interfaced to \texttt{PYTHIA}.
\subsubsection{\jpsi photoproduction}\label{sec:MCjpsiphotoproduction}
As discussed in Section~\ref{sec:theory-photoprod}, %
LO NRQCD does not capture the \pT distribution of photoproduced \jpsi. This is apparent in Fig. \ref{fig:photoTune}, where LO cross sections computed with \texttt{HO} for the colour singlet $^3S_1^{[1]}$ and the colour octet $^1S_0^{[8]}$ states are compared to the H1 data: the former distribution (dotted, teal)
is too steep and the latter (dashed, navy blue) is too flat. 
To get a good agreement, one would need to combine both contributions %
and to change their normalisation, which amounts to changing the LDMEs.

By interfacing \texttt{HO} to \texttt{PYTHIA} (\texttt{HO+PYTHIA}), one indirectly accounts for some radiative QCD corrections, which alter the \pT distributions. The \texttt{HO+PYTHIA} \pT distributions are also shown in \cf{fig:photoTune} with the $^3S_1^{[1]}$ state in solid, teal and  the $^1S_0^{[8]}$ state in dot-dashed, navy blue. {The $\oplus$ symbol implies a rapidity separation between the photon emitter and photoproduced system, \jpsi X}. 
The
difference between \texttt{HO} and \texttt{HO+PYTHIA} \pT spectra can be as large as a factor 6 (4) for the $^3S_1^{[1]}$ ($^1S_0^{[8]}$) state. %
If we had combined contributions and fit LDMEs at the \texttt{HO}-level, which is the common procedure for FO analysis, the same LDMEs could not be used at \texttt{HO+PYTHIA} level to describe the data. 

We highlight two kinematic effects that are relevant for the simulation of photoproduction when generating events at the level of hadrons. Firstly, the integrated cross section for $\pT >1$ GeV and $0.3<z<0.9$ differs for the \texttt{HO} and \texttt{HO}+\texttt{PYTHIA} results. This is due to the fact that the $z$ distribution for both the $^3S_1^{[1]}$ and $^1S_0^{[8]}$ states is peaked at large $z$ and this strong peak in $z$ is smeared by {the} \texttt{PYTHIA} {parton shower}. Secondly, as discussed in Section \ref{sec:theory-photoprod}, to avoid the CO endpoint singularities, we imposed $P^{J/\psi}_T>1$~GeV at the parton level for $^1S_0^{[8]}$. Thus, the bin $0<P^{J/\psi}_T<1$~GeV is empty for the \texttt{HO} result and is filled entirely by the \texttt{PYTHIA} parton shower for the \texttt{HO}+\texttt{PYTHIA} result.

The fact that neither the \texttt{HO} nor the \texttt{HO}+\texttt{PYTHIA} simulations using individual $^3S_1^{[1]}$ or $^1S_0^{[8]}$ contributions describe the experimental data is not particularly problematic since  the MC samples can be tuned to the \pT distribution of existing HERA data. This tuning is then extrapolated to the LHC photoproduction conditions. 

We tune the $^3S_1^{[1]}$ and  $^1S_0^{[8]}$ cross sections individually and generate two corresponding MC samples that exhibit different invariant-mass distributions of \jpsi and the recoiling parton, $M_{J/\psi g}$. The amplitude squared $|\mathcal{M}_{\gamma g \rightarrow ^3S_1^{[1]} g}|^2$ scales like $M_{J/\psi g}^{-4}$, whereas $|\mathcal{M}_{\gamma g \rightarrow ^1S_0^{[8]} g}|^2$  scales like $M_{J/\psi g}^{-2}$.
As a consequence, the particle activity is expected to be more spread out in phase space for the $^1S_0^{[8]}$ sample, and can even extend down to rapidities close to the photon emitter, especially when the quarkonium is produced with backward rapidity.  

Tune factors are fit bin-by-bin in \pT and map the \texttt{HO}+\texttt{PYTHIA} result to a combined fit of the H1 data~\cite{H1:1996kyo,H1:2002voc,H1:2010udv}\footnote{{As in the global-analysis fit of LDMEs performed in~\cite{Butenschoen:2010rq}, we perform our tune to a combined dataset. Data are combined by taking an inverse-variance weighted average per bin, which effectively minimises the $\chi^2$ between the dataset as a whole and our tuned MC. The data used satisfies: $0<\pT<1$~GeV, $30<W_{\gamma p}<150$~GeV, and $z<0.9$~\cite{H1:1996kyo}; $1.0<\pT<7.7$~GeV, $60<W_{\gamma p}<240$~GeV, and $0.3<z<0.9$~\cite{H1:2002voc}; and $1<\pT<10$~GeV; $60<W_{\gamma p}<240$~GeV, and $0.3<z<0.9$~\cite{H1:2010udv}. We neglect the slight difference in coverage of $W_{\gamma p}$ and $z$ within the data.}}. Following ~\cite{Flore:2020jau}, we subtract the expected $b$--feed-down contribution in the three largest \pT bins, see grey points in Fig. \ref{fig:photoTune}.  The \pT of photoproduced \jpsi in H1 data is limited to values below 10~GeV. For \pT$>10$~GeV, the tune factor is assumed to be  $a \times \pT$ for both $^3S_1^{[1]}$ and $^1S_0^{[8]}$ states, where $a$
is fit in the range $5.2< \pT<10.0$~GeV. The {function} 
used for this extrapolation is identical 
for both samples due to parton-shower effects\old{,}{;} however, this would not be the case if using parton-level \texttt{HO} results. Since the bulk of the cross section is located at $\pT<10$~GeV, this extrapolation region will not affect the total yields.
The resulting $^3S_1^{[1]}$ and $^1S_0^{[8]}$ tunes are reported in \ref{appendix:photoprod}. The tuned $^3S_1^{[1]}$ and $^1S_0^{[8]}$ cross sections{,} by definition{,} directly overlap with the data in \cf{fig:photoTune} and thus are not shown.

In our tuning procedure, we assume that {the energy and} longitudinal-momentum {distributions}, which are driven by the proton PDF and the photon flux, are
{correctly accounted for and should not be tuned}, unlike the \pT distribution. This is justified in \cf{fig:photoVal}, which shows a comparison with H1 data~\cite{H1:2010udv} of the cross section as a function of $W_{\gamma p}$ using \texttt{HO} and our tune for both $^3S_1^{[1]}$ and $^1S_0^{[8]}$ states.
The energy dependence is reasonably well described using our tune and the CT18NLO PDF~\cite{Hou:2019qau}. {Comparable agreement is found in \cite{Goncalves:2013ixa}, where the $^3S_1^{[1]}$, using a multiplicative $K$ factor to account for higher-order corrections, is compared to the same H1 data.}

\begin{figure}[hbt!]
    \centering
    \subfloat[]{\label{fig:CMSpTa}\includegraphics[width = 0.75\columnwidth]{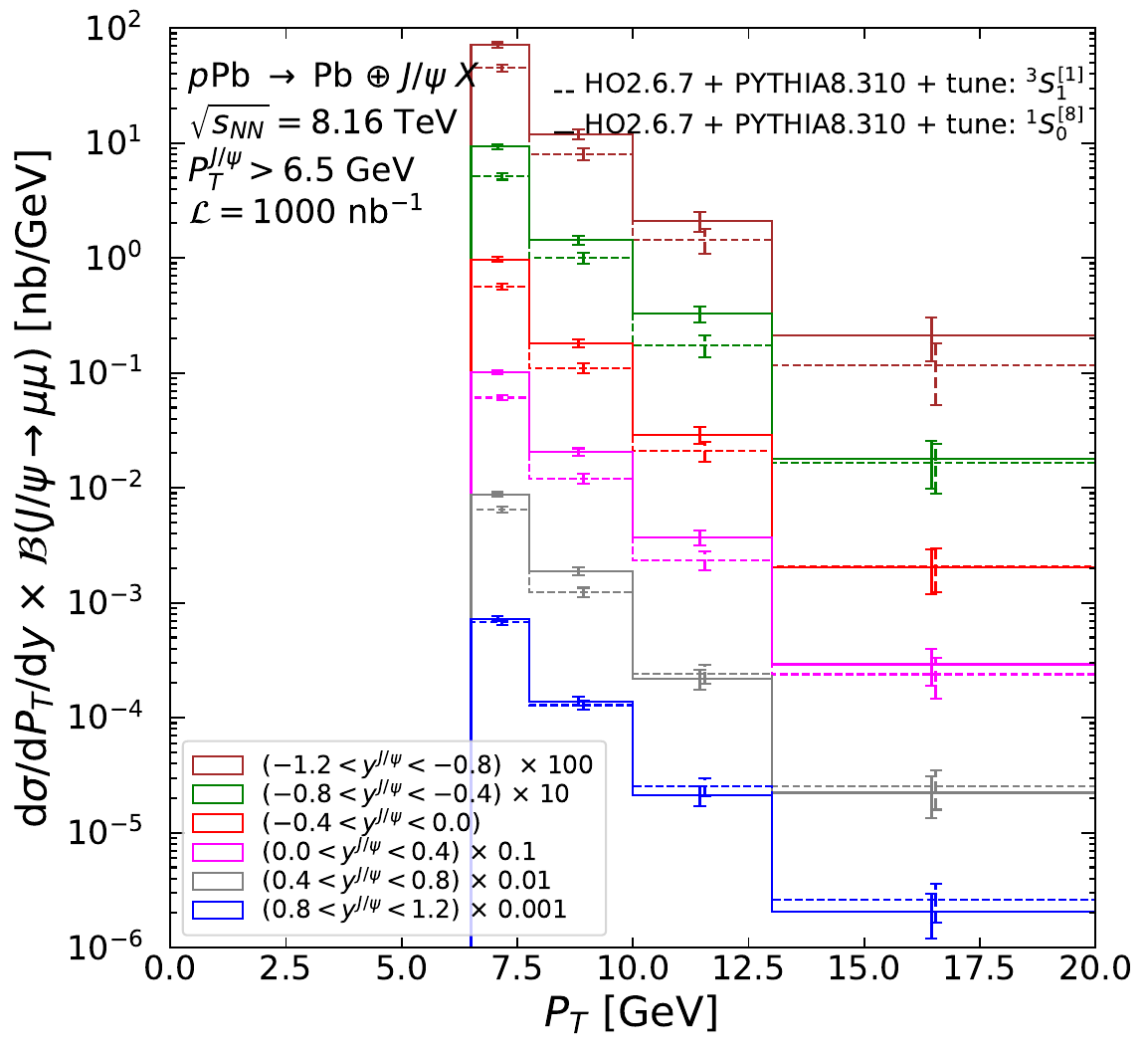}}\\\vspace*{-0.2cm}
    \subfloat[]{\label{fig:CMSpTb}\includegraphics[width = 0.75\columnwidth]{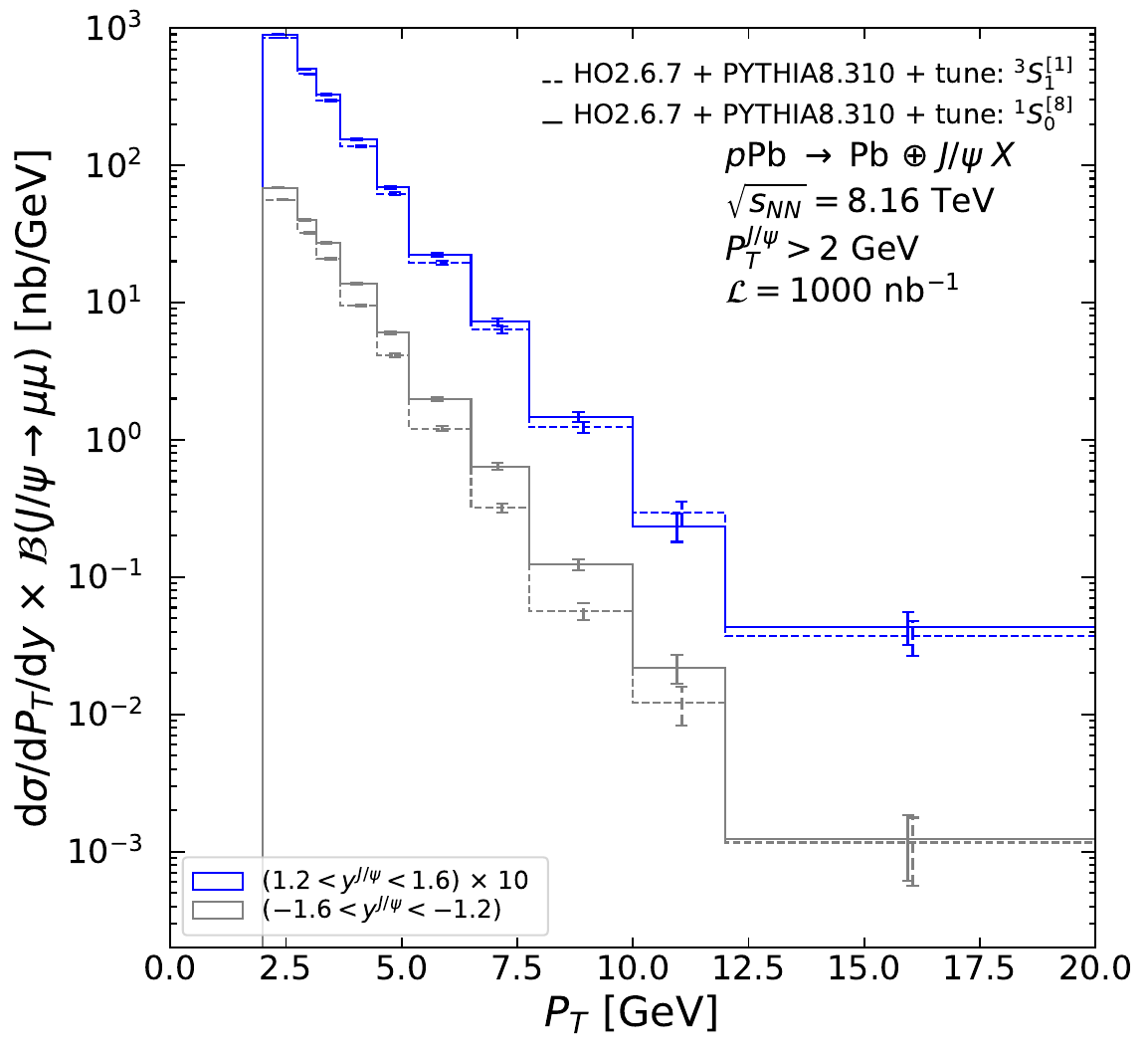}}\\\vspace*{-0.2cm}
    \subfloat[]{\label{fig:CMSpTc}\includegraphics[width = 0.75\columnwidth]{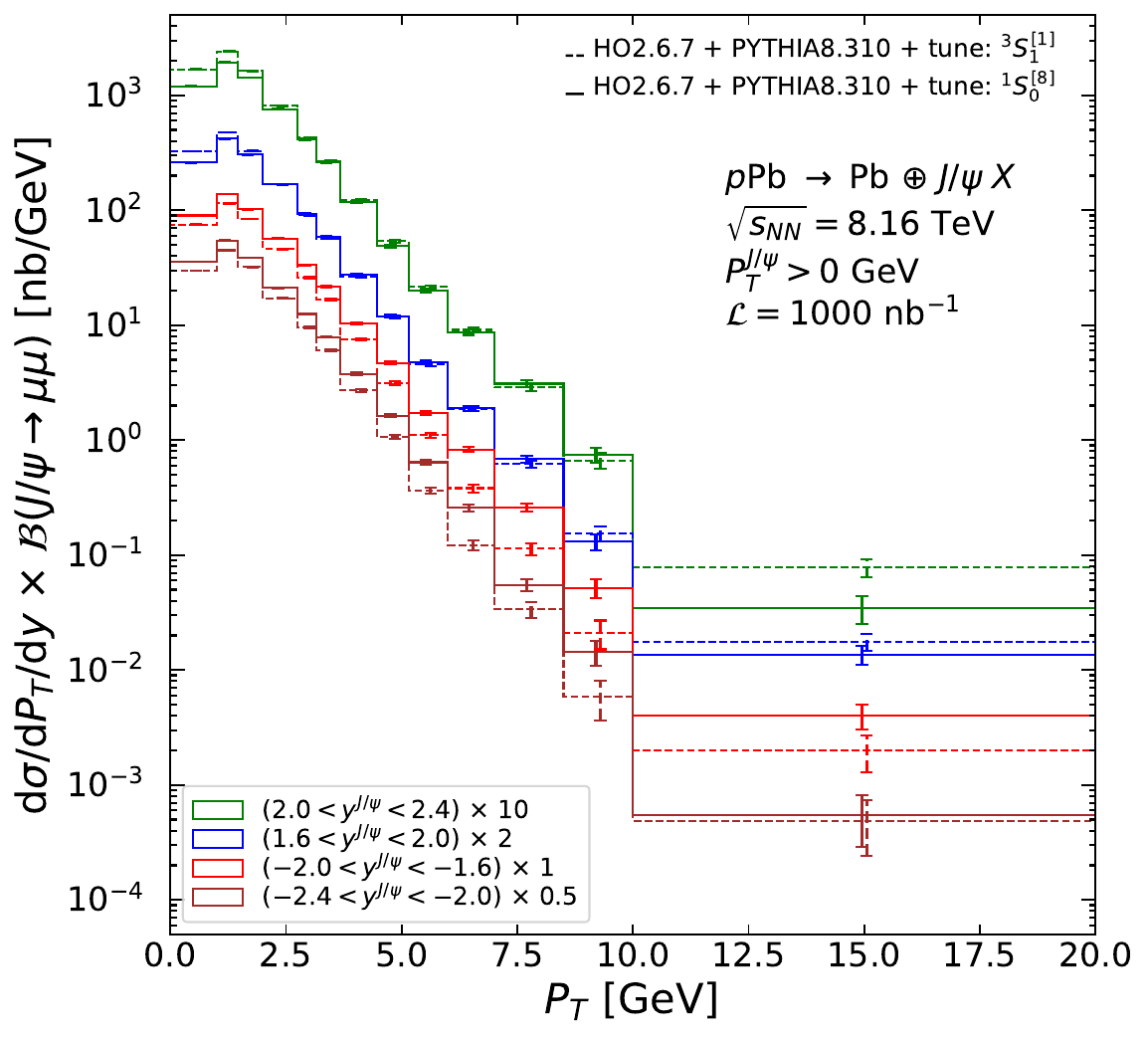}}\vspace*{-0.2cm}
    \caption{{Double-differential cross section, in \pT and $y$, times the branching fraction to dimuons, as a function of \pT} using the $^1S_0^{[8]}$ (solid) and $^3S_1^{[1]}$ (dashed) tunes,
    for photoproduced \jpsi in the CMS acceptance in the rapidity region: (a) $|y^{J/\psi}|<1.2$, (b) $1.2<|y^{J/\psi}|<1.6$, and (c) $1.6<|y^{J/\psi}|<2.4$. The error on the cross section is the signal statistical uncertainty assuming an integrated luminosity of 1000~nb$^{-1}$.}
    \label{fig:CMSpt}
\end{figure}

\begin{figure}[hbt!]
    \centering
    \includegraphics[trim = 0cm 0cm 0cm 0cm,clip,height=0.34\textwidth]{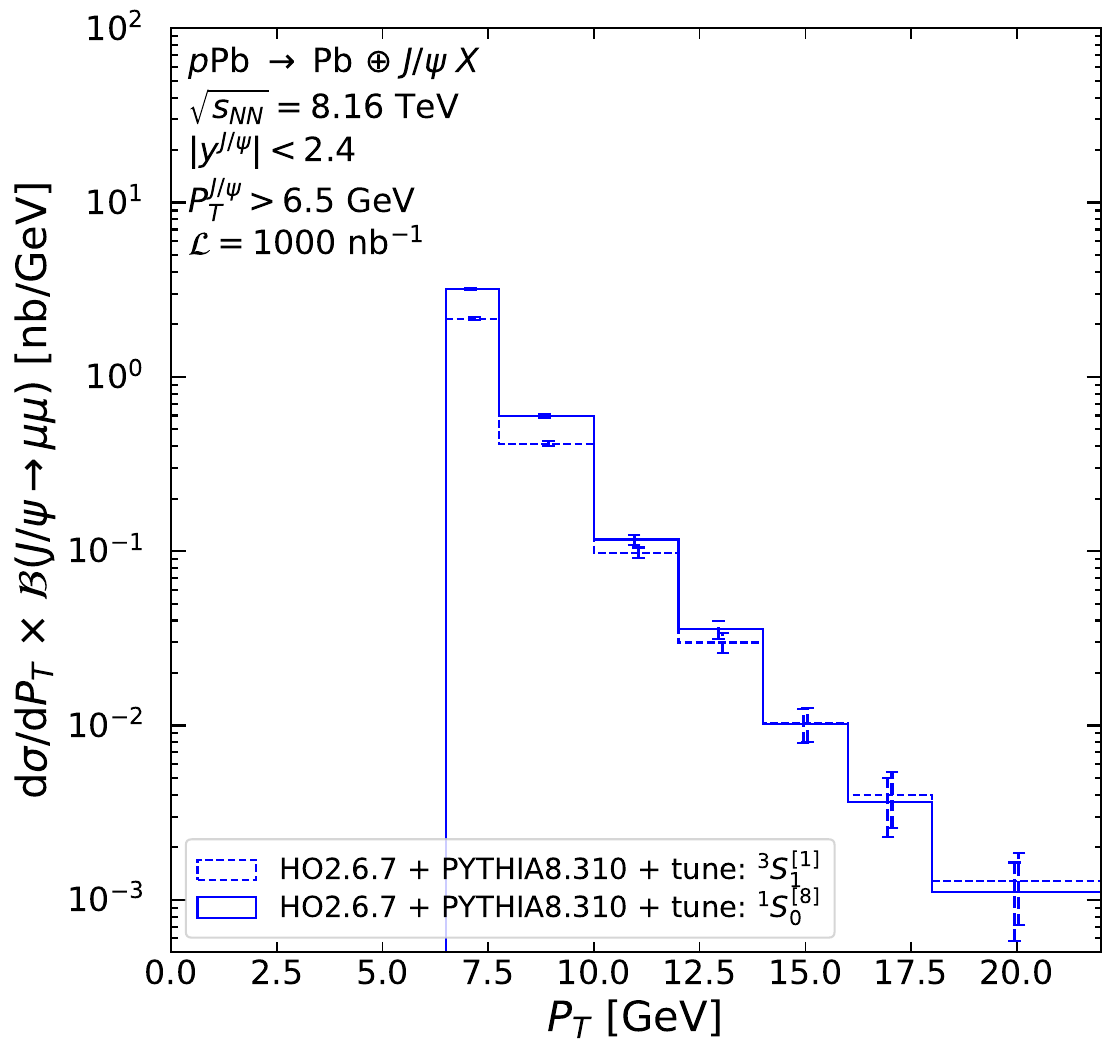}
    \caption{\pT-differential cross section times the branching fraction of \jpsi to dimuons using the $^1S_0^{[8]}$ (solid) and $^3S_1^{[1]}$ (dashed) tunes, for photoproduced \jpsi in the entire rapidity coverage of CMS. The error on the cross section is the signal statistical uncertainty assuming an integrated luminosity of 1000~nb$^{-1}$}
    \label{fig:CMSpt-all}
\end{figure}

\begin{figure}[hbt!]
    \centering
    \subfloat[]{\label{fig:LHCbpPb}\includegraphics[trim = 0cm 0cm 0cm 0cm,clip,height=0.34\textwidth]{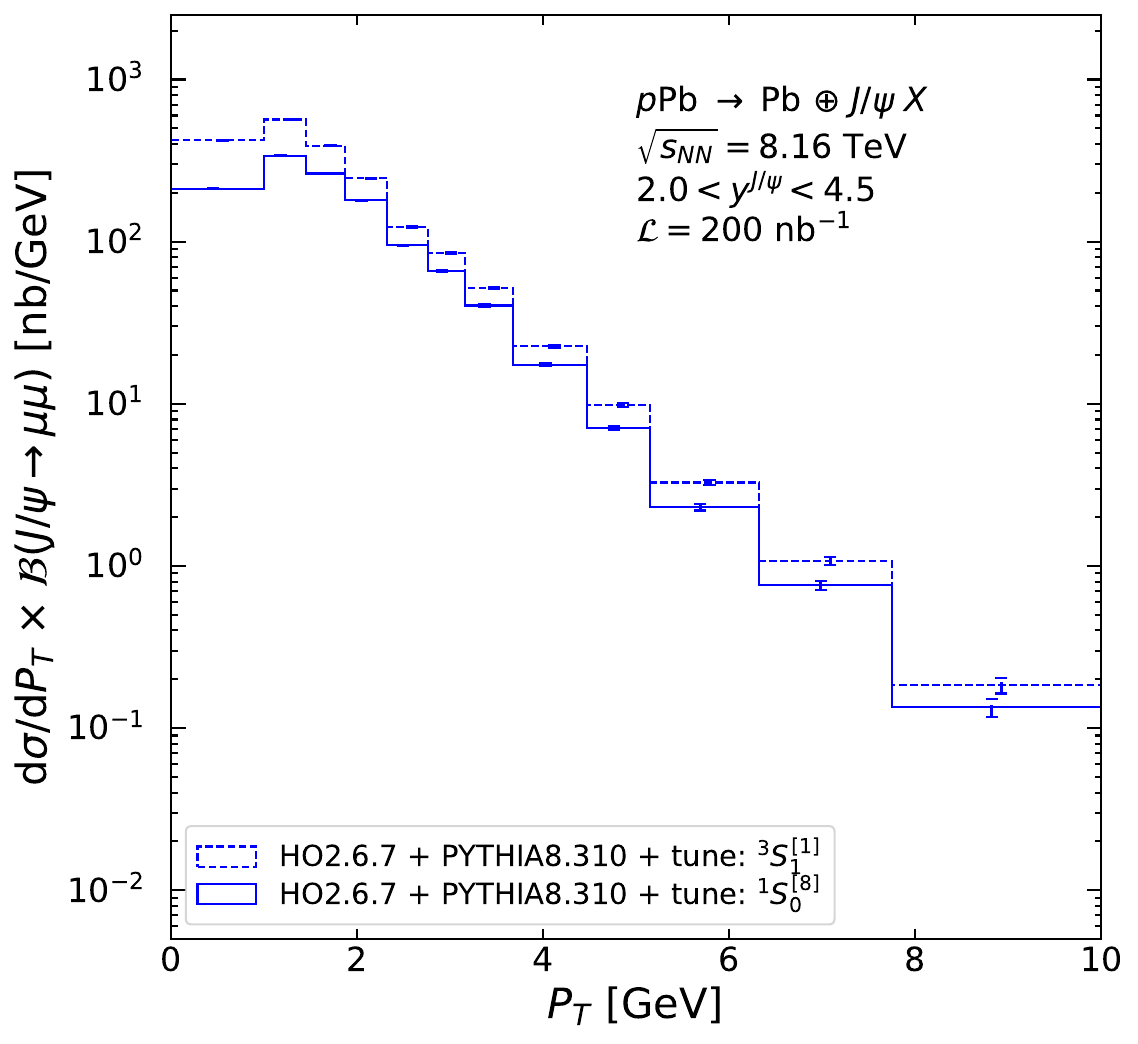}}\\
    \subfloat[]{\label{fig:LHCbPbp}\includegraphics[trim = 0cm 0cm 0cm 0cm,clip,height=0.34\textwidth]{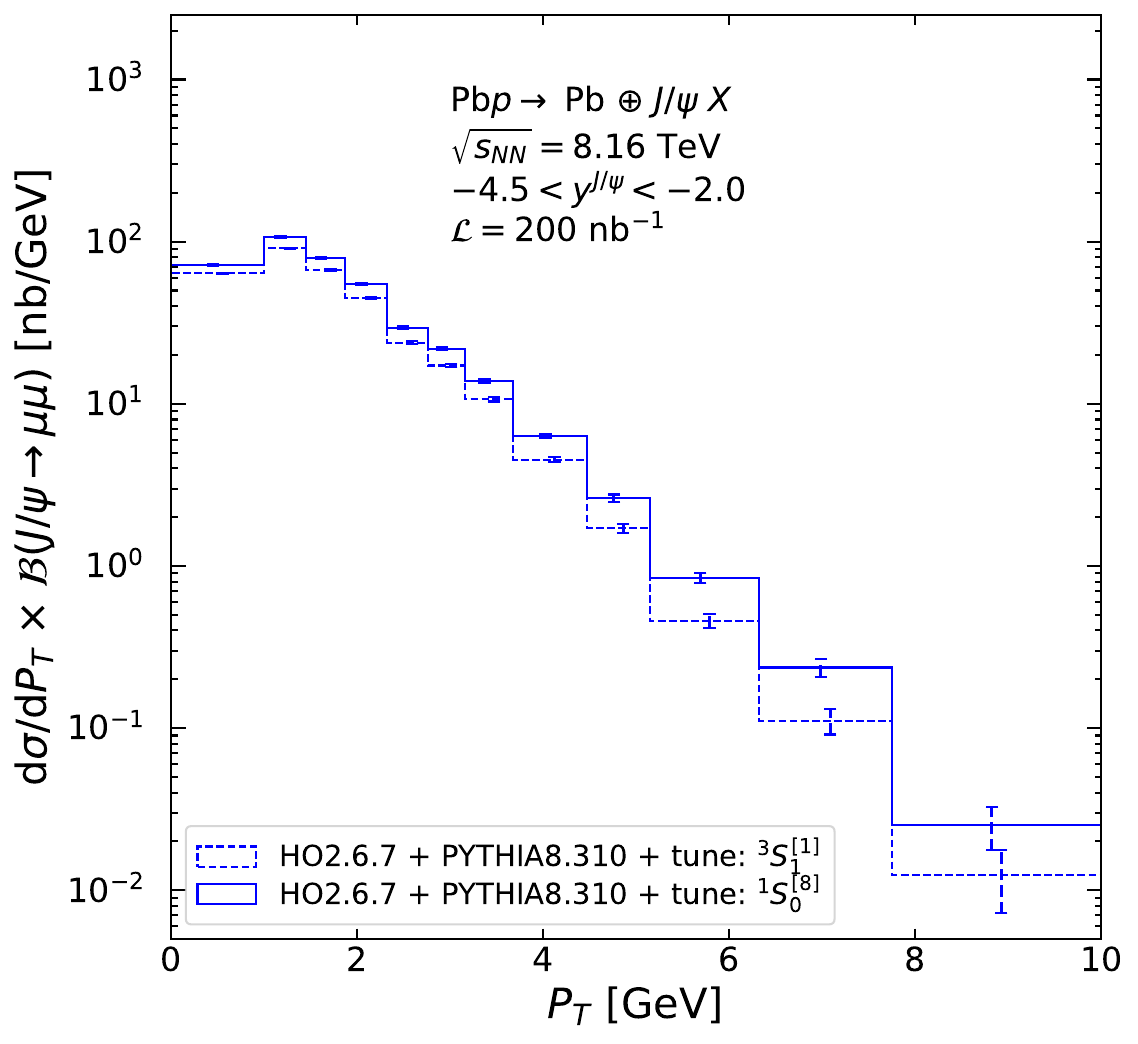}}
    \caption{\pT-differential cross section times the branching fraction of \jpsi to dimuons using the $^1S_0^{[8]}$ (solid) and $^3S_1^{[1]}$ (dashed) tunes, for photoproduced \jpsi in the LHCb acceptance in the (a) $p$Pb and (b) Pb$p$ beam configurations. The error on the cross section is the signal statistical uncertainty assuming an integrated luminosity of 200~nb$^{-1}$.}
    \label{fig:LHCbpt}
\end{figure}

Additionally, \cf{fig:photoVal} shows that the tune uncertainty is small compared to the scale uncertainty: compare the width of the teal band (scale uncertainty) and differences between the solid navy-blue and dot-dashed teal curves (tune uncertainty). The scale uncertainty is evaluated from the envelope of the cross section computed using nine combinations of the factorisation and normalisation scales, $\mu_R$ and $\mu_F$ (9-point scale-variation procedure). More precisely, $\mu_R$ and $\mu_F$ are independently set to $\mu_F=\mu_0\cdot {\zeta_F}$ and $\mu_R=\mu_0\cdot {\zeta_R}$, for $\zeta_{F,R} \in \{1/2,1,2\}$. Here, we use  $\mu_0 = m_T=\sqrt{m_\Q^2 + \pT^2}$.

In order to make predictions for \jpsi yields at the LHC, we generate $^3S_1^{[1]}$ and $^1S_0^{[8]}$ photoproduction samples, with a photon flux from~\cite{Jackson:1998nia}, for proton-lead collisions at 8.16~TeV using \texttt{HO}+\texttt{PYTHIA} and apply our tune in \pT. 
We place rapidity and \pT cuts 
according to the detector acceptances described in Table~\ref{Acc}. 

{Figure \ref{fig:CMSpt} shows resulting double-differential cross sections, in \pT and $y$, times the branching fraction of \jpsi to dimuons, as a function of \pT corresponding to the various CMS acceptance cuts for both the $^3S_1^{[1]}$ (dashed) and $^1S_0^{[8]}$ (solid) tunes, with statistical uncertainties given by the expected Run3+4 luminosity from Table \ref{lumi}. Figure~\ref{fig:CMSpt-all} shows the corresponding single-differential quantity in the entire rapidity coverage of CMS and \cf{fig:LHCbpt} for LHCb in (a) $p$Pb and (b) Pb$p$ collisions. }

{As can be seen on \cf{fig:photoTune}, the latest and most precise H1 data shows a 50\% statistical uncertainty in the last bin from 7.75 to 10~GeV. Similar uncertainties are reached for the double-differential cross section for CMS  in the bin 13 to 20~GeV for  $|y^{J/\psi}|<1.2$ (\cf{fig:CMSpTa}), in the bin 12 to 20~GeV for $1.2<|y^{J/\psi}|<1.6$ (\cf{fig:CMSpTb}), and in the bin 10 to 20~GeV for $1.6<|y^{J/\psi}|<2.4$ (\cf{fig:CMSpTc}). If one considers its entire rapidity coverage,
50\% statistical uncertainty is reached around  $\pT=20$~GeV, twice the H1 value. For LHCb (\cf{fig:LHCbpt}), it corresponds to the bin 8 to 10~GeV.} ATLAS has effectively the same rapidity acceptance as CMS for $\pT>8.5$~GeV, it has the same \pT reach as CMS and is not shown. We do not show results for the ALICE muon arm as the kinematic coverage is similar to LHCb.
We consider the difference between the $^3S_1^{[1]}$ and $^1S_0^{[8]}$ tunes as indicative of a systematic uncertainty. 
Such differences are most pronounced in the low-\pT region in \cf{fig:LHCbpPb}.

\begin{table*}[hbt!]
 \centering
  \caption{Photoproduction cross sections and yields of $J/\psi$ and $\psi(2S)$ satisfying the reconstruction requirements of ALICE and ATLAS in the $p$Pb (Pb$p$) beam configuration. The $J/\psi$ reconstruction requirements from  Table \ref{Acc} and the luminosity values from Table \ref{lumi}  are used.}
\begin{tabular}{lllll}

 &\multicolumn{3}{c}{\textbf{ALICE} } & \textbf{ATLAS} \\
  \hline
  \hline
& $|y^{J/\psi }|<0.9$ & \multicolumn{2}{c}{ $2.5<y^{J/\psi }<4.0$}& $|y^{J/\psi }|<2.1$\\
&&&&$P^{J/\psi }_T > 8.5$~GeV  \\
  $J/\psi$ &  \\
$\sigma$~[nb]&790$\pm$49&530$\pm$130 & (99$\pm$7)&0.8$\pm$0.1 \\
Run 2 yields [$\times 10^{3}$]&17$\pm$1&4.5$\pm$1.1 &(1.3$\pm$0.1)&0.14$\pm$0.02\\
Run3+4 yields [$\times 10^{5}$]&4.0$\pm$0.2&2.7$\pm$0.7 &(0.52$\pm$0.04)&0.008$\pm$0.001\\
\hline 
    $\psi(2S)$ &  \\
  $\sigma$ [nb] &40$\pm$3&27$\pm$7&(4.9$\pm$0.4)&0.04$\pm$0.01\\

Run 2 yields &840$\pm$52&220$\pm$56&(63$\pm$5)&7$\pm$1\\

Run3+4 yields [$\times 10^{2}$] &200$\pm$12&130$\pm$34&(25$\pm$2)&0.4$\pm$0.1\\
\hline
 \end{tabular}

\label{tab:PhotoYields1}
\end{table*}

\begin{table*}[hbt!]
 \centering
  \caption{Photoproduction cross sections and yields of $J/\psi$ and $\psi(2S)$ satisfying the reconstruction requirements of CMS and LHCb in the $p$Pb (Pb$p$) beam configuration. The $J/\psi$ reconstruction requirements from  Table \ref{Acc} and the luminosity values from Table \ref{lumi}  are used.}

\begin{tabular}{lllll}
&  \multicolumn{2}{c}{\textbf{CMS} }   & \multicolumn{2}{c}{\textbf{LHCb}} \\
  \hline
  \hline

 &$|y^{J/\psi }|<2.1$& $\pT^{ J/\psi } > 6.5 $~GeV for $|y^{J/\psi }|<1.2$& \multicolumn{2}{c}{$2.0<y^{J/\psi }<4.5$} \\
&$\pT^{ J/\psi } > 6.5 $~GeV&$P^{ J/\psi }_T > 2$~GeV for $1.2 < |y^{  J/\psi }| < 1.6$  \\
 &  &$ \pT^{ J/\psi }> 0$~GeV for $1.6< |y^{J/\psi}| < 2.4$\\
  $J/\psi$ &  \\
$\sigma$ [nb] &4.8$\pm$0.9&630$\pm$1&880$\pm$210&(200$\pm$17)\\
Run 2 yields [$\times 10^{3}$] &0.9$\pm$0.2&110.0$\pm$0.2&11.0$\pm$2.7&(3.4$\pm$0.3)\\
Run 3+4 yields [$\times 10^{5}$] &0.05$\pm$0.01&6.32$\pm$0.01&1.8$\pm$0.4&(0.44$\pm$0.03)\\

  \hline  
    $\psi(2S)$ &  \\
$\sigma$ [nb] &0.24$\pm$0.05&31.0$\pm$0.1&44$\pm$11&(9.9$\pm$0.9)\\
Run 2 yields &43$\pm$8&5600$\pm$10&550$\pm$130&(170$\pm$15)\\
Run3+4 yields [$\times 10^{2}$] &2.4$\pm$0.5&310$\pm$1&88$\pm$21&(20$\pm$2)\\

 \end{tabular}

\label{tab:PhotoYields2}
\end{table*}

\begin{table*}[hbt!]
 \centering
  \caption{Photoproduction cross sections and yields of $\Upsilon(1S)$, $\Upsilon(2S)$, and $\Upsilon(3S)$ satisfying the reconstruction requirements of ALICE, ATLAS, CMS, and LHCb in the $p$Pb (Pb$p$) beam configuration. The $\Upsilon$ reconstruction requirements from  Table \ref{Acc} and the luminosity values from Table \ref{lumi}  are used.}

 \begin{tabular}{lllllll}
 
& \multicolumn{2}{c}{\textbf{ALICE} } & \textbf{ATLAS}&  {\textbf{CMS} }   & \multicolumn{2}{c}{\textbf{LHCb}} \\
  \hline \hline
 & \multicolumn{2}{c}{$2.5<y^{\Upsilon }<4.0$}&$|y^{\Upsilon }|<2.0$& $|y^{\Upsilon }|<2.4$ &  \multicolumn{2}{c}{$2.0<y^{\Upsilon }<4.5$} \\
$\Upsilon(1S)$\\
$\sigma$~[nb]&0.3&(0.02)&1.2&1.4&0.5&(0.05)\\
Run 2 yields&4&(0.1)&220&250&9.2&(0.7)\\
Run 3+4 yields&160&(8.2)&1200&1400&110&(11)\\ \hline
$\Upsilon(2S)$\\ 
$\sigma$~[nb]&0.1&(0.007)&0.5&0.6&0.2&(0.02)\\
Run 2 yields&1.6&(0.06)&87&99&4&(0.3)\\
Run 3+4 yields&63&(3)&480&550&42&(4)\\ \hline
$\Upsilon(3S)$\\
$\sigma$~[nb]&0.1&(0.005)&0.4&0.4&0.2&(0.02)\\
Run 2 yields&1.2&(0.04)&65&74&3&(0.2)\\
Run 3+4 yields&47&(3)&360&410&32&(3.2)\\

 \end{tabular}
\label{tab:PhotoYields3}
\end{table*}

\subsubsection{$\psi(2S)$ photoproduction}
Existing $\psi(2S)$ measurements are limited to \pT-integrated cross sections  because of the reduced cross section with respect to \jpsi. %
{We assume the same tune factors and \pT shape as that of $J/\psi$ and yields are related by $N_{\psi(2S)} = 0.04 \times N_{J/\psi}$ \footnote{In this scaling \jpsi and $\psi(2S)$ cross sections are related using the H1 determination for \jpsi feed-down contribution from $\psi(2S)$, which is found to be 15\%~\cite{H1:2010udv}. This feed-down fraction is expressed as
\begin{equation}
F^{\psi(2S)}_{J/\psi} = \frac{\sigma_{\psi(2S)} \mathcal{B}(\psi(2S)\rightarrow J/\psi X)}{\sigma_{J/\psi}+\sigma_{\psi(2S)} \mathcal{B}(\psi(2S)\rightarrow J/\psi X)}, 
\end{equation}
with $\mathcal{B}(\psi(2S)\rightarrow J/\psi X)=59.5\pm0.8$~\%~\cite{ParticleDataGroup:2022pth}. The relative yields are then determined by applying the respective branching to dimuons: $\mathcal{B}(J/\psi\rightarrow \mu\mu)=5.961$~\% and $\mathcal{B}(\psi(2S)\rightarrow \mu\mu)=0.77$~\%~\cite{ParticleDataGroup:2022pth}.}.

\subsubsection{$\Upsilon$ photoproduction}
For $\Upsilon(1S)$, there are currently no experimental \pT-differential  photoproduction data. Therefore, we restrict our predictions to LO CSM cross sections computed with \texttt{HO} with the acceptance cuts on the $\Upsilon(1S)$ described in Table~\ref{Acc}. 
Production of $\Upsilon$ mesons is associated with a larger scale than $J/\psi$ production and {therefore has a more convergent pQCD series, but may be more sensitive to parton-shower effects.} As for the $\psi(2S)$ state, we use scaling relations to estimate yields for radially excited states: $N_{\Upsilon(2S)} \simeq 0.4 N_{\Upsilon(1S)}$ and $  N_{\Upsilon(3S)} \simeq 0.3 N_{\Upsilon(1S)}$~\cite{ColpaniSerri:2021bla}, where the cross sections are related by the relative sizes of the radial wave functions at the origin, $|R_{\Upsilon(1S)}(0)|^2=7.5$~GeV$^3$, $|R_{\Upsilon(2S)}(0)|^2=2.89$~GeV$^3$, and $|R_{\Upsilon(3S)}(0)|^2=2.56$~GeV$^3$, and the yields are obtained using the relevant branching ratios: $BR_{\Upsilon(1S)\rightarrow \mu\mu}=2.48$\%, $BR_{\Upsilon(2S)\rightarrow \mu\mu}=1.93$~\%, and $BR_{\Upsilon(3S)\rightarrow \mu\mu}=2.18$~\%~\cite{ParticleDataGroup:2022pth}. %

\subsubsection{Estimated quarkonium yields}

{Tables \ref{tab:PhotoYields1}--\ref{tab:PhotoYields3} give predicted yields for $J/\psi$, $\psi(2S)$, $\Upsilon(1S)$, $\Upsilon(2S)$, and $\Upsilon(3S)$ satisfying the acceptance criteria of ALICE, ATLAS, CMS, and LHCb from  Table~\ref{Acc}, assuming 100\% detector efficiency\footnote{
While this can induce significant corrections at low \pT and central rapidities in the ATLAS and CMS detectors, the corresponding acceptance corrections are systematically smaller than $50\%$~\cite{CMS:2017exb,ATLAS:2011aqv} at the largest \pT we will consider in this study.}, and luminosity values from Table~\ref{lumi}. For comparison, in~\cite{Goncalves:2013ixa}, cross sections for $J/\psi$ and  $\Upsilon(1S)$ were computed in $p$Pb collisions at $\sqrt{s_{NN}}=5.5$~TeV and were found to be 1.6~nb and 0.19~$\mu$b, respectively.}

\subsection{Simulation of the main experimental background: hadroproduced $J/\psi$}\label{sec:background}

The hadroproduction yield dominates over the photoproduction yield by a factor ranging from $10^2$ to $10^4$ with increasing \pT, related to the strong decrease of the photoproduction cross section with increasing \pT, {prompting the need} for data selection requirements that dramatically reduce this hadroproduction background. Since we must show that the background can be reduced%
{ by a factor of as much as $10^{-4}$ in certain \pT regions}, this calls for a reliable description of the hadroproduction background. 

 {We recall that the NRQCD description of hadroproduced \jpsi faces the same issues as those of photoproduced \jpsi, namely tensions in describing world data and large QCD corrections.}
Thus, hadroproduced $J/\psi$ are simulated in a similar way to photoproduced $J/\psi$: octet and singlet partonic-level processes are generated using \texttt{HO}: $g +  g \rightarrow c\bar{c}  \left(^3S_1^{[1]}\right) + g$ and $g +  g \rightarrow c\bar{c}\left(^3S_1^{[8]}\right) + g$. {However, we make use of a different CO contribution than we did for photoproduction. This is because, as demonstrated in LO studies \cite{Cho:1995vh}, the $^3S_1^{[8]}$ is the dominant octet contribution. }

The generated partonic events are passed to \texttt{PYTHIA} and tuning factors are determined using LHCb $pp$ data at $\sqrt{s}=5$~TeV \cite{LHCb:2021pyk} and validated against 
$pp$ data 
differential in both \pT and $y$ \cite{LHCb:2021pyk} and at different centre-of-mass energies \cite{LHCb:2012kaz,LHCb:2015foc}. Results of the tune are reported in \ref{appendix:hadroprod}. These tuning factors are then applied to the events generated at $\sqrt{s_{NN}}=8.16$~TeV in $p$Pb collisions but neglecting nuclear effects\footnote{{This is equivalent to assuming that the nuclear modification factor, $R_{p\text{Pb}} = \sigma_{p\text{Pb}}/(208\, \sigma_{pp})$, is equal to unity. Experimental determinations of  $R_{p\text{Pb}}$ integrated in centrality~\cite{LHCb:2017ygo,ALICE:2018mml,ALICE:2022zig}
for prompt $J/\psi$ production in $p$Pb collisions at $\sqrt{s_{NN}}=8.16$~TeV find that $R_{p\text{Pb}}$ ranges  from 0.6 to unity. %
When the most peripheral events are selected~\cite{ALICE:2020tsj},  $R_{p\text{Pb}}$ gets closer to unity as expected from the scaling of the cross section. This is the reason why we disregard a possible suppression of the hadroproduction background.
}
}. Resulting \pT distributions based on this tuning procedure appear in \ref{appendix:hadroprod}. Note that for photoproduction we had to subtract the non-prompt contribution from the H1 data,
whereas the LHCb hadroproduction tuning data only contain contributions from prompt decays. 

In addition to possible nuclear effects, the number of nucleon-nucleon interactions per collision, $N_\text{coll}$, differs between $pp$ and $p$Pb collisions since for $pp$, $N_\text{coll}=1$, whereas for $p$Pb, $N_\text{coll}\ge1$. Hence, more detector activity can be expected in an inclusive $p$Pb collision than in an inclusive $pp$ collision. In order to take this into account, minimum bias events, generated with \texttt{PYTHIA} and weighted according to an $N_\text{coll}$ distribution extracted from ALICE data~\cite{ALICE:2015kgk}, are folded with the single nucleon-nucleon interaction in which the \jpsi is hadroproduced.

\begin{figure}[hbt!]
    \centering
    \subfloat[]{\includegraphics[height=0.3\textwidth]{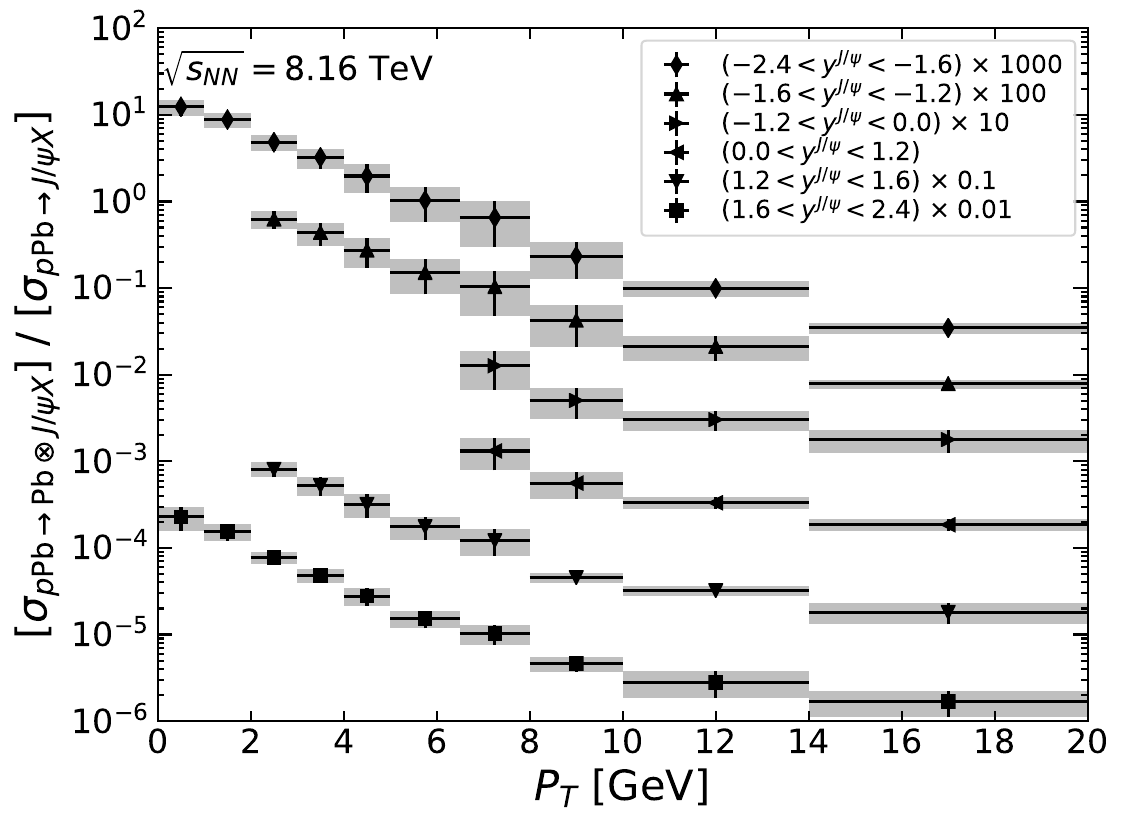}
    \label{fig:cms-ptratio}}\\
    \subfloat[]{\label{fig:LHCbPbp_ptratio}\includegraphics[height=0.3\textwidth]{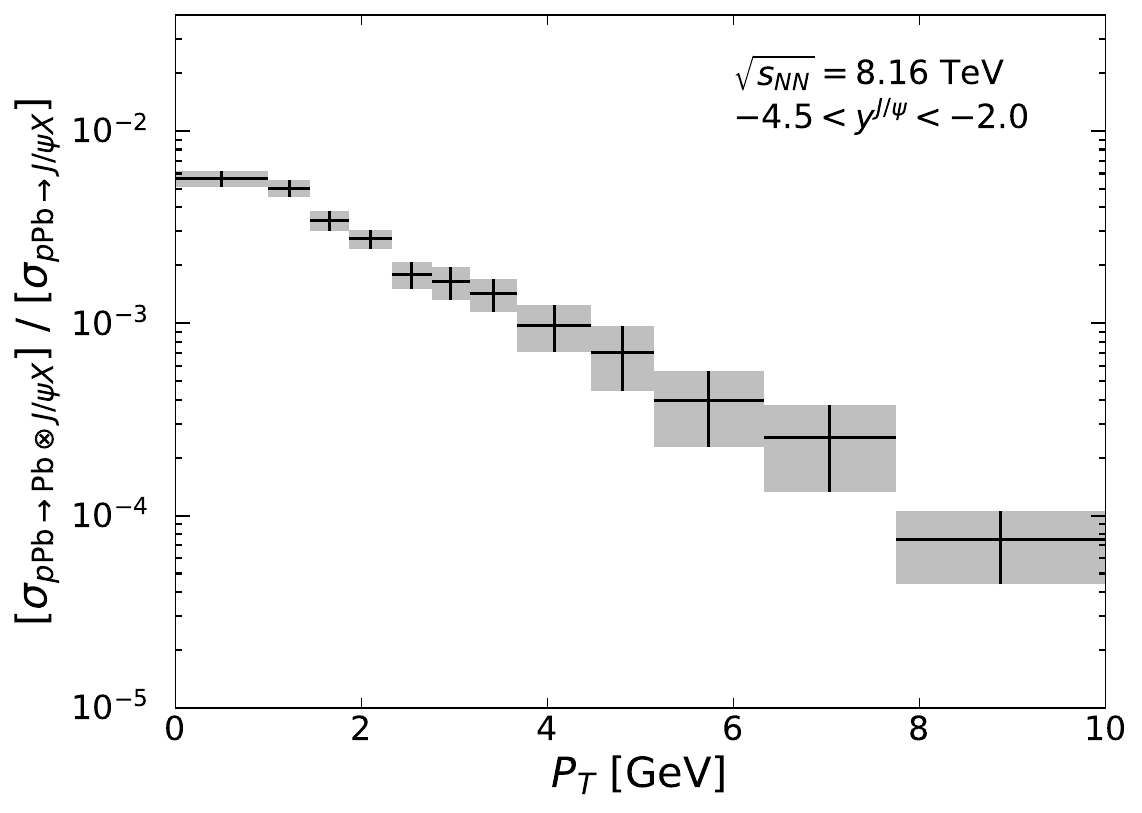}}\\
    \subfloat[]{\includegraphics[trim=0cm 0cm 0cm 0cm, clip,height=0.3\textwidth]{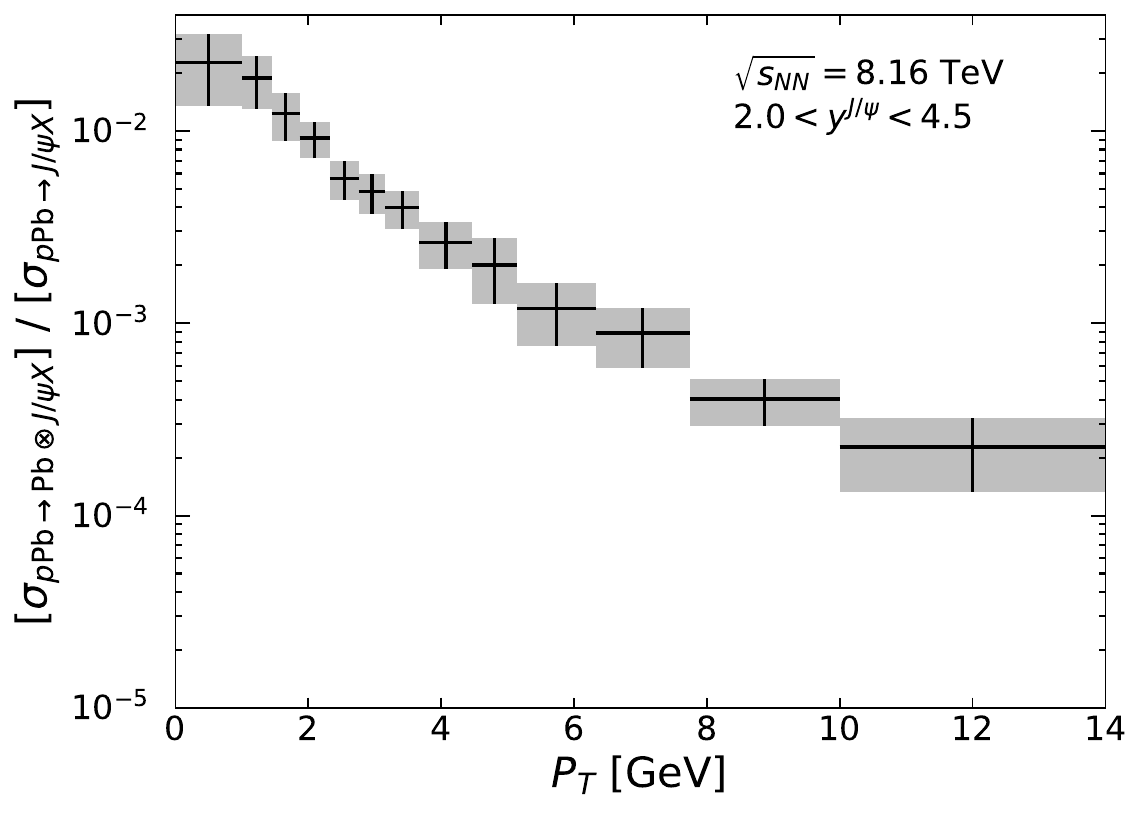}\label{fig:LHCbpPb_ptratio}}
    \caption{Ratio of the photo- and hadroproduced \jpsi cross sections as a function of \pT in the (a) CMS acceptance and in the LHCb acceptance in the (b) Pb$p$ and (c) $p$Pb beam configurations. The grey bands indicate the bin size and tune uncertainty. }
    \label{fig:lhcb-ptratio}
\end{figure}

\section{Experimental selection of direct photoproduced quarkonia at the LHC}\label{sec:expselection}

Sizeable \jpsi yields are anticipated within the acceptance of the four LHC detectors, as shown in Tables~\ref{tab:PhotoYields1} and~\ref{tab:PhotoYields2}, but Figure \ref{fig:lhcb-ptratio} shows that the
hadroproduced background for \jpsi mesons is a factor 
$\mathcal{O}(10^2)$ greater at low \pT and a factor $\mathcal{O}(10^4)$ larger at high \pT. Similar comments can be made for ATLAS and the ALICE muon arm as their acceptances are similar to CMS and LHCb, respectively. Focusing on \jpsi mesons, as these have the highest cross section, we propose three requirements to reduce the hadroproduction background. These criteria exploit  differences between photo- and hadroproduction event topologies.  Photoproduction is characterised by an intact photon emitter and a rapidity separation between the central system and the photon emitter, whereas the hadroproduction background is associated with two broken beam particles and particle-activity spread between the beam remnants on both sides. The proposed selection requirements, described in Sections \ref{sec:rapidity}--\ref{sec:zdc} and summarised in Section \ref{summary}, make use of detectors available at the four LHC experiments. 
{In} Section~\ref{sec:resolved} 
the resolved photon contribution {is discussed}.

{

\begin{figure}[hbt!]
        \centering
              \subfloat[]{\label{fig:CMS5}\includegraphics[trim=0cm 0cm 0cm 0cm,clip,height=0.35\textwidth]{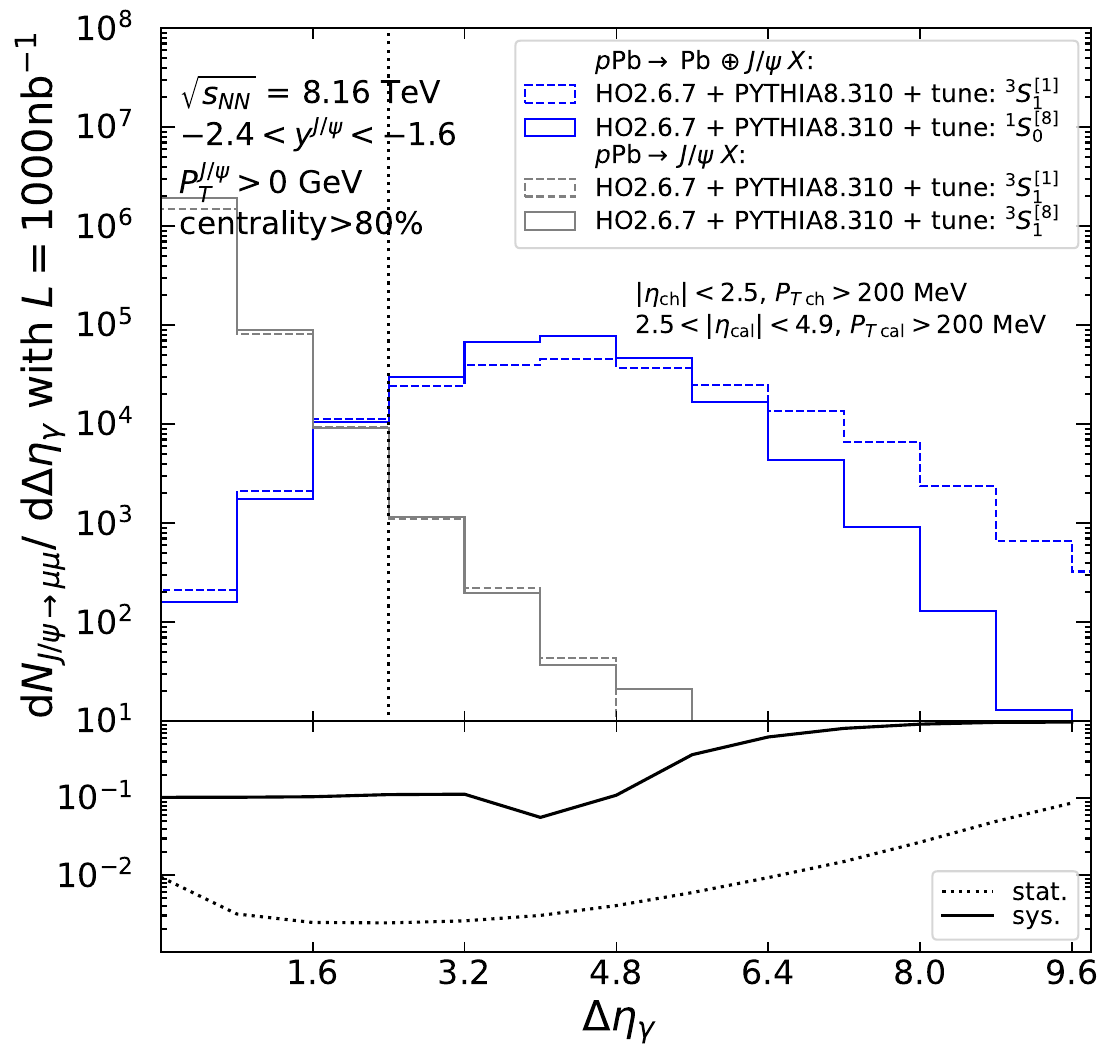}}\\
         \subfloat[]{\label{fig:CMS6}\includegraphics[trim=0cm 0cm 0cm 0cm,clip,height=0.35\textwidth]{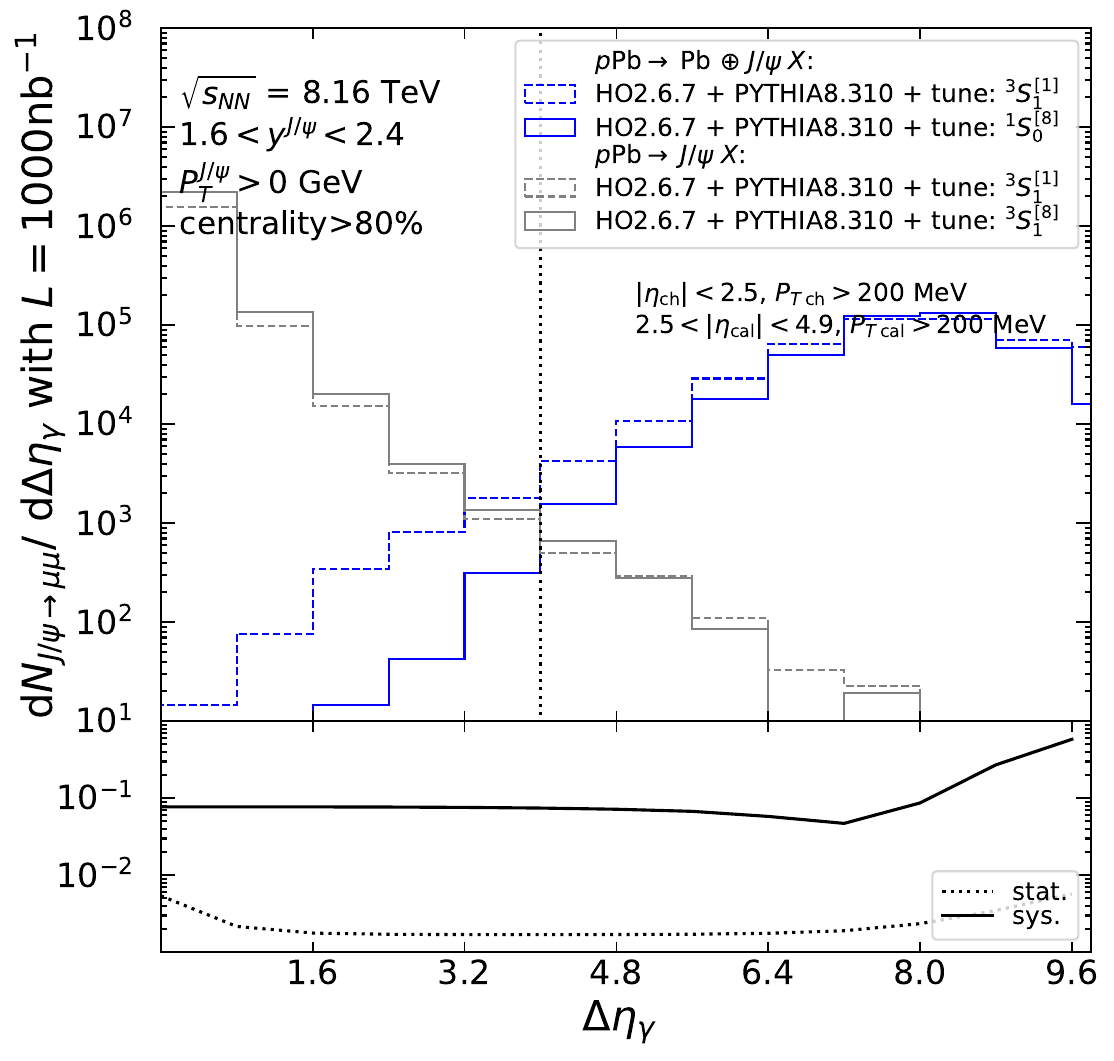}}
    \caption{Differential yield for $J/\psi\rightarrow \mu \mu$ as a function of $\Delta \eta_\gamma$ in the CMS low-\pT acceptance, using the singlet (dashed) and octet (solid) tunes of photoproduction (blue) and hadroproduction (grey) for (a) $-2.4<y^{\jpsi}<-1.6$ and (b) $1.6<y^{\jpsi}<2.4$. The lower panel shows the relative statistical (dotted) and systematic (solid) uncertainties as a function of the cut value on $\Delta \eta_\gamma$. The dotted vertical line indicates the cut value that minimises the statistical uncertainty.  }
 \label{fig:CMS-rapgap-ydiff}
\end{figure}

\begin{figure}[hbt!]
        \centering
     \subfloat[]{\label{fig:gap-cms1}\includegraphics[trim= 0cm 0cm 0cm 0cm,clip, height=0.34\textwidth]{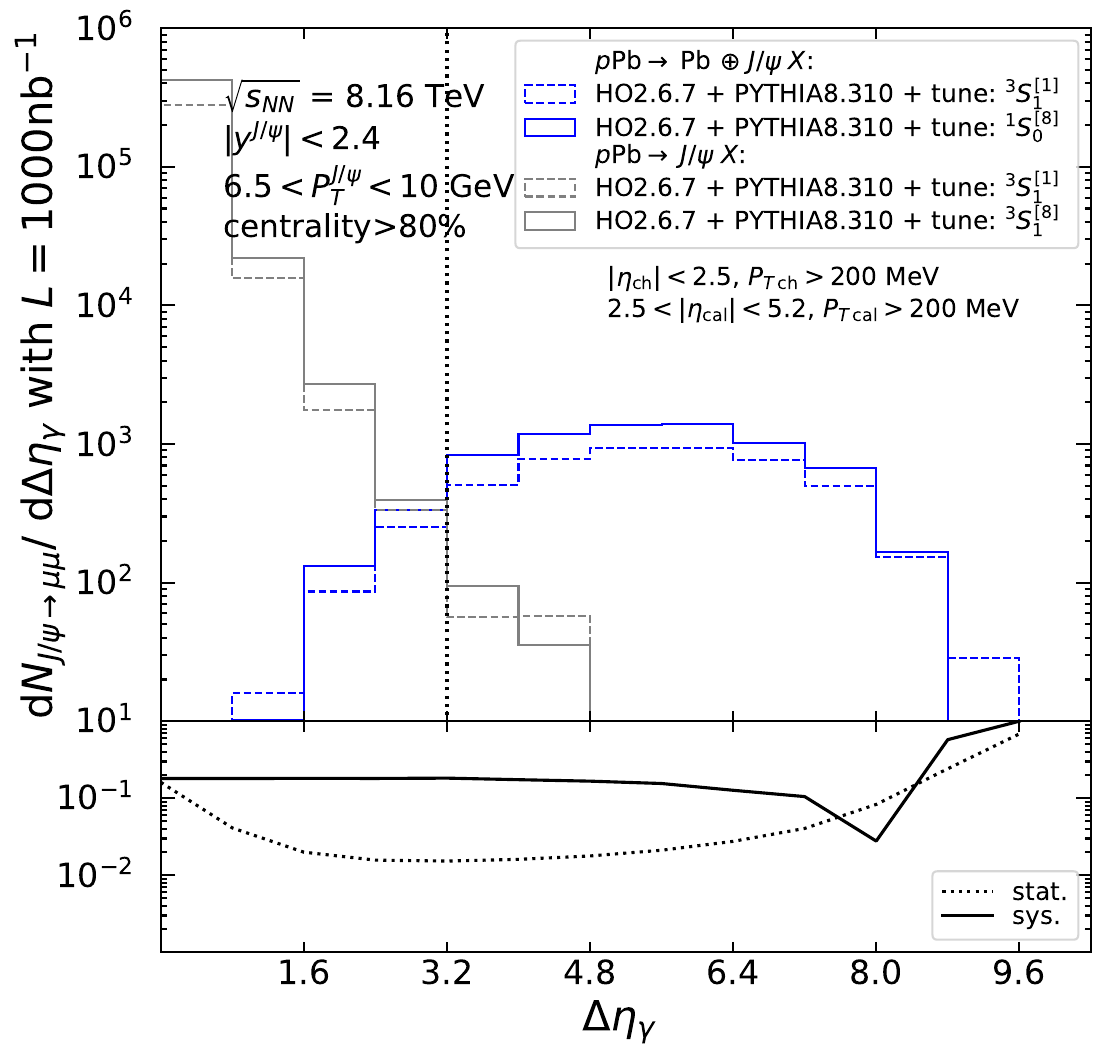}}\\
    \subfloat[]{\label{fig:gap-cms2}\includegraphics[trim= 0cm 0cm 0cm 0cm,clip, height=0.34\textwidth]{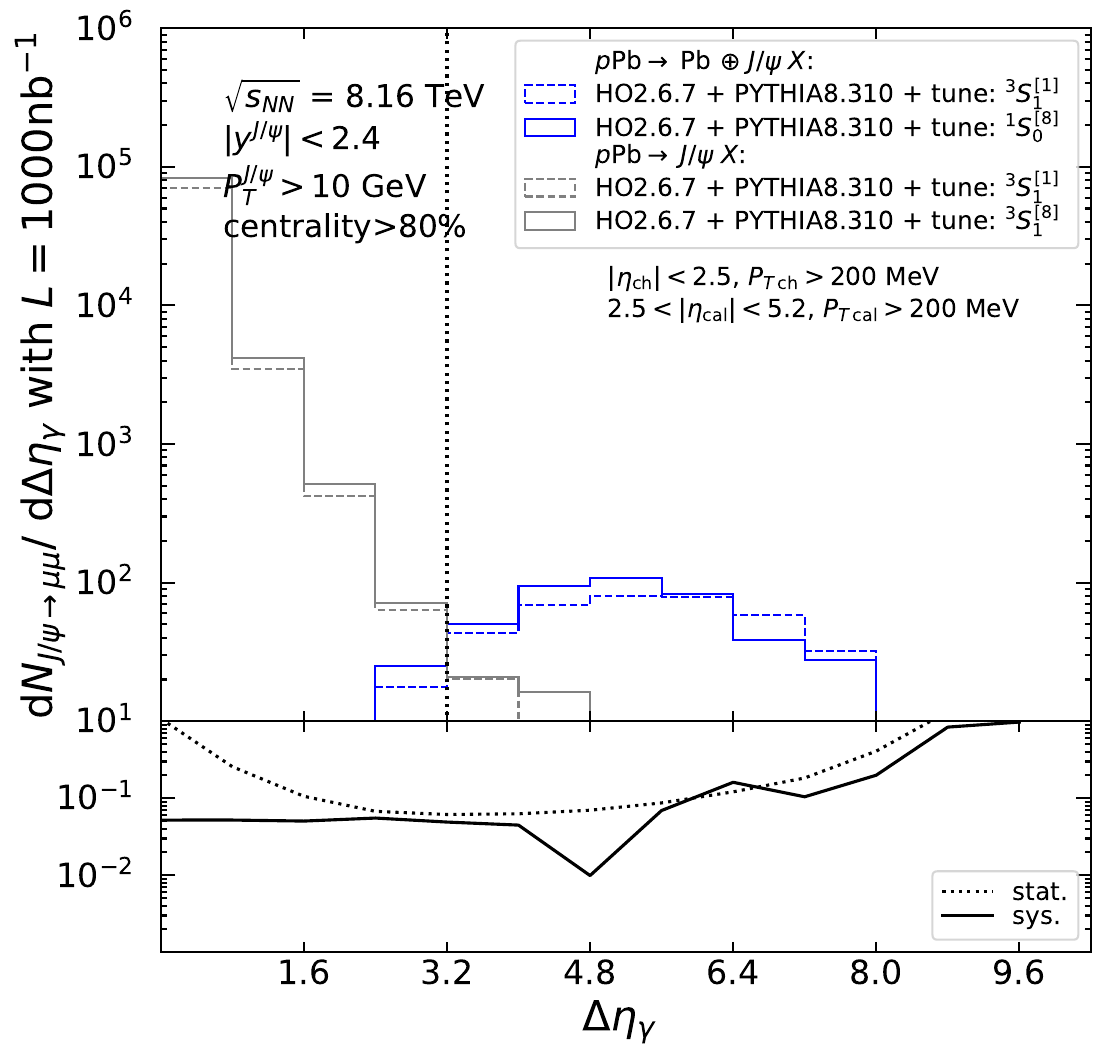}}
         \caption{Differential yield for $J/\psi\rightarrow \mu \mu$ as a function of $\Delta \eta_\gamma$ in the CMS acceptance, using the singlet (dashed) and octet (solid) tunes of photoproduction (blue) and hadroproduction (grey) for (a) $6.5<\pT<10$~GeV and (b) $\pT>10$~GeV. The lower panel shows the relative statistical (dotted) and systematic (solid) uncertainties as a function of the cut value on $\Delta \eta_\gamma$. The dotted vertical line indicates the cut value that minimises the statistical uncertainty.  }
 \label{fig:CMS-rapgap-ptdiff}
\end{figure}

\begin{figure}[hbt!]
    \centering

\subfloat[]{\label{fig:RapLHCbpPb}\includegraphics[trim= 0cm 0cm 0cm 0cm,clip, height=0.34\textwidth]{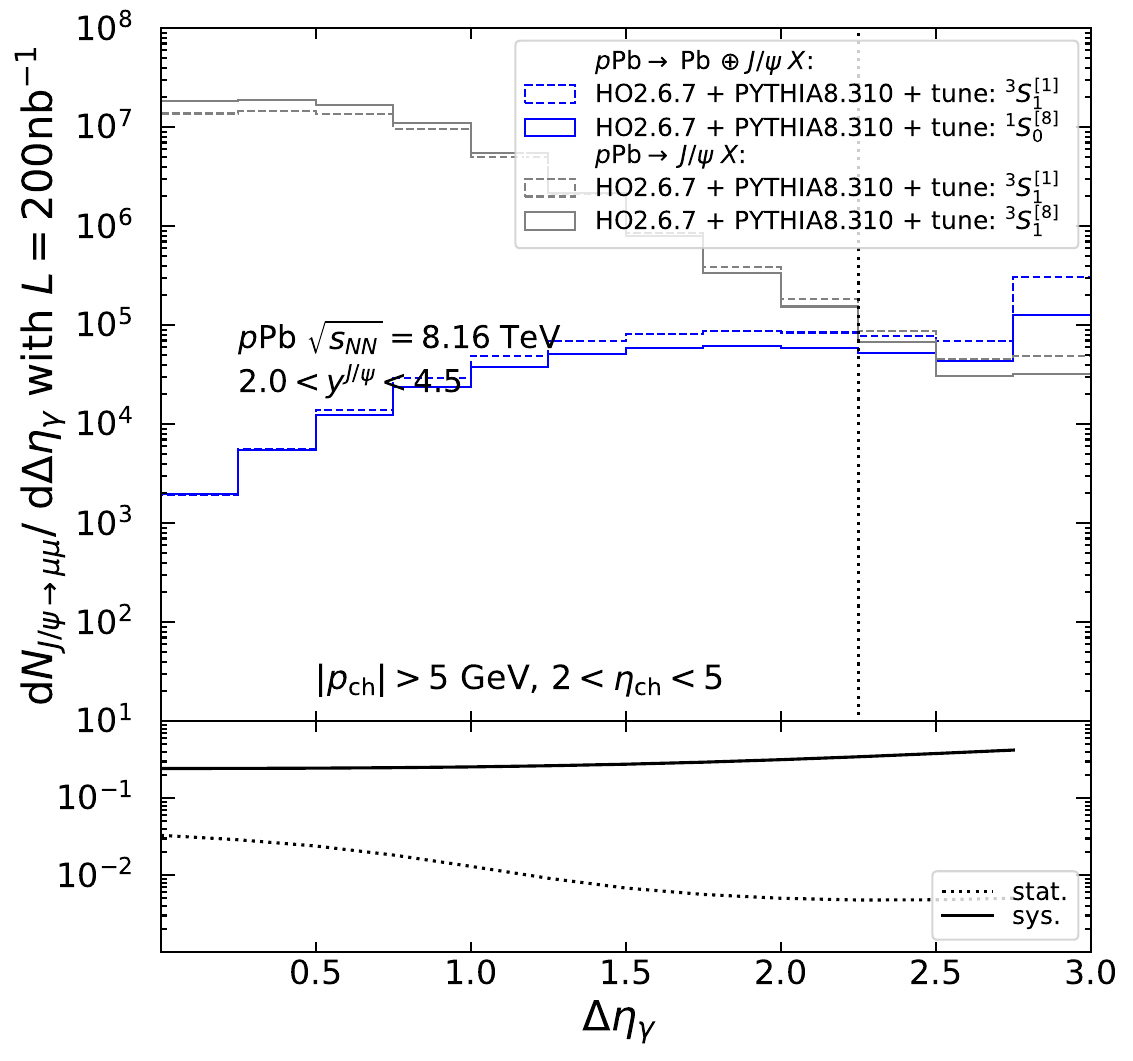}}\\
    \subfloat[]{\label{fig:RapLHCbPbp}\includegraphics[trim= 0cm 0cm 0cm 0cm,clip, height=0.34\textwidth]{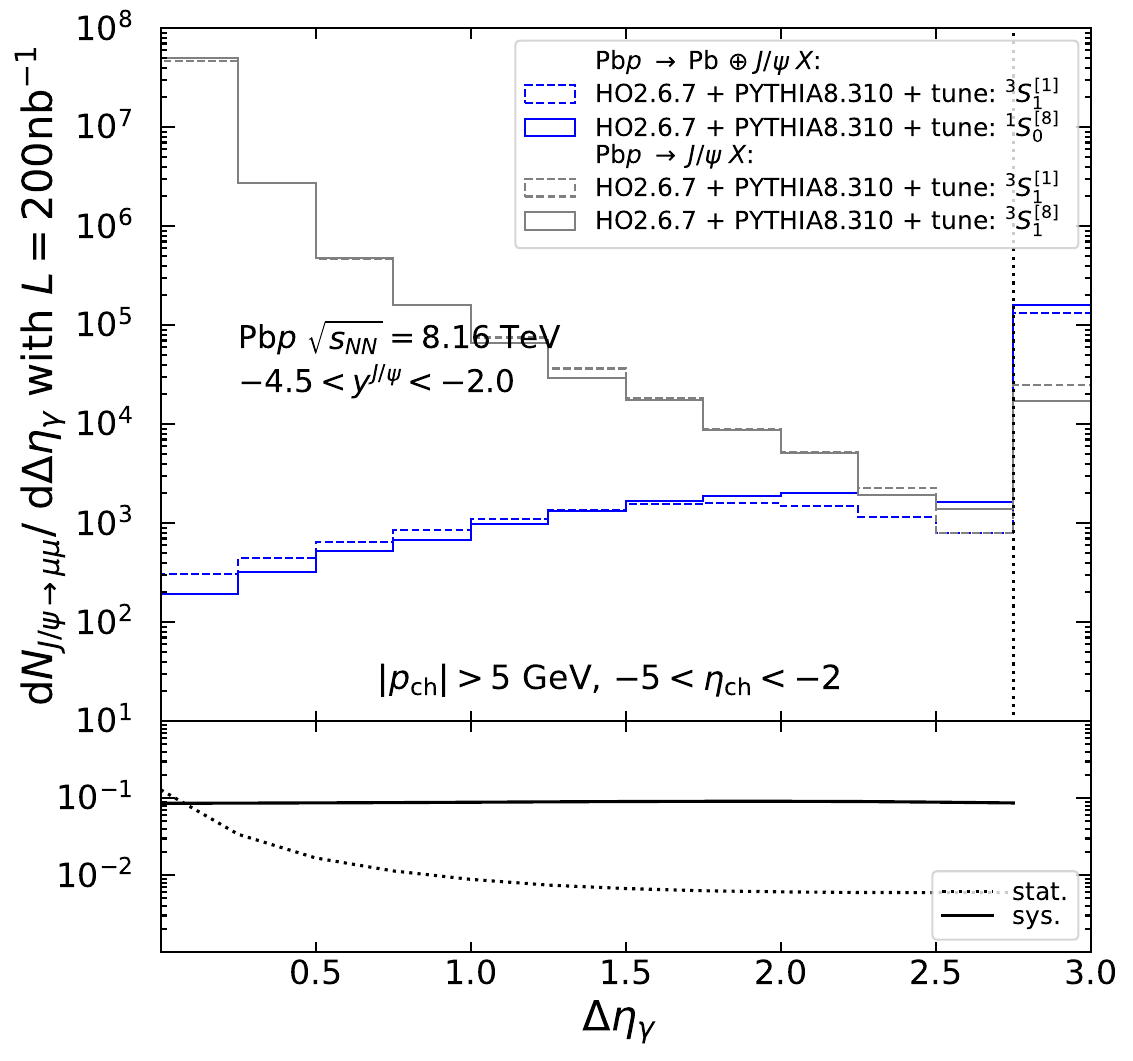}}
\caption{Differential yield for $J/\psi\rightarrow \mu \mu$ as a function of $\Delta \eta_\gamma$ in the LHCb acceptance, using the singlet (dashed) and octet (solid) tunes of photoproduction (blue) and hadroproduction (grey) for the (a) $p$Pb and (b) Pb$p$ beam configurations. The lower panel shows the relative statistical (dotted) and systematic (solid) uncertainties as a function of the cut value on $\Delta \eta_\gamma$. The dotted vertical line indicates the cut value that minimises the statistical uncertainty.   }
 \label{fig:LHCb-rapgap}
\end{figure}

\subsection{Rapidity gaps}\label{sec:rapidity}

Experimentally, rapidity gaps can be defined in a variety of ways. %
Here, we define $\Delta \eta_\gamma$ as the difference in pseudorapidity between the edge of the detector on the lead-going side and the particle detected closest to this edge {(similar to what is employed in~\cite{ATLAS:2012djz,Nurse:2011ttl})}. Figures \ref{fig:CMS-rapgap-ydiff}--\ref{fig:LHCb-rapgap} show the $J/\psi$ yield differential in $\Delta \eta_\gamma$ within the CMS and LHCb acceptances in proton-lead collisions, as obtained from the MC simulations described in Sections \ref{sec:signal} and \ref{sec:background} for the singlet  (dashed) and  octet  (solid) tunes of photoproduction (blue) and hadroproduction (grey). Particles entering the rapidity-gap algorithm are required to pass the acceptance cuts summarised in Table~\ref{Acc} and detailed on the figures. {To enrich the photoproduction signal purity, the \jpsi yield within the CMS acceptance (Figs. \ref{fig:CMS-rapgap-ydiff} and \ref{fig:CMS-rapgap-ptdiff}) is restricted to the 80--100\% centrality class, as explained in Section~\ref{sec:zdc}.}

Inclusive photoproduction is characterised by having one empty region on the side of the intact photon emitter and another region containing the particle activity from the break up of the other beam particle and the hard scattering. This activity %
{extends} from the rapidity region of the particle of interest (here the \jpsi) to that of the broken beam particle. The largest rapidity gap tends to be produced in the region between the \jpsi and the intact photon emitter, with the probability for a large gap decreasing as the difference in rapidity between the two decreases. As a result, the distribution of the \jpsi yield is shifted to smaller values of $\Delta \eta_\gamma$: compare the blue histograms in Fig.~\ref{fig:CMS5} (backward \jpsi closer to the photon emitter) with those in Fig.~\ref{fig:CMS6} (forward \jpsi further from the photon emitter).

The hadroproduction background, on the other hand, is characterised by more particle activity that is more uniformly distributed with rapidity and thus typically has smaller gaps.
In practice, one observes a steadily decreasing yield with $\Delta \eta_\gamma$ (grey histograms in Fig.~\ref{fig:CMS-rapgap-ydiff}) that is similar for forward and backward \jpsi. For a given detector set-up, better hadroproduction-background reduction is thus expected for \jpsi with most forward rapidities.

As seen in Fig.~\ref{fig:lhcb-ptratio}, at large \pT the hadroproduction background is a factor $\mathcal{O}(10,000)$ greater than the signal, and so, its reduction becomes critical. Figure~\ref{fig:gap-cms1} for $6.5<\pT<10$~GeV and \cf{fig:gap-cms2} for $\pT>10$~GeV show that the hadroproduction background and photoproduction signal can be separated using $\Delta \eta_\gamma$ in the CMS acceptance. 
The same conclusion can be drawn for the ATLAS detector.

For LHCb, the rapidity-gap coverage is narrower ($\Delta \eta_\gamma^\text{max}\simeq3$ vs. $\Delta \eta_\gamma^\text{max}\simeq10$) and the gap is necessarily near the \jpsi as the rapidity coverage of particles entering the $\Delta \eta_\gamma$ algorithm is similar to that of the \jpsi. As a result, distinguishing photo- from hadroproduction based only on $\Delta \eta_\gamma$ is less efficient. This is shown in Fig.~\ref{fig:LHCb-rapgap}, where the \jpsi yield is plotted as a function of $\Delta \eta_\gamma$  for (a) $p$Pb (forward going \jpsi) and  (b) Pb$p$ (backward going \jpsi) beam configurations. The hadroproduction-background reduction is probably sufficient to derive \pT-integrated cross sections. However, to reach $\pT\gtrsim5$~GeV one needs to additionally employ the method discussed in the next section}\footnote{Selection based on $\Delta \eta_\gamma$ could be improved for the $p$Pb beam configuration by including the information of backward VELO tracks, covering $-3.5<\eta<-1.5$ in the laboratory frame~\cite{LHCb:2022syj}.}.

The lower panels of Figs. \ref{fig:CMS-rapgap-ydiff}--\ref{fig:LHCb-rapgap} show the relative statistical (dashed) and systematic (solid) uncertainties as a function of the selection requirement $\Delta \eta_\gamma>X$. The relative systematic uncertainty, which characterises the model dependence, is given by the relative difference between the number of (photoproduction) signal events (blue lines) in the selection region modelled by the $^3S_1^{[1]}$ and $^1S_0^{[8]}$ tunes: $|S_{^3S_1^{[1]}} - S_{^1S_0^{[8]}}|/(S_{^3S_1^{[1]}} + S_{^1S_0^{[8]}})$, where $S_{^{2S+1}L_J^{[c_f]}}$ is the number of signal events in the selection region. We assume that hadroproduction (grey lines) will also be measured and as a result the photoproduction measurement can be made by subtracting the small (hadroproduction) background contribution in the selection region using a template fit. The relative statistical uncertainty for the signal extracted by such a subtraction is given by ${\sqrt{\overline{S}+2\overline{B}}}/{\overline{S}}$, where $\overline{S}$ ($\overline{B}$) is the average of the tunes for the number of signal (background) events in the selection region. %
We propose a selection requirement that minimises the statistical uncertainty, indicated by dotted, vertical lines in Figs.~\ref{fig:CMS-rapgap-ydiff}--\ref{fig:LHCb-rapgap}. \ref{Appendix:AdditionalRapidityGap} shows \jpsi yields differential in $\Delta \eta_\gamma$ both for ALICE covering $|y^{\jpsi}|<0.9$, $-4.0<y^{\jpsi}<-2.5$, and $2.5<y^{\jpsi}<4.0$ and for CMS, covering $|y^{\jpsi}|<1.6$ in four rapidity bins.

\subsection{HeRSCheL}\label{sec:Hersh}
The HeRSCheL\footnote{High-Rapidity Shower Counters at LHCb.} detector, which was installed in LHCb during Run 2, can be used to reduce the hadroproduction background. It consisted of five plastic-scintillator panels that were sensitive to charged-particle showers in the forward and backward regions, $5<|\eta|<10$~\cite{Akiba_2018}, without segmentation in rapidity. This dual coverage aids in the identification of double-diffractive, single-diffractive, and inelastic contributions. 

For the present study, we believe that our \texttt{Pythia}-based set-up is not reliable enough to properly simulate both the activity and particle transport in the far-forward region, in order to estimate the response of HeRSCheL to the photoproduction signal and the hadroproduction background. An indication of the potential of using HeRSCheL is given in~\ref{appendix:hershel}, where it is assumed that any charged particle with $5<|\eta|<10$ \old{are} {is} detected. Our \texttt{PYTHIA}-based simulation shows that, if we put the threshold on the minimal number of charged particles in the acceptance of the HeRSCheL detector to retain 100\% of the photoproduction signal, the hadroproduction-background contamination is at the level of 3\% in the $p$Pb beam configuration and 20\% in the Pb$p$ beam configuration. Both of these numbers depend weakly on the quarkonium-production model but have a strong dependence on the modelling of charged-particle multiplicities in the far-forward region.

\subsection{Zero-degree calorimeters and neutron emission}\label{sec:zdc}

{The ALICE, ATLAS, and CMS experiments have zero-degree calorimeters (ZDCs) installed on both sides of the interaction point, covering pseudorapidities $|\eta|\gtrsim 8$. The ZDCs are calorimeters capable of detecting neutral particles produced along the beam direction. Bending and focusing magnets sweep charged particles away from the path of the ZDC, while letting neutral particles, in particular neutrons, pass through it. The main source of these neutrons are Pb ions broken during hadronic collisions and therefore the activity in the ZDC can be used to classify the centrality of heavy-ion collisions. The ZDC detectors have excellent resolution: the 1, 2, 3, and 4 neutron emission peaks are clearly visible and contained within the 90--100\% centrality class~\cite{Suranyi:2019bfk}. 

Neutrons reaching the ZDCs can also come from the de-excitation of a Pb ion, after excitation through the absorption of a photon emitted by the other beam particle. This de-excitation can result in the emission of one or {more} %
neutrons. Lower neutron multiplicities are typically associated with a softer photon exchange and thus larger impact parameters.%

In PbPb collisions, a substantial fraction of quarkonium photoproduction events can be accompanied by neutron emissions, $\mathcal{O}(20\%)$ (\ref{appendix:NeutronEmissions})\footnote{In \cite{ATLAS:2022cbd}, it is estimated from experimental data that the probability for inclusive photoproduction of dijets in PbPb collisions with the absence of neutron emission is of the order of $\mathcal{O}(0.5)$. },  coming from additional photonuclear interactions. Placing requirements on the number of forward neutrons emitted in PbPb collisions biases the impact-parameter dependence of the cross section. In fact, these forward neutrons have been used in association with diffractive \jpsi production in PbPb collisions to lift the ambiguity in the identity of the photon emitter \cite{CMS:2023snh,ALICE:2023jgu}. For a $p$Pb collision system, on the other hand, due to the reduced photon flux of the proton, the probability of a photoproduced $J/\psi$ accompanied by at least one neutron emission is $\mathcal{O}(0.01\%)$ (\ref{appendix:NeutronEmissions}).

{The \jpsi yield within the CMS acceptance is restricted to the 80--100\% centrality class in order to enrich the signal purity. The probability for hadroproduction increases with decreasing impact parameter and $N_\text{coll}$ increases with decreasing impact parameter. Based on simulations from ALICE for $N_\text{coll}$ as a function of centrality~\cite{ALICE:2015kgk}, we estimate that the (0--20\%,20--40\%,40--60\%,60--80\%,80--100\%) centrality classes contain (33\%, 29\%, 20\%, 11\%, 6\%) of hadroproduced \jpsi events. Thus vetoing the 0--80\% centrality class is expected to remove 94\% of the hadroproduced background\footnote{The $J/\psi$ yield was measured as a function of centrality of proton-lead collisions~\cite{ALICE:2015kgk,ALICE:2020tsj}. In particular in ~\cite{ALICE:2015kgk}, the 80--100\% centrality class is measured and thus already provides an enriched-photoproduction sample. }.}

In the selection of UPC events, a tighter constraint than a veto of the 0--80\% centrality class (discussed in Section~\ref{sec:rapidity}) may be considered, such as in \cite{ATLAS:2021jhn,ATLAS:2022cbd}, where photonuclear events in absence of neutron emission are selected. In practice, this was achieved by ATLAS by requiring that the energy deposition in the ZDC on the photon-going side was {smaller than} $1$~TeV, which rejects any neutron emission as the single-neutron peak is at 2.5~TeV. Such a tight cut on the ZDC is ideal for $p$Pb collisions, as it would not affect the photoproduction signal at all (including resolved-photon contributions) and it would remove essentially all of the hadroproduction background associated with the exchange of a coloured parton where at least one of the nucleons of the lead ion breaks up.} Only extremely rare non-photonic exchanges could pass this cut with cross sections certainly much smaller than for photoproduction. %

{A quantitative estimate of the hadroproduction-background--reducing power of a no-neutron--selection requirement would require modelling beyond the scope of this work. We restrict}
{the discussion to presenting results with a veto of the 0--80\% centrality class.}

\subsection{Predicted inclusive $J/\psi$ photoproduction \pT spectra at the LHC }\label{summary}
{We now discuss the differential spectra in \pT after applying the requirements on $\Delta\eta_\gamma$ discussed in Section \ref{sec:rapidity}, and in the case of CMS an additional veto on the 0--80\% centrality class.}
The differential photoproduction cross sections in \pT using Run3+4 luminosity are shown in \cf{fig:ptaftercuts} as a function of \pT, for the singlet (dashed) and octet (solid) tunes within (a) the CMS acceptance and (b) the LHCb acceptance in the Pb$p$ beam configuration. 
We only plot the photoproduction cross section because we assume that the hadroproduction background will be measured and subtracted. 
Correspondingly, the rapidity-gap requirement is optimised per bin in \pT in order to minimise the relative statistical uncertainty, $\sqrt{S+2B}/S$, as discussed in Section \ref{sec:rapidity}. The resulting $\Delta \eta_\gamma$ values are indicated on the figures. Differences between the dashed and solid histograms show our estimated systematic uncertainty.

For CMS (\cf{fig:finalCMS}), the statistical uncertainty after the hadroproduction-background subtraction is similar to the pure photoproduction statistical uncertainty (see Section~\ref{sec:MCjpsiphotoproduction}). The \pT reach after background subtraction extends up to 20~GeV. Our systematic uncertainty at large \pT is much smaller than the statistical uncertainty, meaning a model independent measurement can be made. {A no-neutron requirement is expected to significantly improve this result.} Similar conclusions can be made for ATLAS.

\begin{figure}[!hbt]
    \centering
     \subfloat[]{\label{fig:finalCMS}\includegraphics[width=0.35\textwidth]{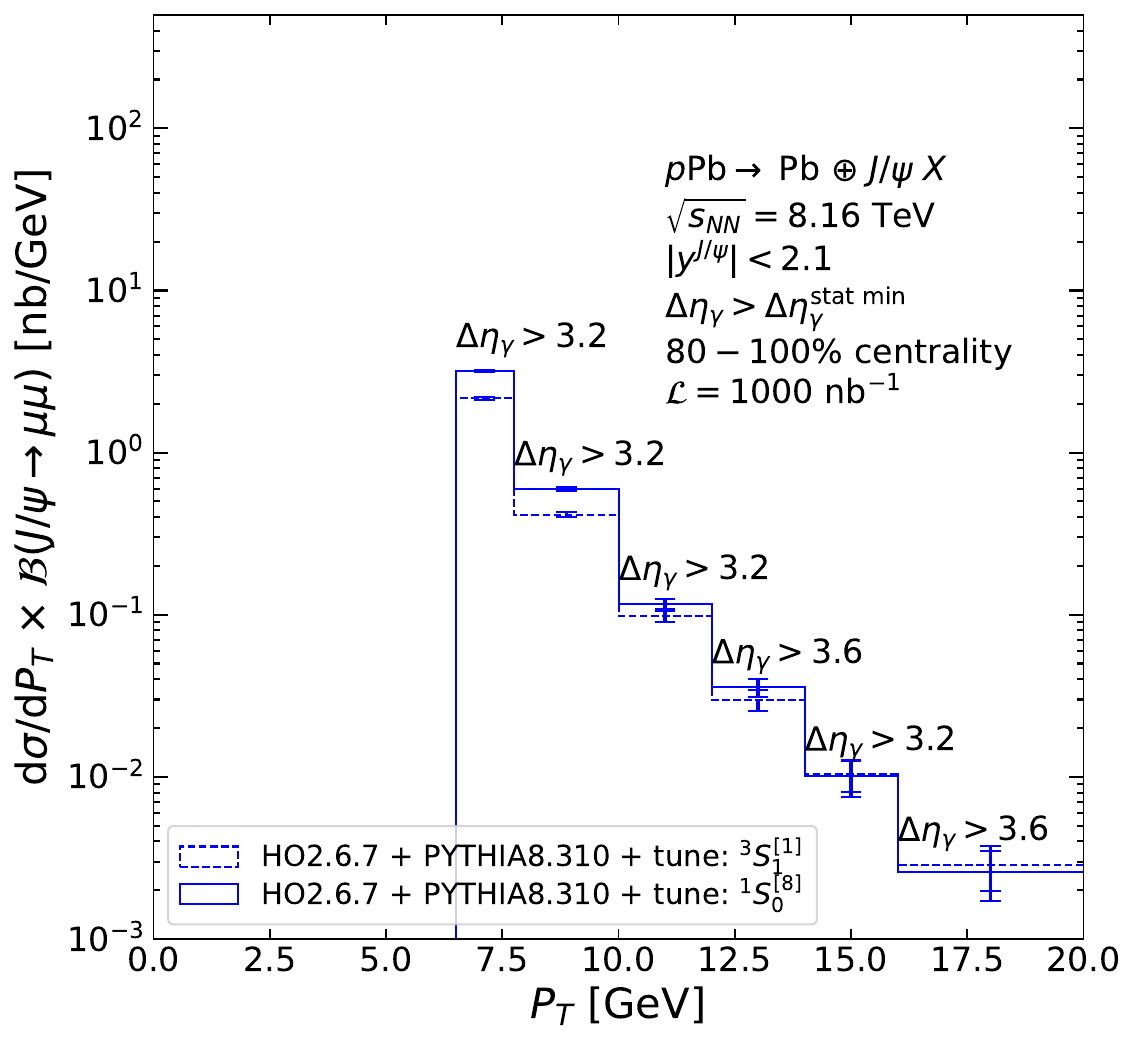}}\\
  \subfloat[]{\label{fig:finalLHCbPbp}\includegraphics[width=0.35\textwidth]{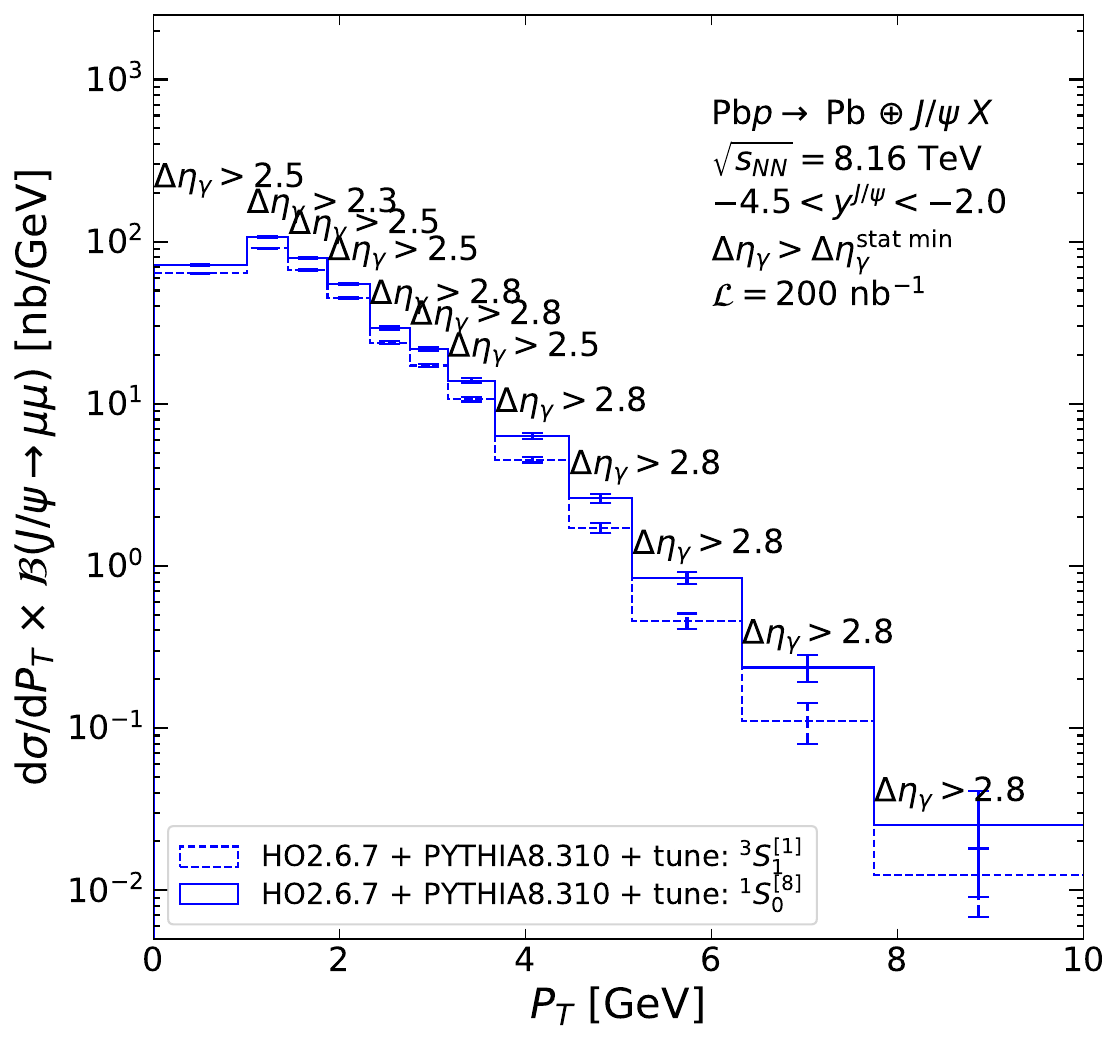}}
    \caption{
    {$\pT$-differential $J/\psi$ photoproduction cross section times dimuon branching as a function of \pT using the $^1S_0^{[8]}$ (solid) and $^3S_1^{[1]}$ (dashed) tunes within (a) the CMS acceptance for $|y^{J/\psi}|<2.1$ and (b) the LHCb acceptance in the Pb$p$ beam configuration. The associated statistical uncertainty after background subtraction computed with Run3+4 luminosity is also show and, for each  bin, the optimal value of $\Delta \eta_\gamma$ minimising the statistical uncertainty is indicated.}}
    \label{fig:ptaftercuts}
\end{figure}

For LHCb, we quote results for the \pT-differential cross section based on $\Delta\eta_\gamma$ alone and  restrict ourselves to the more favourable Pb$p$ beam configuration (Fig.~\ref{fig:finalLHCbPbp}). In this configuration, the statistical uncertainty after background subtraction for $7.7<\pT<10$~GeV is three times larger than the pure photoproduction statistical uncertainty (compare to Fig.~\ref{fig:LHCbPbp}). We do not show results for the $p$Pb beam configuration as the statistical uncertainties are even larger and the \pT reach would be limited to $3$~GeV. The inclusion of HeRSChel in the analysis is expected to substantially improve these results for both beam configurations.

For ALICE, just as for ATLAS and CMS, selecting the 80--100\% centrality class should remove 94\% of the hadroproduction background. Note that for backward produced \jpsi in ALICE, rapidity-gap requirements have no effect (see discussion in \ref{Appendix:AdditionalRapidityGap}). 
{For forward and central \jpsi, combining centrality and rapidity-gap requirements results in a relative statistical uncertainty of $\mathcal{O}(10^{-3})$.}

\subsection{Discussion of the resolved-photon contribution}\label{sec:resolved}

\begin{figure}[bt!]
\centering
\subfloat[]{\includegraphics[scale=.35,draft=false]{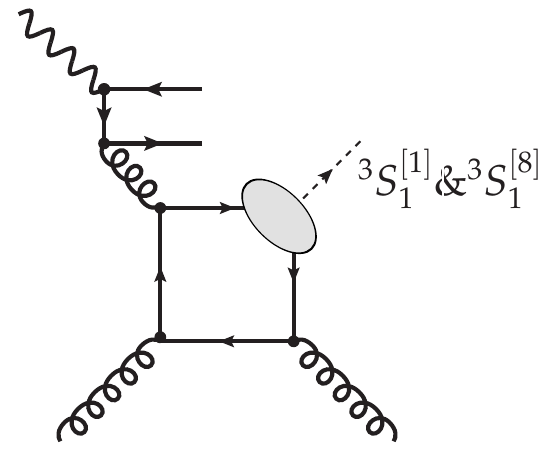} \label{fig:resolvedphotonFEYN}}
\subfloat[]{\includegraphics[scale=.35,draft=false]{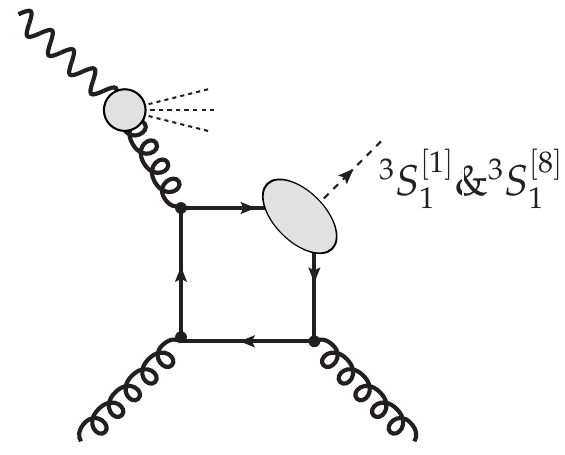} \label{fig:resolvedphotonRESUM}}
\caption{Representative Feynman diagram of (a) collinearly enhanced $^3S_1^{[1]}$ and $^3S_1^{[8]}$ photoproduction  at order $\alpha\alpha^4_s$ and (b) resolved-photon contribution from gluon-gluon fusion.} 
  \label{fig:resolvedphoton}
\end{figure}

In the above we have only considered direct-photon contributions and disregarded resolved-photon ones. Despite being in essence of non-perturbative origin, it is instructive to analyse resolved-photons from a perturbative view point. Among NNLO corrections to direct photoproduction, topologies like Fig.~\ref{fig:resolvedphotonFEYN} exhibit collinear divergences and are similar to reactions induced by the resolved non-perturbative content of the photons, for instance by a gluon, as in Fig.~\ref{fig:resolvedphotonRESUM}.
Effectively, the resolved-photon contribution, or equivalently these specific NNLO corrections, have the same dependence on the NRQCD LDMEs as hadroproduction. 
{They are also expected to be associated with an increased particle activity, intermediate between photo- and hadroproduction.}

Like the direct-photon contribution, the resolved-photon contribution keeps the photon emitter intact and can be considered part of the photoproduction signal. Selection requirements based on far-forward activity, in the ZDCs, %
would fully retain the resolved contribution. As what regards the central-activity criteria based on rapidity gaps, the consideration of resolved photons should in principle allow us to test the robustness of our MC projection of the signal selection against the effects of higher-order perturbative corrections. 

{In order not to reject the resolved-photon contribution from their selection sample, the ATLAS UPC analyses~\cite{ATLAS:2021jhn,ATLAS:2022cbd} introduced a cumulative rapidity-gap criterion, which looks at the difference in pseudorapidity between pairs of adjacent particles and sums them if $\Delta \eta>0.5$. A comparison between the $\Delta \eta_\gamma$ and the cumulative--rapidity-gap, $\sum \Delta \eta_\gamma$, distributions is given in \ref{Appendix:AdditionalRapidityGap} for direct photoproduction and hadroproduction in the peripheral limit where $N_\text{coll}=1$. The cumulative-gap definition has an improved efficiency for retaining resolved-photon contributions but a reduced efficiency for rejecting hadroproduction background. }

Contrary to the ATLAS choice of $\sum \Delta \eta_\gamma$, we opt for the use of $\Delta \eta_\gamma$. Our reasoning is twofold: first, it features better hadroproduction-background--reduc{tion} capability and second, we propose that the resolved-photon contribution can be isolated through a determination of the elasticity $z$, as discussed in the  section~\ref{sec:reco}. We checked{, through simulation of the resolved photon contribution,} that placing a constraint on $\Delta \eta_\gamma$ and, at the same time on the reconstructed $z$ value{, is} an unbiased means to remove the resolved-photon contribution.

Clearly the question of measuring the resolved photon should be addressed as it is fundamental when dealing with photon-induced processes. However, this is not the purpose of our work. When considering cross sections at moderate to large values of $z$, the resolved-photon contribution is small, but it certainly cannot be ignored when extracting cross sections at small values of $z$ ($z\lesssim 0.3$).

\section{Assessment of the reconstruction of $z$ and $W_{\gamma p}$}
\label{sec:reco}

In order to reconstruct $W_{\gamma p}$ and $z$ we propose to use the Jacquet-Blondel method \cite{Amaldi:1979qp,Pawlak:1999ph}, also referred to as the $E-p_z$ method, where the photon kinematics are reconstructed not through the detection of the photon emitter but through the detection of final-state particles produced in the collision. This method has been used extensively by the HERA collider experiments, see e.g.~\cite{H1:2010udv,Pawlak:1999ph}. The key {to this} method is that these reconstructed variables are insensitive to produced particles moving near and collinear to the target beam, while they are very sensitive to produced particles (with high enough momentum) moving in the direction of the photon source. Because of the rapidity gap induced by the colourless exchange, i.e., the photon emission, the majority of these particles which should be detected to determine $W_{\gamma p}$ and $z$ are relatively well contained within the detector acceptance. 

For $\gamma p \rightarrow \jpsi X$, the photon momentum 
can be written as  $P_\gamma = P_{\jpsi} + P_X -P_p$, where $P_X$ represents the momenta of all final-state particles excluding the $\jpsi$ and the scattered {photon emitter}. We recall the definitions of 
\begin{align}
W_{\gamma p }= \sqrt{(P_\gamma + P_p)^2} \simeq\sqrt{2P_p \cdot P_\gamma } && \& && z =\frac{P_p \cdot P_{\jpsi}}{P_p \cdot P_\gamma},
\end{align}
where in our approximation we neglect the virtuality of the quasi-real photon and, owing to the ultra-relativistic velocities of beam particles circulating in the LHC, the mass of the proton. We note that the photon momentum {that needs} to be reconstructed appears only in a scalar
product with the proton momentum.

An alternative representation of the four-momentum, $P^\mu = (E,p_x,p_y,p_z)$, is the light-cone representation, $P^\mu = (p^+,p^-,\bm{p}_T)$. This representation expresses momenta in terms of two scalar quantities, which are parallel to light-like vectors, $n_\pm^\mu$, with motion (anti)parallel to the $z$ direction, $p^\pm=P \cdot n_\pm$, as well as a transverse component, $\bm{p}_T = (p_x,p_y)$. In the laboratory frame these light-like vectors can be written explicitly as $n_-^\mu = (1,0,0,1)/\sqrt{2}$ and $n_+^\mu = (1,0,0,-1)/\sqrt{2}$. It follows that $p^\pm = (E\pm p_z) /\sqrt{2}$.
Momenta can be written explicitly in terms of these light-like vectors $P^\mu= p^+ n_-^\mu + p^-n_+^\mu + p_T^\mu$. By neglecting the mass of the proton we can write its momentum as $P_p^\mu = (p_p^+, 0, \bm{0}_T)$. Consequently, a scalar product\footnote{ The scalar product between two momenta $u$ and $v$ is $u \cdot v = u^+ v^- + u^- v^+ - \bm{u}_T\cdot \bm{v}_T$.} with the proton momentum, $P_p \cdot P_i = p_p^+ p_i^-$, is only dependent on the minus component, $p^-_i$, of the momentum of $i$, which in the laboratory frame is proportional to $(E-p_{z})_i$.

Using this notation and $P_\gamma = P_{\jpsi} + P_X -P_p$, we obtain
\begin{align}
W_{\gamma p } \simeq\sqrt{ (p^-_X + p^-_{\jpsi})  p^+_p} && \& && z \simeq \frac{ p^-_{\jpsi}}{ p^-_{\jpsi}+ p^-_X}.
 \label{zcalc}
\end{align}
In order to measure $W_{\gamma p}$ and $z$, the only unknown quantity is $p^-_X=\sum_i(p^-_i)$. 
Of all the particles in $X$, those collinear to the proton are not relevant since they have $p^-\simeq0$. %
Only those flying in the direction of the Pb ion will contribute. Table~\ref{Acc} gathers the acceptance cuts of the various detectors {that} can be used to detect these particles to reconstruct $W_{\gamma p}$ and $z$.

We use our MC simulation of the signal described in Section~\ref{sec:signal} to estimate the reconstruction potential of $z$ and $W_{\gamma p}$ at the LHC. Because of the larger acceptance and consequently superior reconstruction precision, the discussion is focused on experiments with a central-rapidity coverage, such as CMS. The reconstruction capability of the CMS detector is shown in Fig. \ref{fig:recoCMStyp}, where the medians of the reconstructed (rec) and generated (gen) values of $W_{\gamma p}$ and $z$, $\med{W_{\gamma p}^\text{gen,rec}}$ and $\med{z^\text{gen,rec}}$, are plotted using the tuned
$^3S_1^{[1]}$ (teal circle) and $^1S_0^{[8]}$ (navy-blue cross) simulation samples. The reconstruction bias per bin is the distance between the line of perfect reconstruction (red, dotted line where $W_{\gamma p}^\text{rec}=W_{\gamma p}^\text{gen}$ and $z^\text{rec}=z^\text{gen}$) and the teal circle or navy-blue cross. The variance of the reconstructed values per bin can be estimated as the difference between the $16^{\rm th}$ and $84^{\rm th}$ percentile values, as represented by the vertical error bars, and the model dependence of the reconstruction can be seen as the difference between values obtained using the $^3S_1^{[1]}$ and $^1S_0^{[8]}$ tunes.

\begin{figure}[hbt!]
\centering
\subfloat[]{\includegraphics[width=0.5\textwidth]{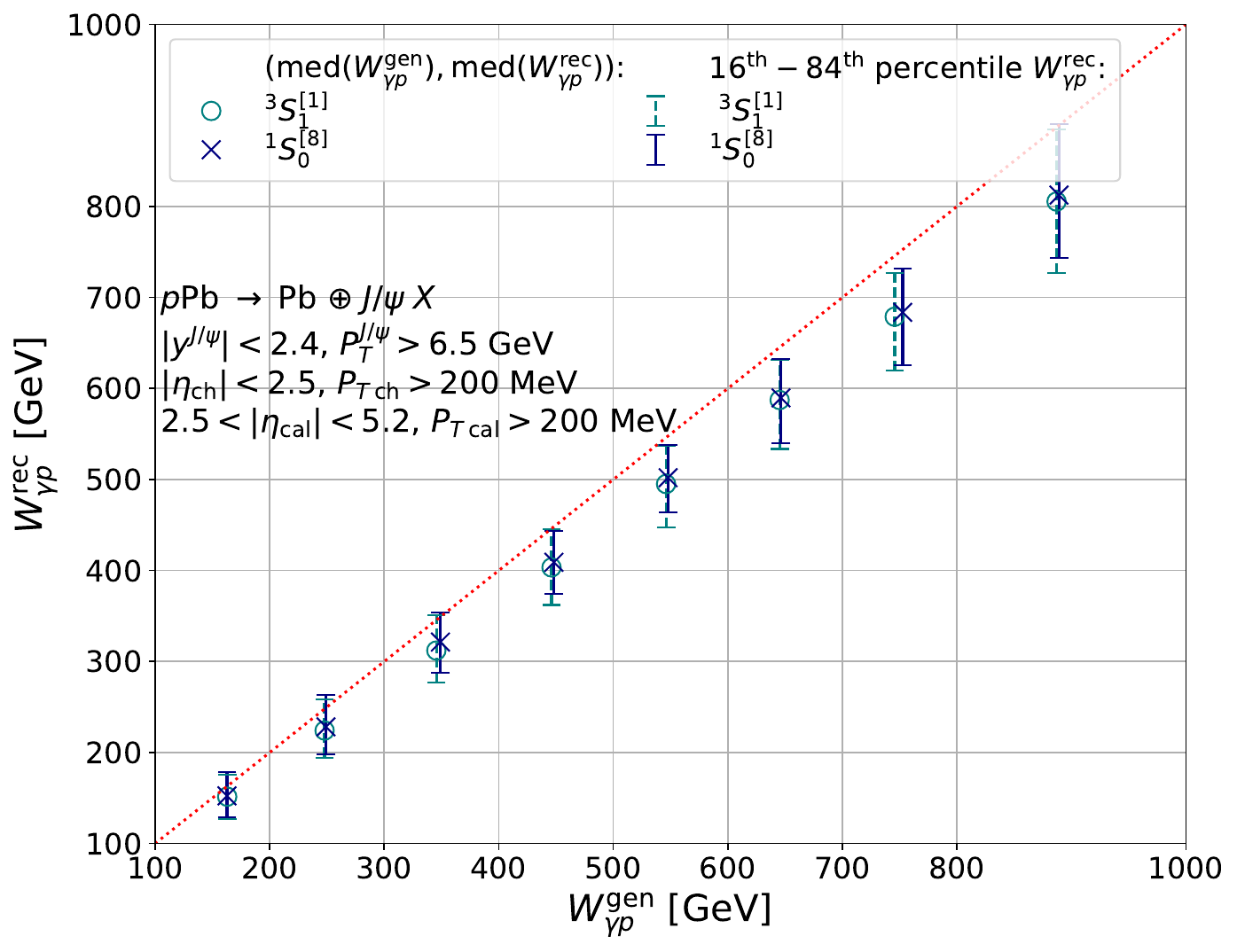}}\\
\subfloat[]{\includegraphics[width=0.5\textwidth]{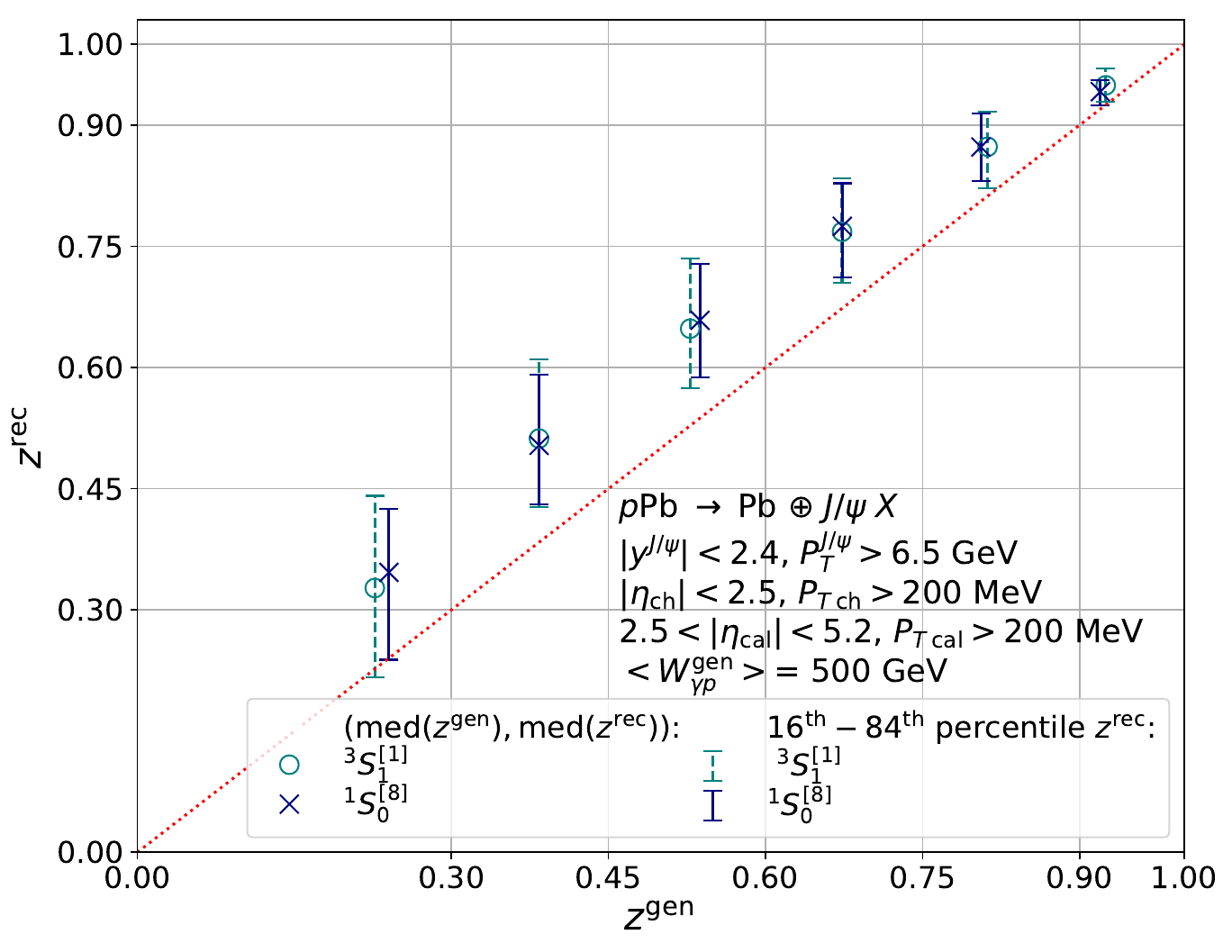}}
      \label{fig:recoCMStyp}
\caption{The median reconstructed (rec) values as a function of the median generated (gen) values of $W_{\gamma p}$ (a) and $z$ (b), using the tuned $^3S_1^{[1]}$ (teal circle) and $^1S_0^{[8]}$ (navy-blue cross), for \jpsi reconstructed within the CMS acceptance. The lower and upper bounds on the error bars indicate the $16^{\rm th}$ and $84^{\rm th}$ percentile on the reconstructed values and the grid lines indicate the chosen binning.}
\label{fig:recoCMStyp}
\end{figure}

The bias on $z$ and $W_{\gamma p}$ originates from particles that are not reconstructed by the detector. This results in $z^\text{rec}>z^\text{gen}$ and $W_{\gamma p}^\text{rec}<W_{\gamma p}^\text{gen}$, i.e., $z^\text{rec}$ and $W_{\gamma p}^\text{rec}$ are respectively systematically above and below the diagonal, red line. At largest $z$ and lowest $W_{\gamma p}$, where the kinematics are dominated by the \jpsi, the variables are {best} reconstructed. 
Conversely, at low $z$ and large $W_{\gamma p}$, where the kinematics are dominated by the $X$ state, the limited detector coverage results in an increasingly large variance. Additional results are given in~\ref{appendix:reconstructionCMS} in five bins of \jpsi rapidity for the CMS acceptance and in both beam configurations for the LHCb acceptance. 

The reconstruction accuracy can also be evaluated through the variable $F$, introduced by the FNAL-E-0516 experiment~\cite{Denby:1983az} as $z^\text{gen}= p^{-\,\text{rec}}_{\jpsi}/(p^{-\,\text{rec}}_{\jpsi}+ F p^{-\,\text{rec}}_{X})$. It accounts for the loss in reconstruction of $p_X^-$ by the detector. The value of $F$ was found to be $1.84 \pm 0.51$. We find a value of the same order: $F=1.55^{+0.87}_{-0.35}$ and $F=1.60^{+0.61}_{-0.34}$ for CMS with \jpsi in the low-\pT and $\pT>6.5$~GeV acceptances, respectively. 

The determination of the bias {on} the $F$ value in principle allows for correction factors to be applied to reconstructed distributions in order to obtain the underlying $z$ and $W_{\gamma p}$ distributions. At the ATLAS and CMS experiments, a determination of $z$ and $W_{\gamma p}$ is possible with similar binning to that of HERA.

 \section{Conclusion and Outlook}\label{outlook}
In the present paper, we have examined the potential to extend the LHC from a hadron-hadron to a photon-hadron collider to perform inclusive-production studies of quarkonia. {For} 
more than 20 years, hadron beams have been used extensively as effective photon beams, but mostly for the study of exclusive processes. We have demonstrated that it is also possible to use hadron beams as photon beams for inclusive processes, where we have focused on the study of photoproduced quarkonia in $p$Pb collisions. More precise inclusive photoproduction data will provide the opportunity to better constrain the quarkonium-production mechanism, which to date remains little understood, and {then} to improve our knowledge on the nucleon structure{~\cite{ColpaniSerri:2021bla}}. 

We have shown that the inclusive photoproduction signal can be isolated with respect to the large competing hadroproduction background in a model-independent way using 
{several methods}. The most powerful method to select photoproduced events involves the use of a ZDC, since the photoproduction of a low--invariant-mass system, i.e., a quarkonium and some recoiling particles, in a $p$Pb collision has a negligible probability for neutron emission from the Pb ion. On the contrary, the probability of zero neutron emission in hadroproduction is certainly below the per mil level. This allows for essentially all of the photoproduction signal to be kept, while rejecting an extremely large proportion of the hadroproduction background. Quantifying this proportion, however, involves advanced simulations, which are beyond the scope of this paper. For this reason, we have restricted our estimates to the selection of the 80--100\% centrality for which experimental measurements exist in $p$Pb collisions. We have estimated that this removes 94\% of the hadroproduction background.

Photoproduction may also be selected based on a rapidity-gap criterion. Using %
simulations with \texttt{PYTHIA} with spectra of quarkonia tuned to data, we have shown that ATLAS and CMS, due to their broad rapidity coverage, offer sufficient discriminating power between photo- and hadroproduction to perform photoproduction cross-section measurements at energies and transverse momenta never reached before. The situation is slightly less favourable for ALICE and LHCb, but will allow for the measurement of $P_T$-integrated cross sections. From our simulations we anticipate that we can obtain a signal-over-background{,} integrated {within their} \pT acceptance{s}, of 8, 30, 160, and 20 for ALICE, ATLAS, CMS {(in the low \pT configuration)}, and LHCb, respectively. 

In the LHCb experiment, we advocate the use of the HeRSCheL detector {but} because of a lack of modelling of the detector response, we have not tried to quantify its reduction power. However, simulation qualitatively shows that the charged-particle distribution in the HeRSCheL region is highly discriminant between photo- and hadroproduction.

We have shown that both the elasticity, $z$, and the photon-proton centre-of-mass energy, $W_{\gamma p}$, can be determined from the momenta of the final-state particles detectable in the ATLAS and CMS detectors in lieu of that of the photon emitter as usually done in lepton-proton experiments.

Overall, the study of inclusive photoproduction of quarkonia in $p$Pb collisions at the LHC has the potential to extend the existing photoproduced-quarkonium measurements in lepton-proton collisions at HERA {and earlier fixed-target experiments} with improved statistical accuracy and by increasing the reach in $W_{\gamma p}$ from 240~GeV to 1.4~TeV and in \pT from 10 GeV to 20 GeV.

While the present study concentrates on quarkonium production in $p$Pb collisions, the techniques discussed here do not need to be limited to either inclusive quarkonium production or $p$Pb collisions, but may be extended to a host of photoproduction processes and other collisions systems.

\section*{Acknowledgments}
We thank 
V.~Bertone,
J.R.~Cudell,
Z.~Conessa del Valle, 
D.~d'Enterria, 
C.~Flett,
C.~Flore, 
B.~Gilbert,
C.~Hadjidakis, 
L.~Massacrier,
M.~Nefedov,
A.~Penzo, 
H.S.~Shao,
M.~Strikman,
L.~Sudit, 
L.~Szymanowski,
and M.~Winn
for useful discussions and inputs.

The research conducted in this publication was funded by the Irish Research Council under grant number GOIPG/2022/478.

This project has also received funding from the European Union's Horizon 2020 research and innovation programme under grant agreement No.~824093 in order to contribute to the EU Virtual Access {\sc NLOAccess} and the JRA Fixed-Target Experiments at the LHC.

R.M.N. acknowledges the hospitality and the financial support of Universit\'e Paris-Saclay through a senior Jean d'Alembert fellow.

This project has also received funding from the Agence Nationale de la Recherche (ANR) via the grant ANR-20-CE31-0015 (''PrecisOnium'') and via the IDEX Paris-Saclay "Investissements d'Avenir" (ANR-11-IDEX-0003-01) through the GLUODYNAMICS project funded by the "P2IO LabEx (ANR-10-LABX-0038)" and through the Joint PhD Programme of Universit\'e Paris-Saclay (ADI).
This work  was also partly supported by the French CNRS via the IN2P3 projects "GLUE@NLO" and "QCDFactorisation@NLO". 

C.V.H. has received funding from the programme Atracci\'on de Talento, Comunidad de Madrid (Spain), 
under the grant agreement No 2020-T1/TIC-20295.

\clearpage
\appendix
\section[]{Photoproduction tune factors}\label{appendix:photoprod}

Table \ref{tab:phototune} reports the photoproduction tune factors 
for  the $^3S_1^{[1]}$ and $^1 S_0^{[8]}$ states, 
$\mathcal{T}^\text{ph}_{^{2S+1}L^{[c_f]}_J}(P_T)$ { , } defined as $(d\sigma_\text{H1}/d\pT) /( d\sigma_{\texttt{HO+PYTHIA}: ^{2S+1}L^{[c_f]}_J}/d\pT)$, described in Section~\ref{sec:signal}, normalised to the corresponding LDME value. The tune fixes \texttt{HO+PYTHIA} results for $^3S_1^{[1]}$ and $^1S_0^{[8]}$ states to H1 data \cite{H1:1996kyo,H1:2002voc,H1:2010udv}
 using multiplicative factors in bins of \pT up to 10~GeV and a scale factor, of the form $a\times \pT$, for $\pT>10$~GeV. {We use the CT18NLO~\cite{Hou:2019qau} parameterisation of the PDF with $\alpha_s(M_Z=91.187$~GeV$)=0.118$, 
 $m_c=1.5$~GeV for the $^3S_1^{[1]}$ state, and $m_c=1.6$~GeV for the $^1S_0^{[8]}$ state.} 
 {Using $m_c=1.6$~GeV}
 is necessary when dealing with octet states, as \texttt{PYTHIA} assigns a larger mass to octet states in order to account for soft radiation emitted by the $c\bar{c}$ pair when transitioning into the physical quarkonium state.  

\begin{table}[h!]
    \caption{Tune parameters normalised to the LDME value
    computed using $J/\psi$ photoproduction data collected by H1 \cite{H1:1996kyo,H1:2002voc,H1:2010udv} in lepton-proton collisions and \texttt{HO+PYTHIA} cross sections for $^3S_1^{[1]}$ and $^1S_0^{[8]}$ states. }
    \centering
    \begin{tabular}{|c|cc|}
    \hline
    \pT bin [GeV]& $\mathcal{T}^\text{ph}_{^3S_1^{[1]}}/\langle \mathcal{O}_{J/\psi} (^3 S_1^{[1]})\rangle$ & $\mathcal{T}^\text{ph}_{^1S_0^{[8]}}/\langle \mathcal{O}_{J/\psi} (^1 S_0^{[8]})\rangle$\\ \hline 
0.0 $<P_T<$ 1.0 & 1.5 $\pm$ 0.3 GeV$^{-3}$& 12.6 $\pm$ 2.1 GeV$^{-3}$\\
1.0 $<P_T<$ 1.4 & 1.2 $\pm$ 0.1 GeV$^{-3}$& 9.3 $\pm$ 1.0 GeV$^{-3}$\\
1.4 $<P_T<$ 1.9 & 0.9 $\pm$ 0.1 GeV$^{-3}$& 6.8 $\pm$ 0.7 GeV$^{-3}$\\
1.9 $<P_T<$ 2.3 & 0.8 $\pm$ 0.1 GeV$^{-3}$& 5.6 $\pm$ 0.6 GeV$^{-3}$\\
2.3 $<P_T<$ 2.8 & 0.6 $\pm$ 0.1 GeV$^{-3}$& 4.1 $\pm$ 0.5 GeV$^{-3}$\\
2.8 $<P_T<$ 3.2 & 0.6 $\pm$ 0.1 GeV$^{-3}$& 4.5 $\pm$ 0.6 GeV$^{-3}$\\
3.2 $<P_T<$ 3.7 & 0.7 $\pm$ 0.1 GeV$^{-3}$& 4.8 $\pm$ 0.6 GeV$^{-3}$\\
3.7 $<P_T<$ 4.5 & 0.8 $\pm$ 0.1 GeV$^{-3}$& 5.0 $\pm$ 0.6 GeV$^{-3}$\\
4.5 $<P_T<$ 5.1 & 1.1 $\pm$ 0.2 GeV$^{-3}$& 6.0 $\pm$ 1.0 GeV$^{-3}$\\
5.1 $<P_T<$ 6.3 & 1.2 $\pm$ 0.3 GeV$^{-3}$& 6.5 $\pm$ 1.4 GeV$^{-3}$\\
6.3 $<P_T<$ 7.7 & 1.8 $\pm$ 0.5 GeV$^{-3}$& 10.3 $\pm$ 2.7 GeV$^{-3}$\\
7.7 $<P_T<$ 10.0 & 1.7 $\pm$ 1.2 GeV$^{-3}$& 9.8 $\pm$ 7.0 GeV$^{-3}$\\
$P_T>10$ & 0.221 $\times P_T$ GeV$^{-3}$& 1.286 $\times P_T$ GeV$^{-3}$ \\
\hline
    \end{tabular}
    \label{tab:phototune}
\end{table}

\section[]{Hadroproduction tune factors}\label{appendix:hadroprod}

Table \ref{tab:hadrotune} reports the hadroproduction tune factors for the $^3S_1^{[1]}$ and $^3 S_1^{[8]}$ states, $\mathcal{T}^\text{had}_{^{2S+1}L^{[c_f]}_J}(P_T)$ defined as $(d\sigma_\text{LHCb}/d\pT) /( d\sigma_{\texttt{HO+PYTHIA}: ^{2S+1}L^{[c_f]}_J}/d\pT))$, described in Section \ref{sec:background}, normalised to the corresponding LDME values. The tune fixes \texttt{HO+PYTHIA} results for $^3S_1^{[1]}$ and $^3S_1^{[8]}$ states to LHCb $pp$ data at $\sqrt{s}=5$~TeV~\cite{LHCb:2021pyk} using multiplicative factors in bins of \pT up to 20~GeV. As for photoproduction, {we use the CT18NLO~\cite{Hou:2019qau} parameterisation of the PDF with $\alpha_s(M_Z=91.187$~GeV$)=0.118$ and 
 $m_c=1.5$~GeV for the $^3S_1^{[1]}$ state. For the $^3S_1^{[8]}$ state we use $m_c=1.6$~GeV.}

\begin{table}[hbt!]
    \caption{
    Tune parameters normalised to the LDME values
    computed using $J/\psi$ hadroproduction data from proton-proton collisions at $\sqrt{s}=5$~TeV collected by LHCb~\cite{LHCb:2021pyk} and \texttt{HO+PYTHIA} results for $^3S_1^{[1]}$ and $^3S_1^{[8]}$ states. }
    \centering
    \begin{tabular}{|c|cc|}
    \hline
  \pT bin [GeV]& $\mathcal{T}^\text{had}_{^3S_1^{[1]}}/\langle \mathcal{O}_{J/\psi} (^3 S_1^{[1]})\rangle$ & $\mathcal{T}^\text{had}_{^3S_1^{[8]}}/\langle \mathcal{O}_{J/\psi} (^3 S_1^{[8]})\rangle$\\ \hline 
0.0 $<P_T<$ 1.0 & 1.2 $\pm$ 0.1 GeV$^{-3}$& 19.4 $\pm$ 1.1 GeV$^{-3}$\\
1.0 $<P_T<$ 2.0 & 1.0 $\pm$ 0.0 GeV$^{-3}$& 15.6 $\pm$ 0.9 GeV$^{-3}$\\
2.0 $<P_T<$ 3.0 & 0.9 $\pm$ 0.0 GeV$^{-3}$& 11.6 $\pm$ 0.7 GeV$^{-3}$\\
3.0 $<P_T<$ 4.0 & 1.1 $\pm$ 0.0 GeV$^{-3}$& 9.1 $\pm$ 0.5 GeV$^{-3}$\\
4.0 $<P_T<$ 5.0 & 1.7 $\pm$ 0.1 GeV$^{-3}$& 7.9 $\pm$ 0.4 GeV$^{-3}$\\
5.0 $<P_T<$ 6.0 & 3.1 $\pm$ 0.1 GeV$^{-3}$& 7.2 $\pm$ 0.4 GeV$^{-3}$\\
6.0 $<P_T<$ 7.0 & 5.9 $\pm$ 0.2 GeV$^{-3}$& 6.6 $\pm$ 0.4 GeV$^{-3}$\\
7.0 $<P_T<$ 8.0 & 10.1 $\pm$ 0.4 GeV$^{-3}$& 6.5 $\pm$ 0.3 GeV$^{-3}$\\
8.0 $<P_T<$ 10.0 & 18.6 $\pm$ 0.7 GeV$^{-3}$& 6.3 $\pm$ 0.5 GeV$^{-3}$\\
10.0 $<P_T<$ 14.0 & 36.9 $\pm$ 1.4 GeV$^{-3}$& 6.0 $\pm$ 0.6 GeV$^{-3}$\\
14.0 $<P_T<$ 20.0 & 80.5 $\pm$ 4.2 GeV$^{-3}$& 5.7 $\pm$ 0.4 GeV$^{-3}$\\
\hline
    \end{tabular}
    \label{tab:hadrotune}
\end{table}

\begin{figure}[hbt!]
     \centering
\subfloat[]{\includegraphics[width=0.37\textwidth]{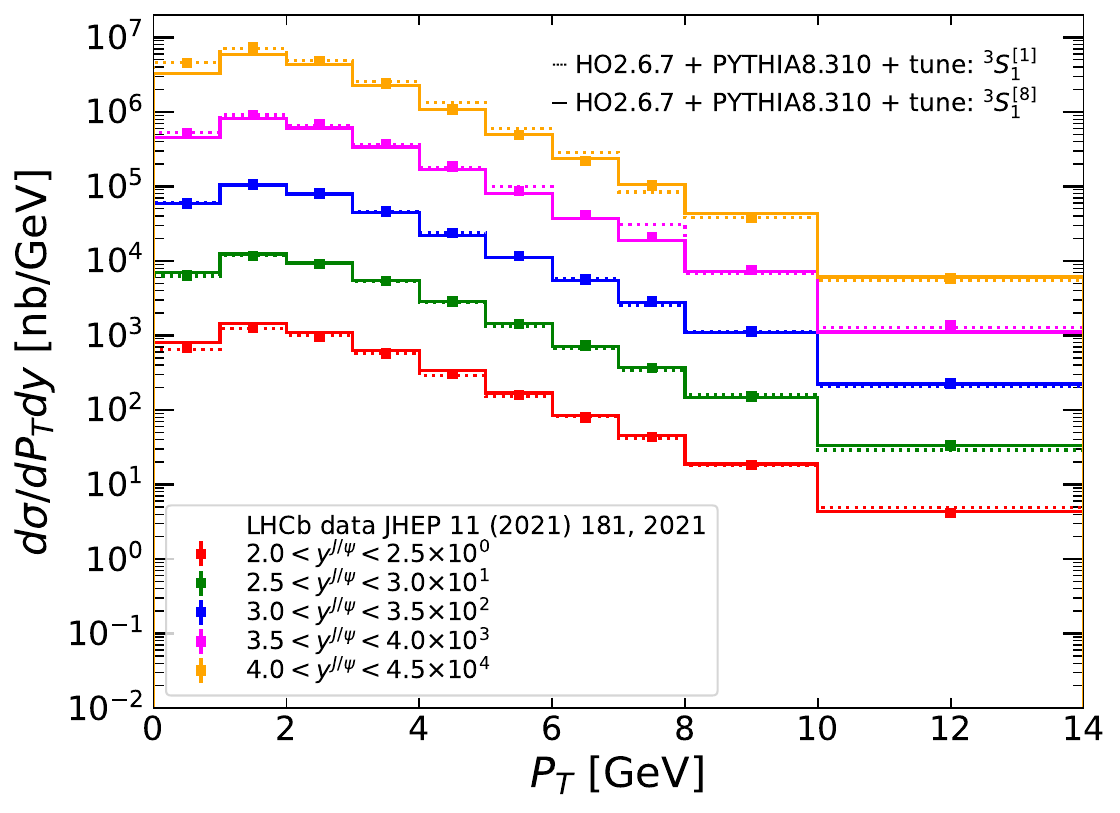}}\vspace{-0.2cm}\\
\subfloat[]{\includegraphics[width=0.37\textwidth]{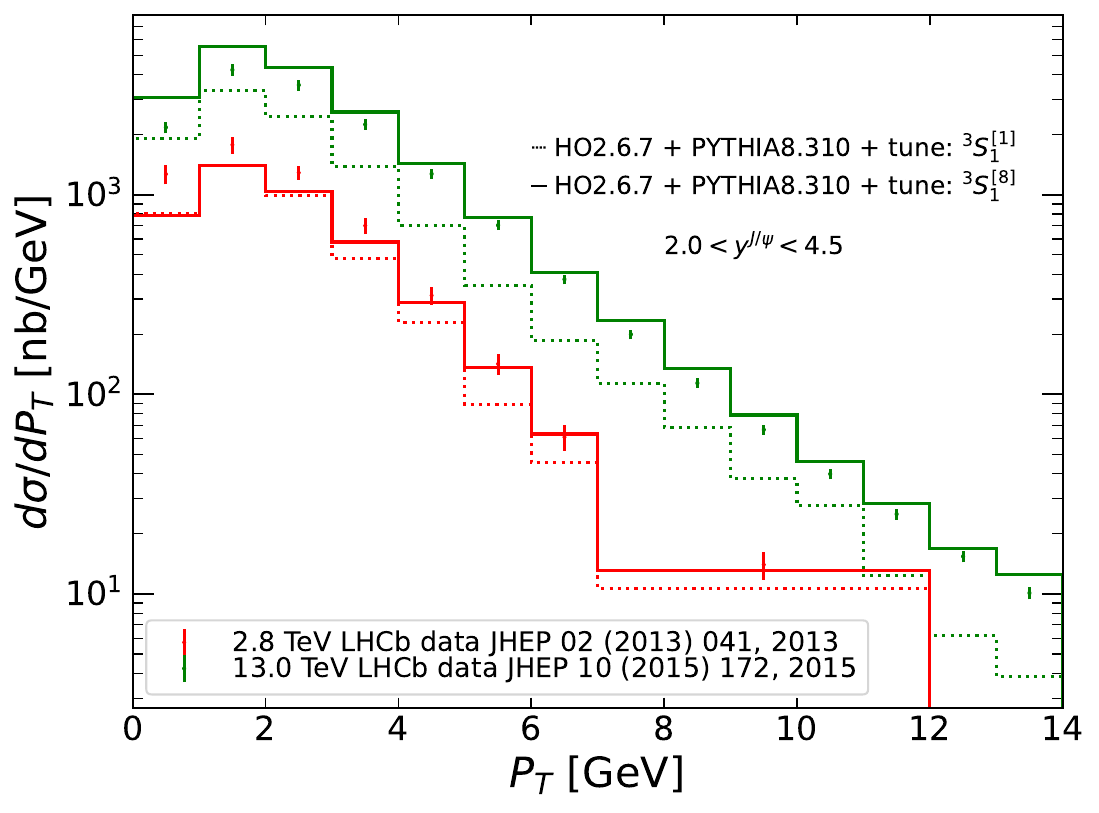}}
\caption{Comparison of the cross sections of the $^3S_1^{[1]}$ (dashed) and $^3S_1^{[8]}$ (solid) hadroproduction tunes to $pp$ data (a) differential in both \pT and $y^{J/\psi}$ at $\sqrt{s}=5.0$~TeV~\cite{LHCb:2021pyk} and (b) differential in \pT at $\sqrt{s}=2.8$~TeV (red) and $\sqrt{s}=13.0$~TeV (green)~\cite{LHCb:2012kaz,LHCb:2015foc}.}
\label{fig:bckval}
\end{figure}

In our tune we assume that the $y^{\jpsi}$ and centre-of-mass energy dependence are correctly accounted for by the PDF and photon flux. These assumptions are validated in figure \ref{fig:bckval}, which compares the tuned hadroproduced MC to (a) {double-differential} data in \pT and $y$~\cite{LHCb:2021pyk} and (b) data at different centre-of-mass energies~\cite{LHCb:2012kaz,LHCb:2015foc}. The tune agrees reasonably well with the data.   

Based on the tuning procedure described in Section \ref{sec:background}, Figs.~\ref{fig:CMSpt-atyp-hadro} and \ref{fig:LHCbpt-hadro} show \pT-differential distributions for hadroproduced $J/\psi$ at $\sqrt{s_{NN}}=8.16$~TeV in $p$Pb collisions in the CMS and LHCb detector acceptance, respectively.
\begin{figure}[hbt!]
    \centering
    \subfloat[]{\includegraphics[scale=0.32]{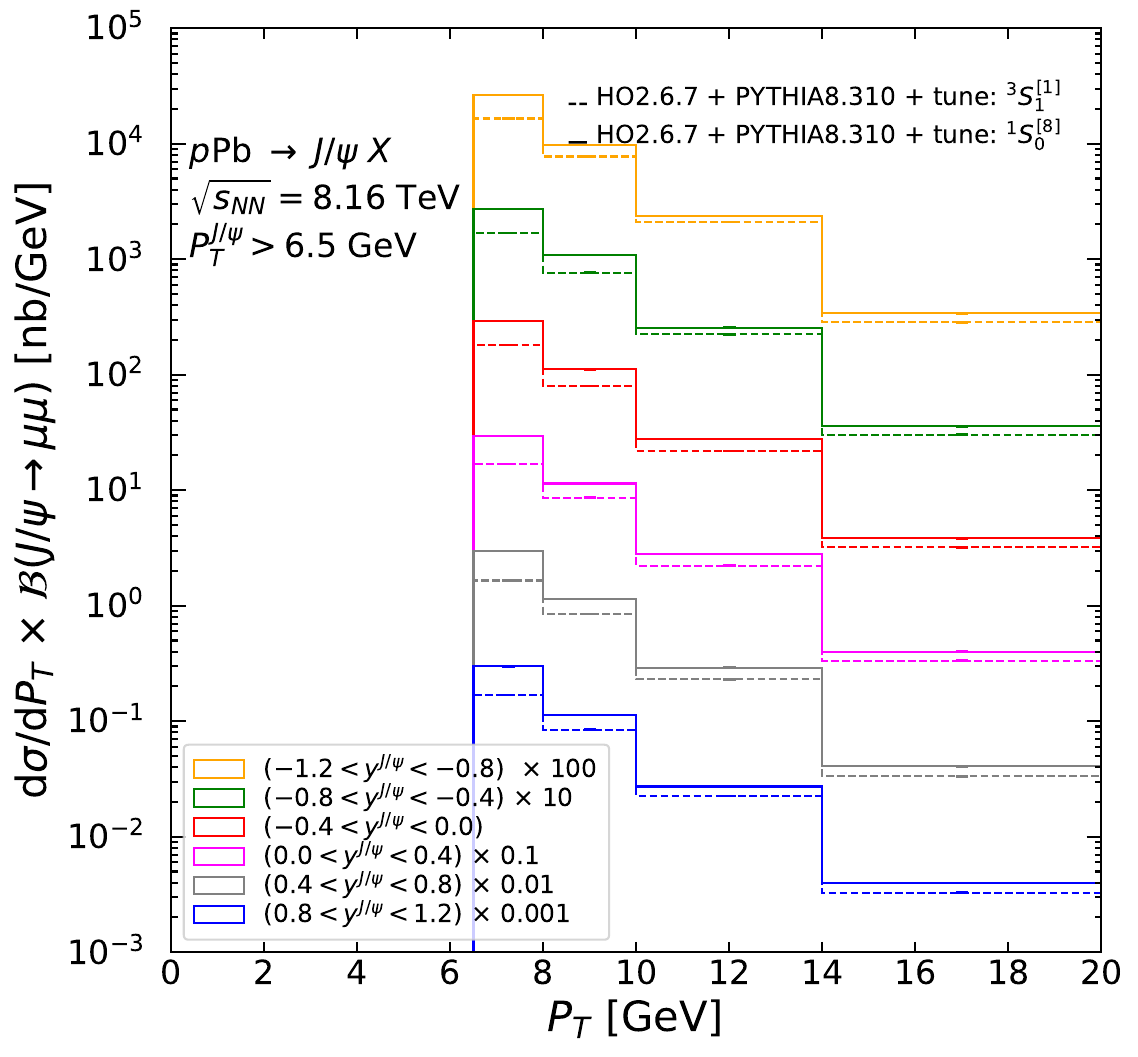}}\vspace*{-0.4cm}\\
    \subfloat[]{\includegraphics[scale=0.32]{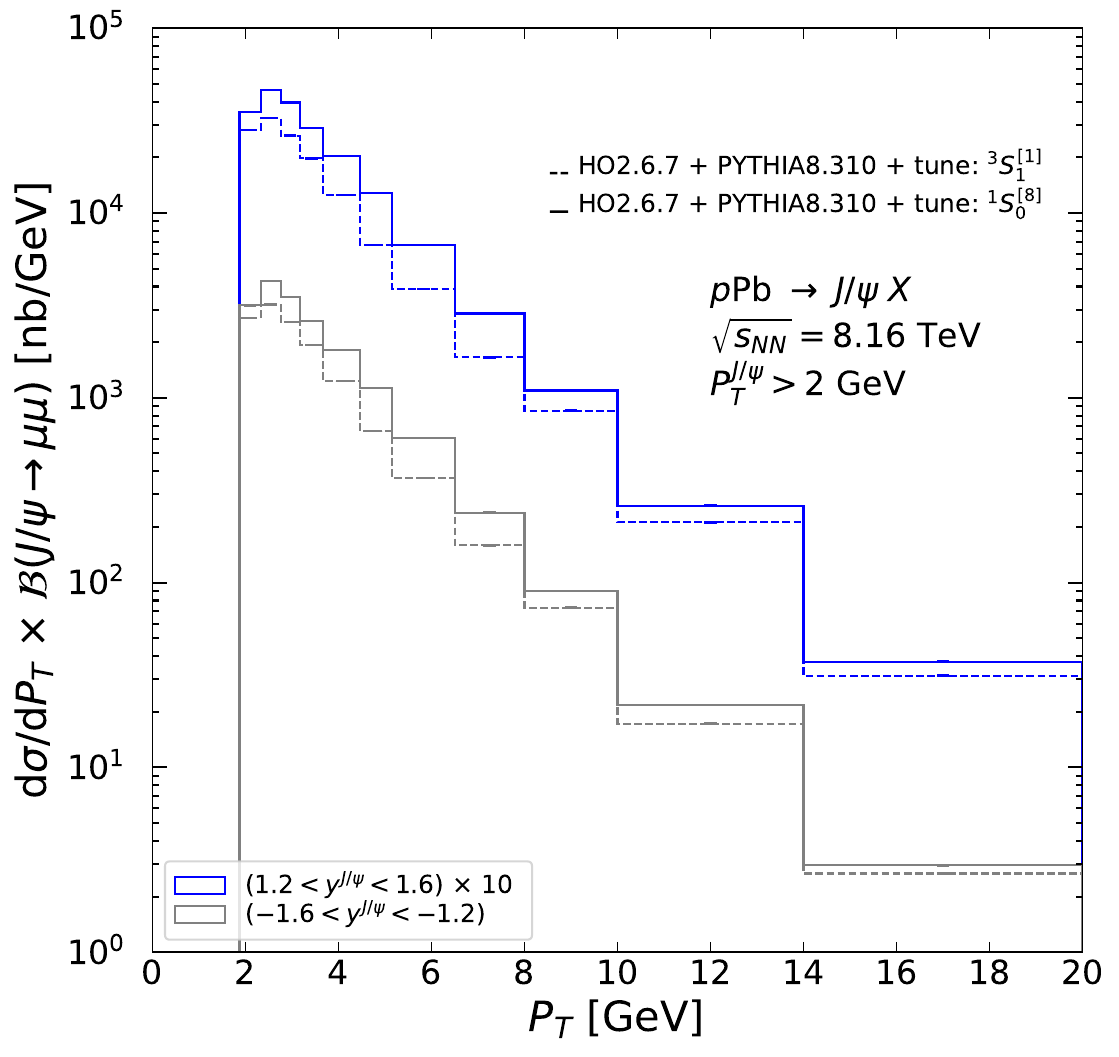}}\vspace*{-0.4cm}\\
    \subfloat[]{\includegraphics[scale=0.32]{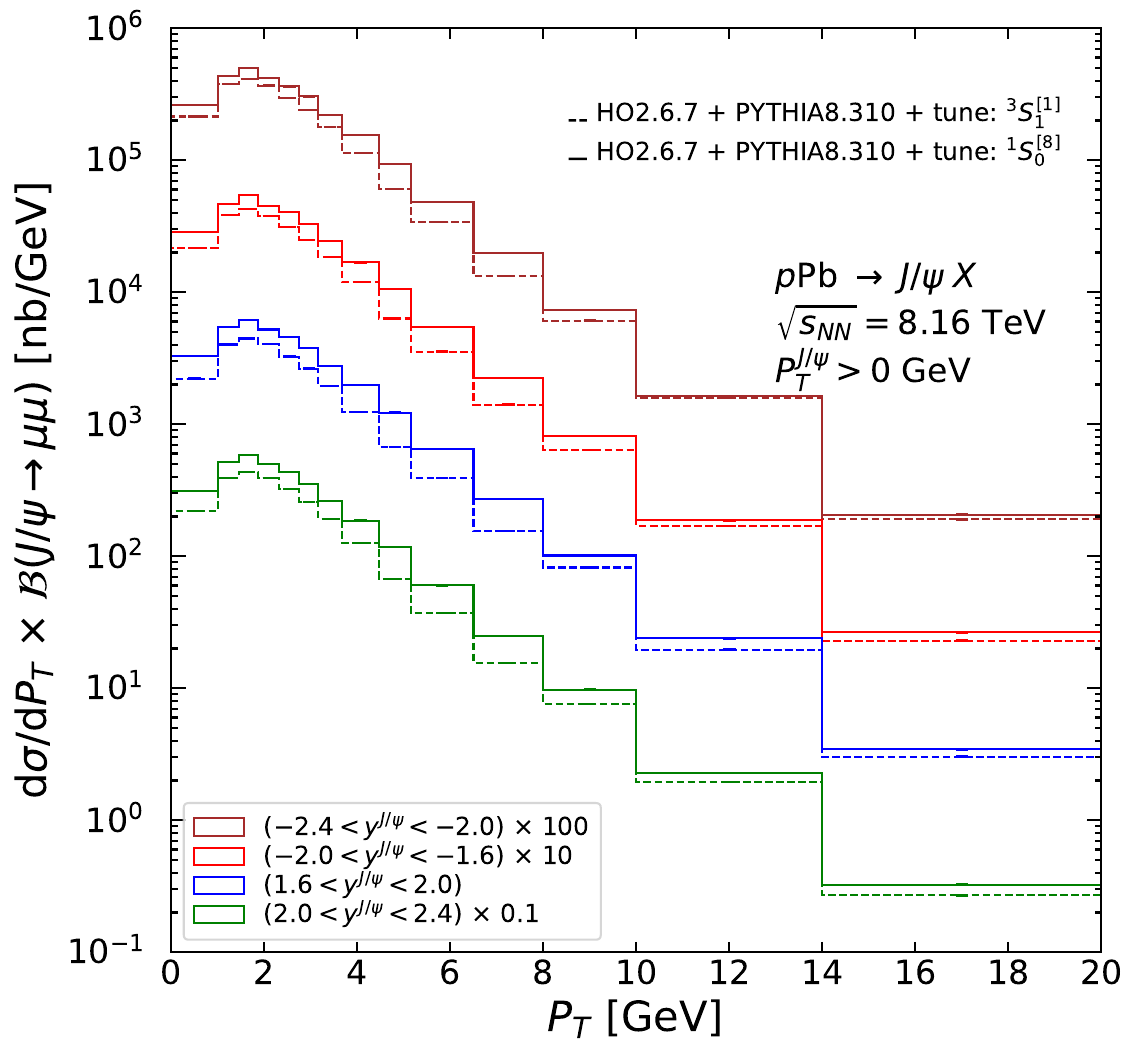}}\vspace*{-0.4cm}
    \caption{\pT-differential cross sections times the branching fraction of \jpsi to dimuons using the $^3S_1^{[8]}$ (solid) and $^3S_1^{[1]}$ (dashed) tunes, for hadroproduced \jpsi in the CMS acceptance for: (a) $|y^{J/\psi}|<1.2$, (b) $1.2<|y^{J/\psi}|<1.6$, and (c) $1.6<|y^{J/\psi}|<2.4$. The error on the cross section is the statistical uncertainty assuming an integrated luminosity of 1000~nb$^{-1}$.}
    \label{fig:CMSpt-atyp-hadro}
\end{figure}

\begin{figure}[hbt!]
    \centering
    \subfloat[]{\includegraphics[trim = 0cm 0cm 0cm 0cm,clip,scale=0.32]{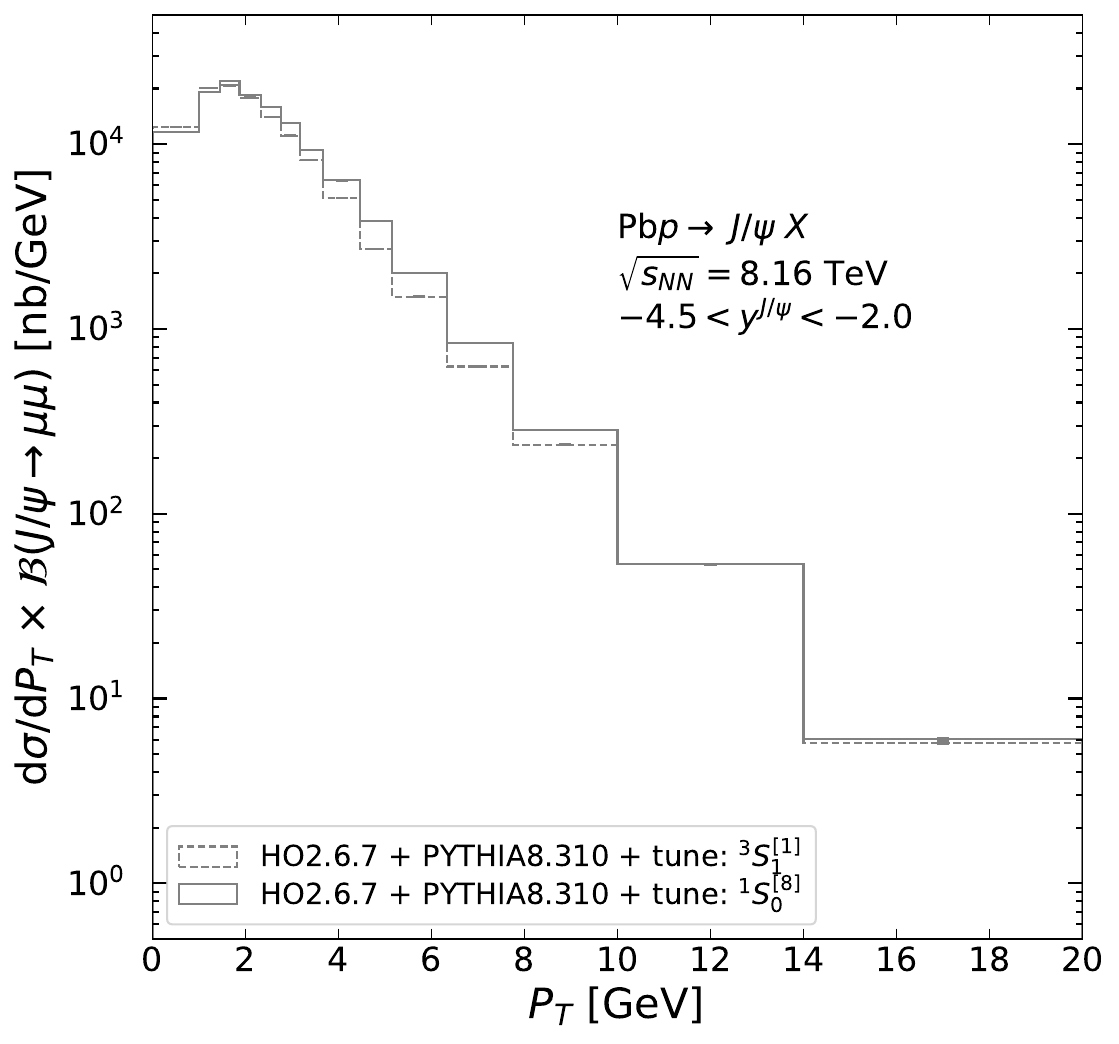}}\vspace*{-0.4cm}\\\subfloat[]{\includegraphics[trim = 0cm 0cm 0cm 0cm,clip,scale=0.32]{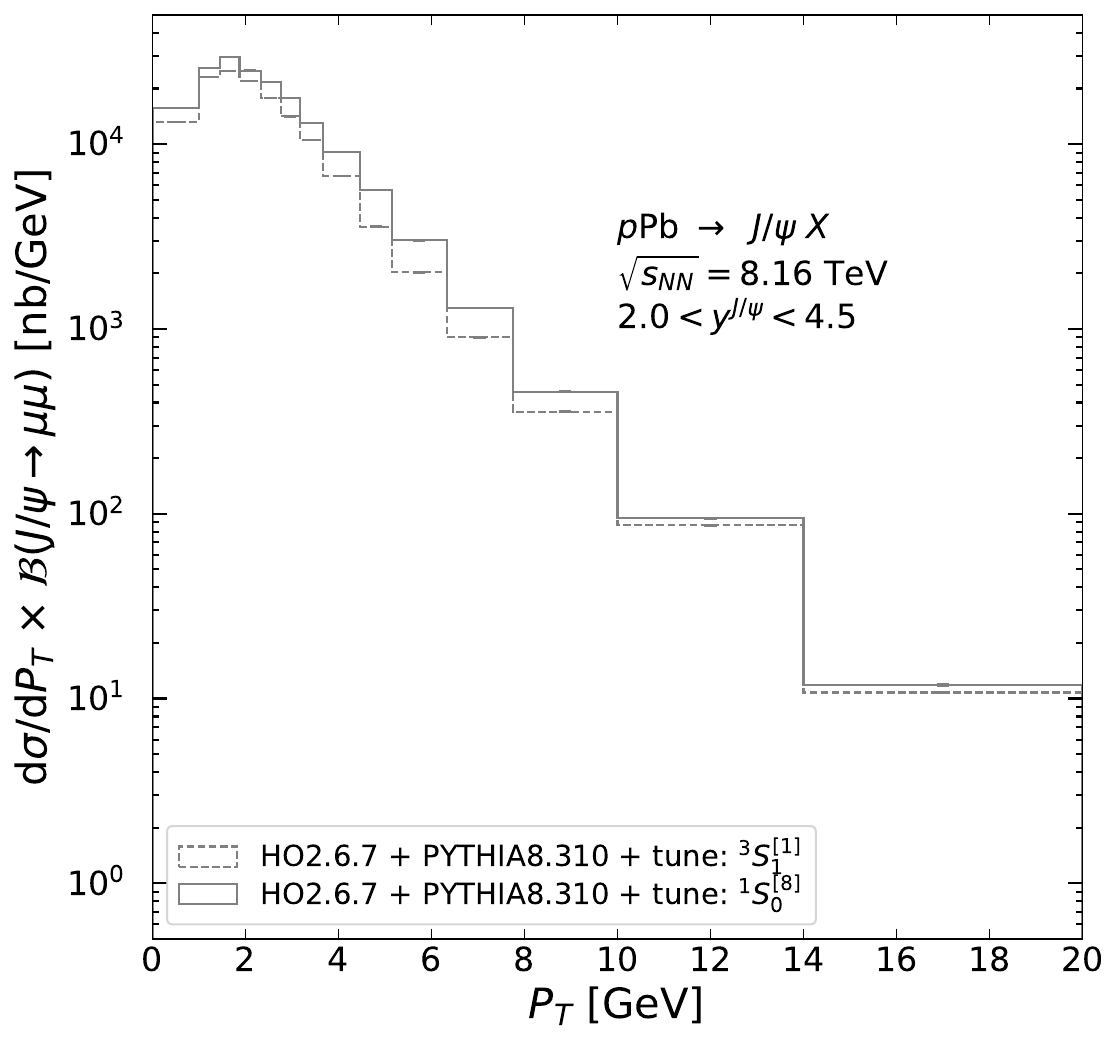}}\vspace*{-0.4cm}
    \caption{\pT-differential cross sections times the branching fraction of \jpsi to dimuons using the $^3S_1^{[8]}$ (solid) and $^3S_1^{[1]}$ (dashed) tunes for hadroproduced \jpsi in the LHCb acceptance in the (a) $p$Pb and (b) Pb$p$ beam configurations. The error on the cross section is the statistical uncertainty assuming an integrated luminosity of 200~nb$^{-1}$. }
    \label{fig:LHCbpt-hadro}
\end{figure}

\section[]{Modelling activity in HeRSCheL}
\label{appendix:hershel}

Fig.~\ref{fig:hershel} shows the \jpsi yield differential in the number of charged particles in the HeRSCheL acceptance on the Pb-going side, $N_\text{ch}$, for $J/\psi$ produced in proton-lead collisions, as obtained from the MC simulations described in Sections \ref{sec:signal} and \ref{sec:background}, as well as the probability of finding a number of charged particles less than $N_\text{ch}$, $P(N_\text{ch})$, for the singlet (dashed) and  octet (solid) tunes of photoproduction (blue) and hadroproduction (grey) in both LHCb beam configurations. As can be seen, photoproduction corresponds to an absence of charged particles, while hadroproduction corresponds to a large 
{number} of charged particles within the HeRSCheL acceptance. Hence, one 
{can} veto hadroproduction by requiring an absence of activity in the HeRSCheL detector. In principle, HeRSCheL is sensitive to a single charged particle. However, with the electronic settings employed at the start of 2015, between two and five charged particles need to cross the detector in order to generate a visible signal~\cite{Akiba:2018neu}. The vertical lines in Fig.~\ref{fig:hershel} show the background reducing potential of the HeRSCheL detector if it is indeed {only} sensitive to either two or five charged particles.

\begin{figure}[htb!]
\centering
\subfloat[]{\includegraphics[trim=0cm 0cm 0cm 0cm,clip,scale=0.32]{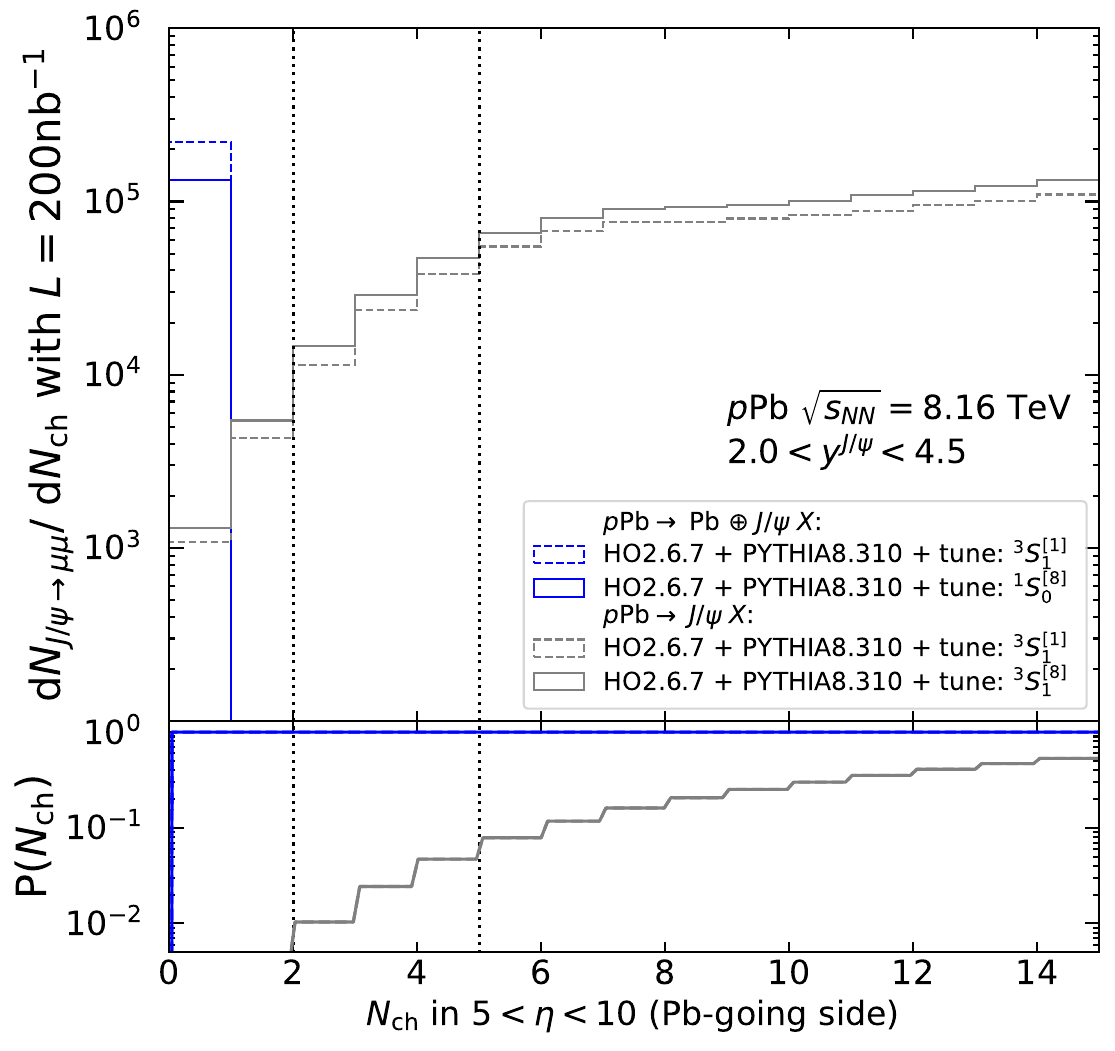}}\\\subfloat[]{\includegraphics[trim=0cm 0cm 0cm 0cm,clip,scale=0.32]{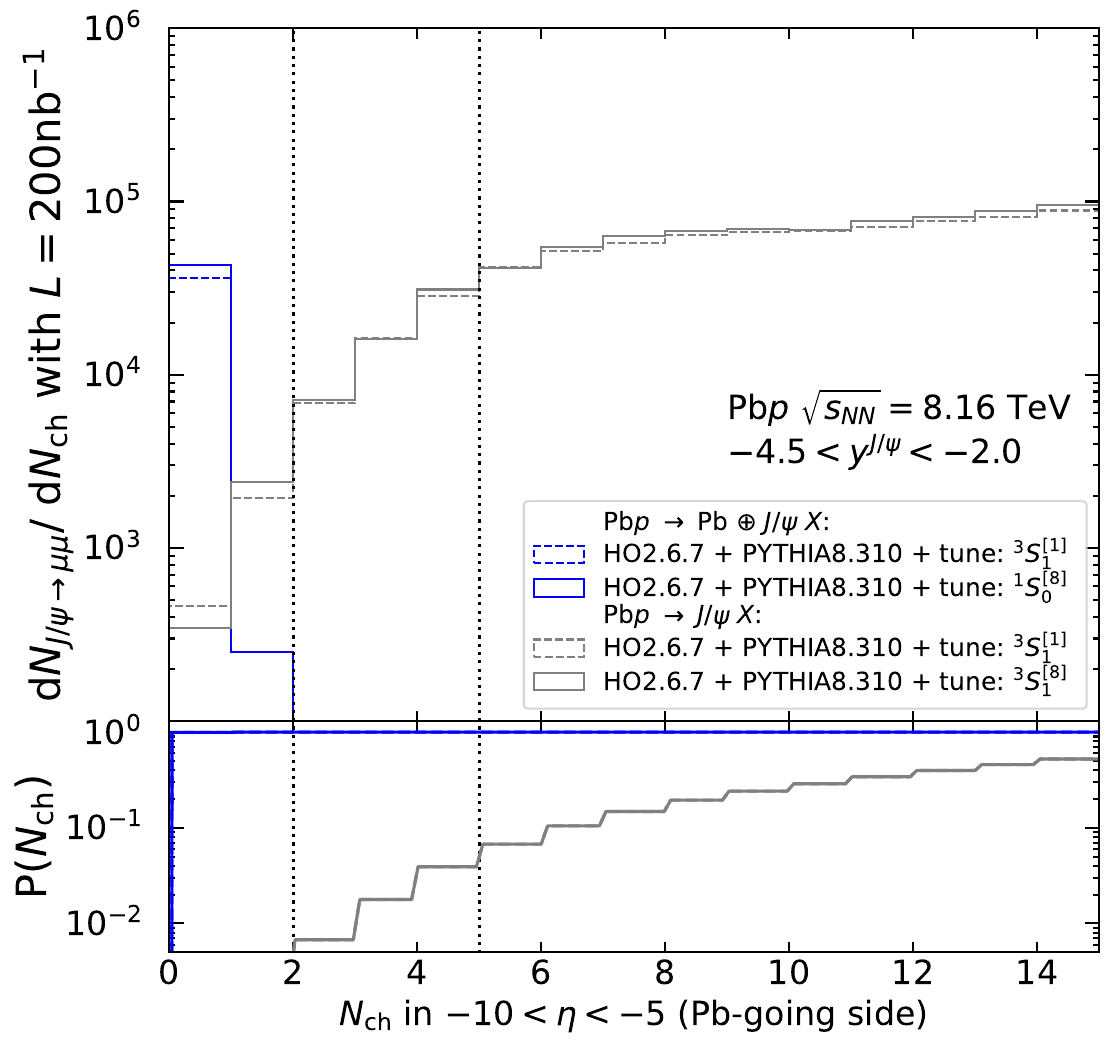}}

  \caption{Differential yield for $J/\psi\rightarrow \mu \mu$ as a function of the number of charged particles within the HeRSCheL acceptance, using the singlet (dashed) and octet (solid) tunes of photoproduction (blue) and hadroproduction (grey) for $J/\psi$ produced in the LHCb acceptance in the (a) $p$Pb and (b) Pb$p$ beam configurations. The lower panel shows the probability for finding a number of charged particles less than $N_\text{ch}$, $P(N_\text{ch})$. The dotted, vertical lines indicate the expected detector sensitivity.  } 
   \label{fig:hershel}
\end{figure}

\section[]{Additional rapidity-gap results}\label{Appendix:AdditionalRapidityGap}

\begin{figure*}[hbt!]
        \centering
\subfloat[]{\includegraphics[trim=0cm 0cm 0cm 0cm,clip,height=0.4\textwidth]{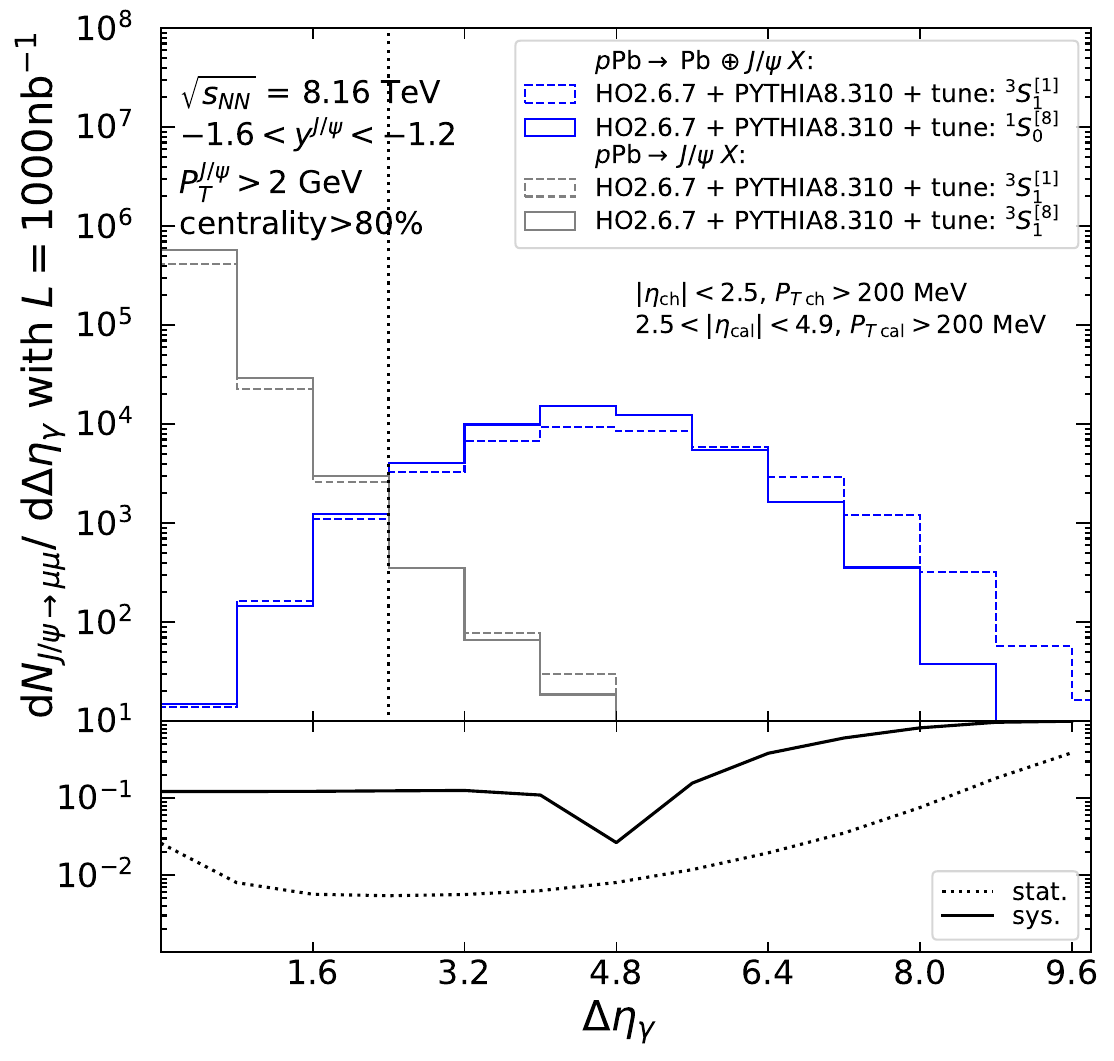}}
\subfloat[]{\includegraphics[trim=2.5cm 0cm 0cm 0cm,clip,height=0.4\textwidth]{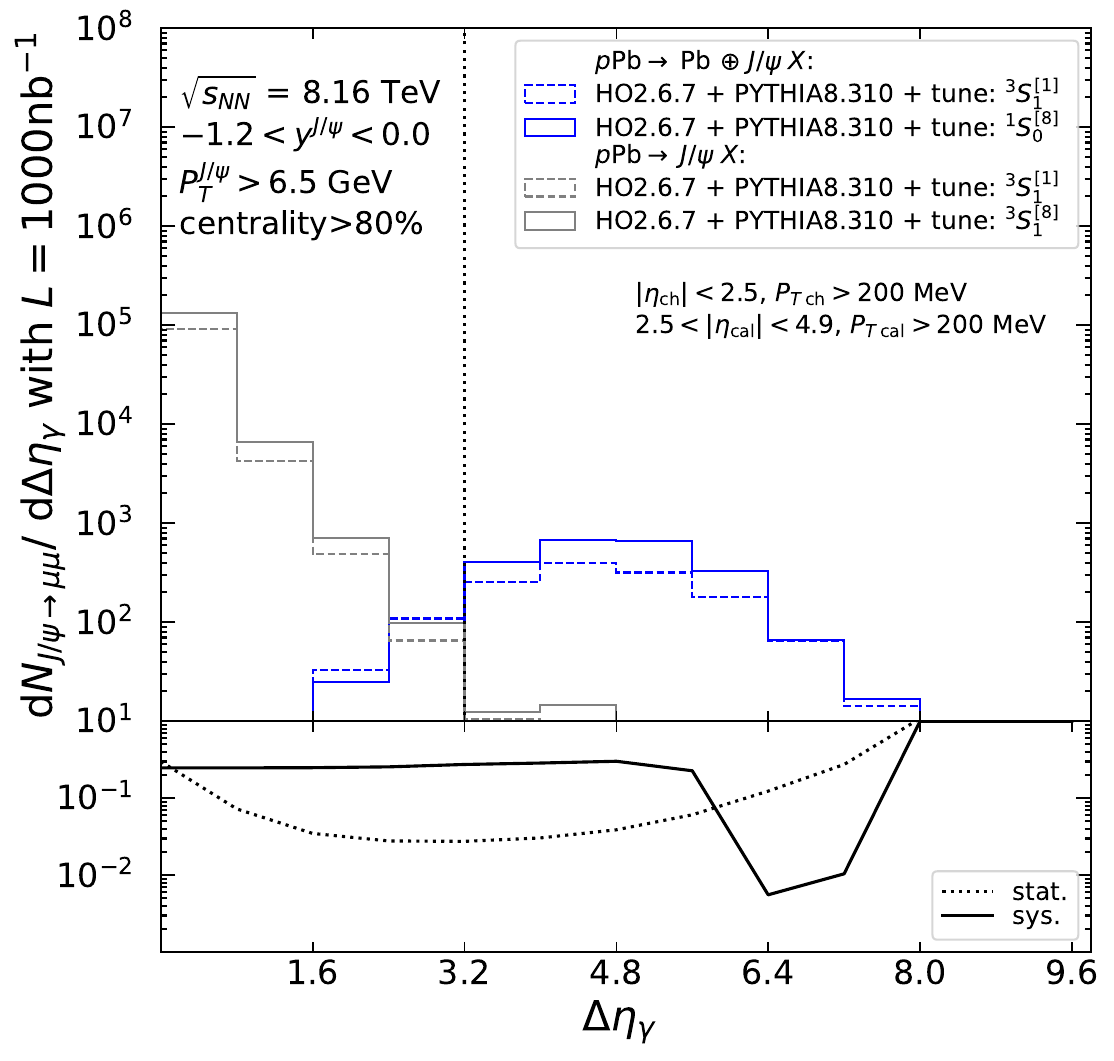}}\\
\subfloat[]{\includegraphics[trim=0cm 0cm 0cm 0cm,clip,height=0.4\textwidth]{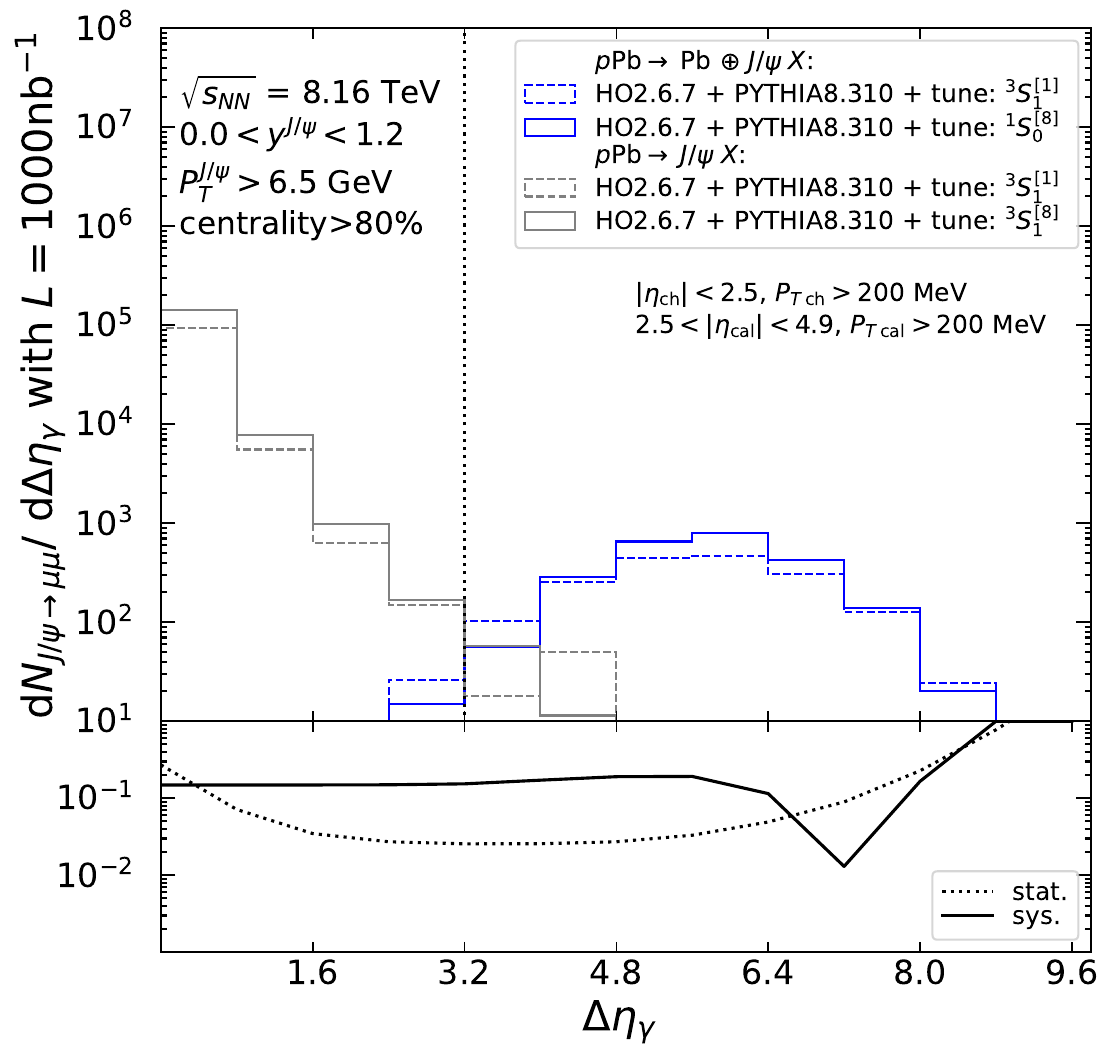}}
\subfloat[]{\includegraphics[trim=2.5cm 0cm 0cm 0cm,clip,height=0.4\textwidth]{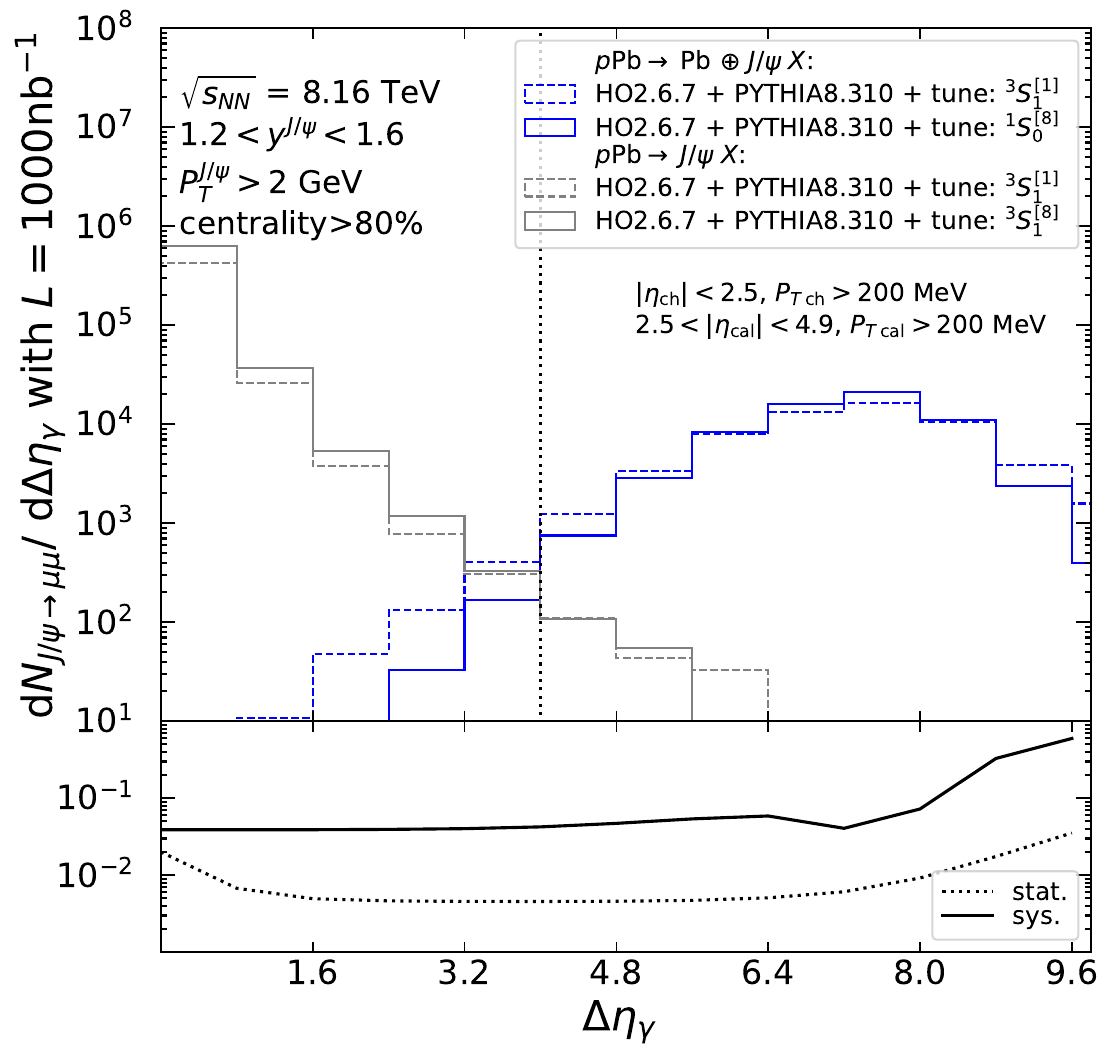}}

    \caption{Differential yield for $J/\psi\rightarrow \mu \mu$ as a function of $\Delta \eta_\gamma$ in the CMS low-\pT acceptance, using the singlet (dashed) and octet (solid) tunes of photoproduction (blue) and hadroproduction (grey) for (a) $-1.6<y^{\jpsi}<-1.2$, (b) $-1.2<y^{\jpsi}<0$, (c) $0<y^{\jpsi}<1.2$, and (d) $1.2<y^{\jpsi}<1.6$. The lower panel shows the relative statistical (dotted) and systematic (solid) uncertainties as a function of the cut value on $\Delta \eta_\gamma$. The dotted vertical line indicates the cut value that minimises the statistical uncertainty. }
 \label{fig:CMS-rapgap-ydiffappendix}
\end{figure*}

Figure~\ref{fig:CMS-rapgap-ydiffappendix} shows the \jpsi yield differential in $\Delta\eta_\gamma$ in the CMS acceptance, supplementary to those presented in Figs.~\ref{fig:CMS-rapgap-ydiff} and~\ref{fig:CMS-rapgap-ptdiff}.

Figure \ref{fig:ALICEATLAS-rapgap-ydiffappendix} shows the \jpsi yield differential in $\Delta \eta_\gamma$ in the ALICE acceptance, where $\Delta \eta_\gamma$ is determined using the central barrel alone. The acceptances of the central barrel and $\jpsi\rightarrow ee$ (resp. $\jpsi\rightarrow \mu \mu$) coincide (resp. differ). Separation between photo- and hadroproduction can be achieved based on a $\Delta \eta_\gamma$ requirement for \jpsi detected in the central barrel (Fig. \ref{fig:ALICE1}) and in the muon arm in the $p$Pb beam configuration (Fig. \ref{fig:ALICE2}). However, for \jpsi in the muon arm in the Pb$p$ beam configuration (Fig. \ref{fig:ALICE3}), the activity produced during the photoproduction process, which lies between the \jpsi and the broken proton, is captured by the central barrel. This results in %
{a similar dependence}  of the \jpsi yield as a function of $\Delta \eta_\gamma$ for both photo- and hadroproduced \jpsi, and so, no separation can be made. It is clear that 
using {only} the central barrel to detect the presence of rapidity gaps is not competitive with the CMS and ATLAS detectors.

\begin{figure}[hbt!]
        \centering
\subfloat[]{\label{fig:ALICE1}\includegraphics[trim=0cm 0cm 0cm 0cm,clip,scale=0.3]{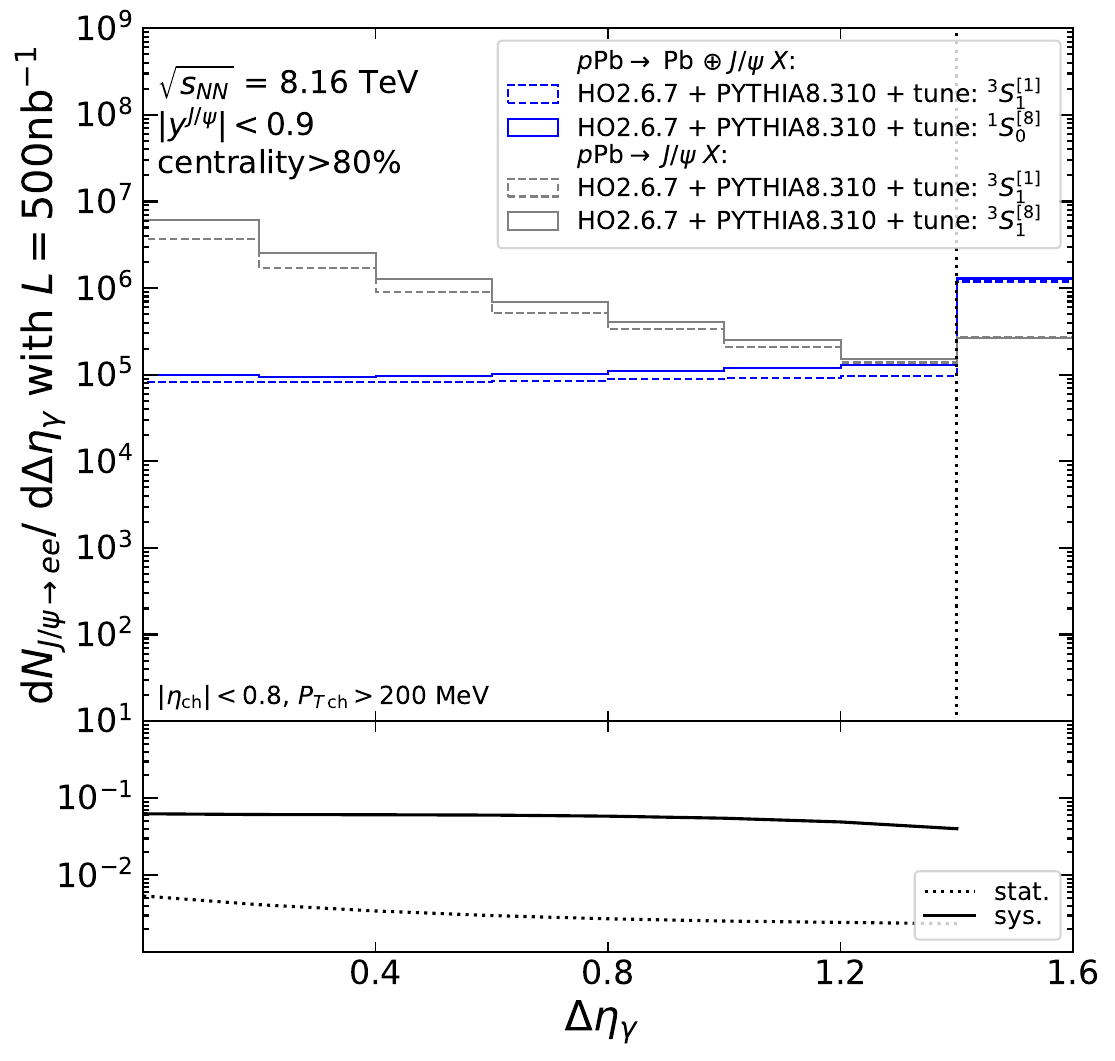}}\vspace*{-0.3cm}\\
\subfloat[] {\label{fig:ALICE2}\includegraphics[trim=0cm 0cm 0cm 0cm,clip,scale=0.3]{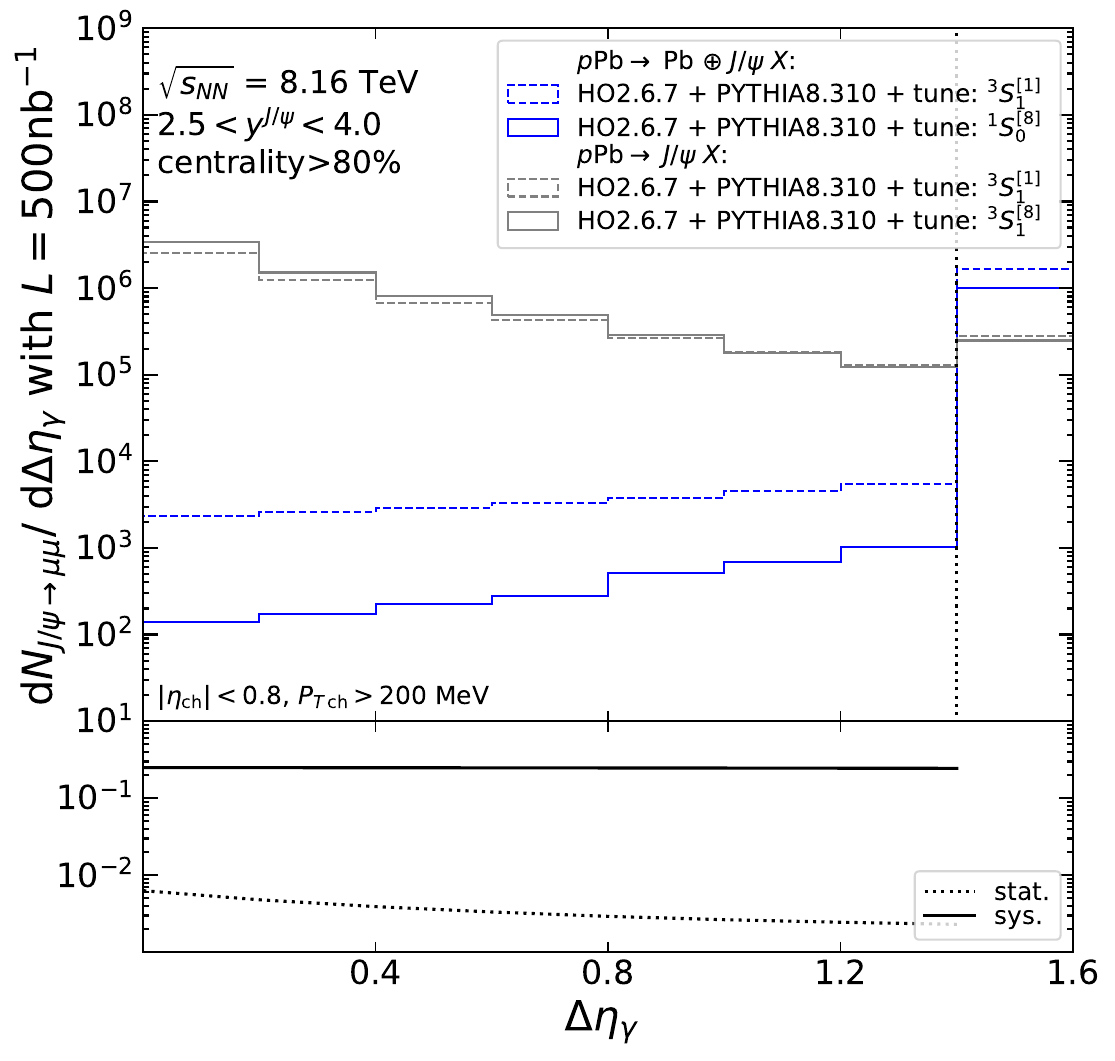}}\vspace*{-0.3cm}\\
\subfloat[]{\label{fig:ALICE3}\includegraphics[trim=0cm 0cm 0cm 0cm,clip,scale=0.3]{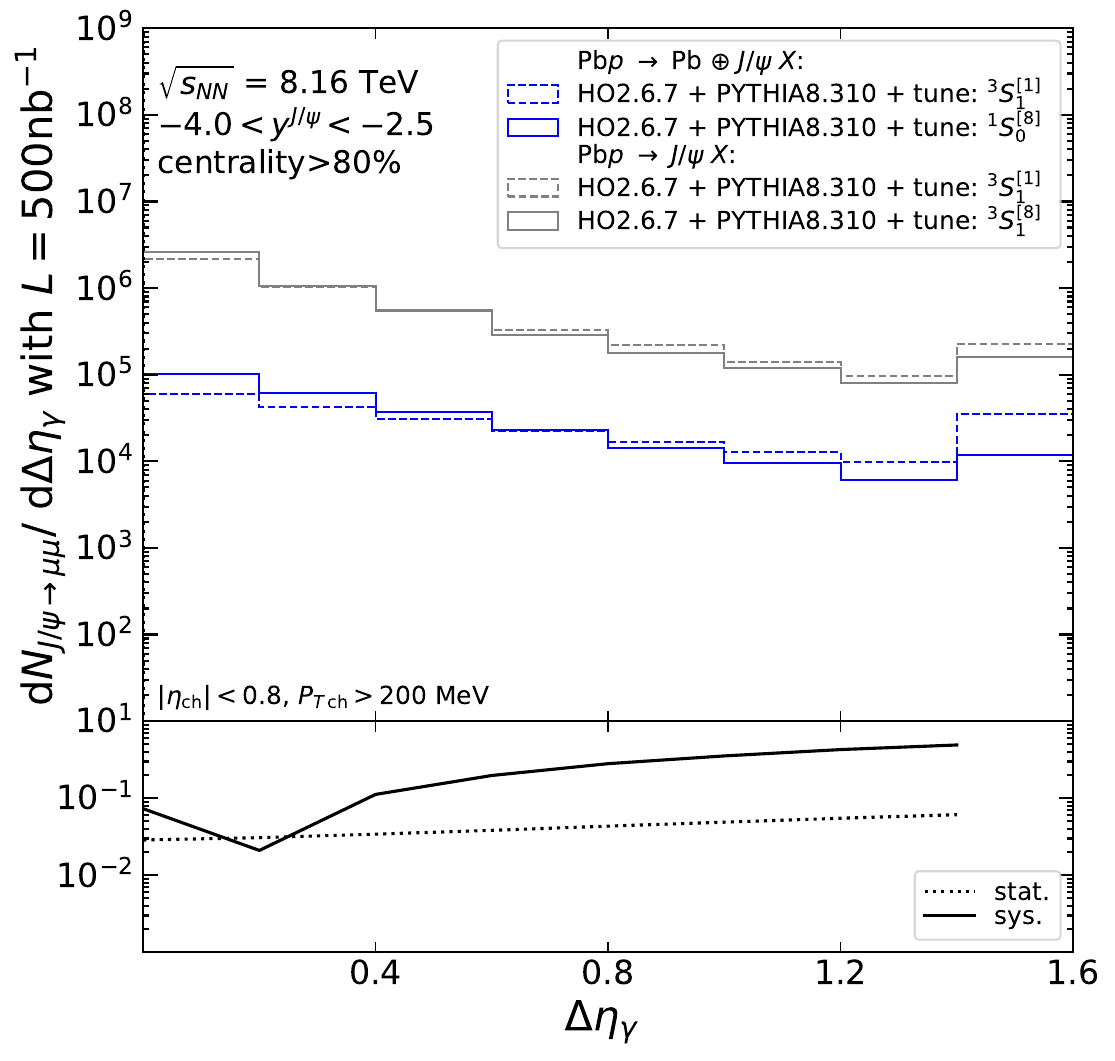}}\vspace*{-0.3cm}
    \caption{Differential yield for (a) $J/\psi\rightarrow ee$ and (b,c) $J/\psi\rightarrow \mu \mu$, as a function of $\Delta \eta_\gamma$ in the ALICE acceptance, using the singlet (dashed) and octet (solid) tunes of photoproduction (blue) and hadroproduction (grey) for the (a,b) $p$Pb and (c) Pb$p$ beam configurations. The lower panel shows the relative statistical (dotted) and systematic (solid) uncertainties as a function of the cut value on $\Delta \eta_\gamma$. The dotted vertical line indicates the cut value that minimises the statistical uncertainty.   }

 \label{fig:ALICEATLAS-rapgap-ydiffappendix}
\end{figure}

Figure \ref{fig:CMS-rapgap-Ncoll1} shows, within the CMS acceptance, the differential \jpsi yield distributions with respect to rapidity gaps based on (a) $\Delta \eta_\gamma$ and (b) $\sum \Delta \eta_\gamma$ for direct photo- and hadroproduction in the peripheral limit ($N_\text{coll}=1$), as discussed in Section~\ref{sec:resolved}. The standard rapidity-gap definition is considerably more efficient at removing background than the cumulative gap definition. 

\begin{figure*}[hbt!]
        \centering
\subfloat[]{\label{fig:standardgap}\includegraphics[trim=0cm 0cm 0cm 0cm,clip,scale=0.45]{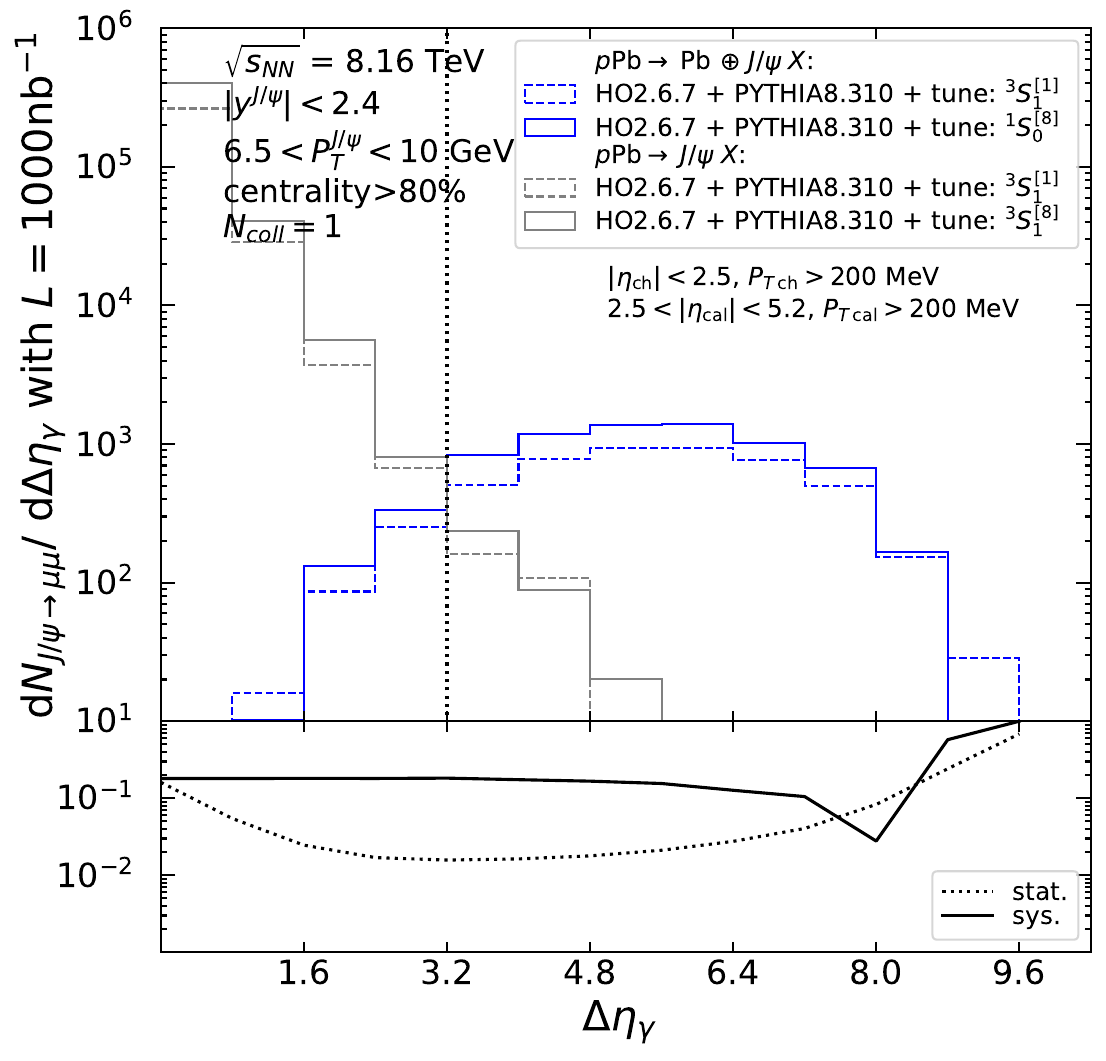}}\subfloat[]{\label{fig:sumofgaps}\includegraphics[trim=0cm 0cm 0cm 0cm,clip,scale=0.45]{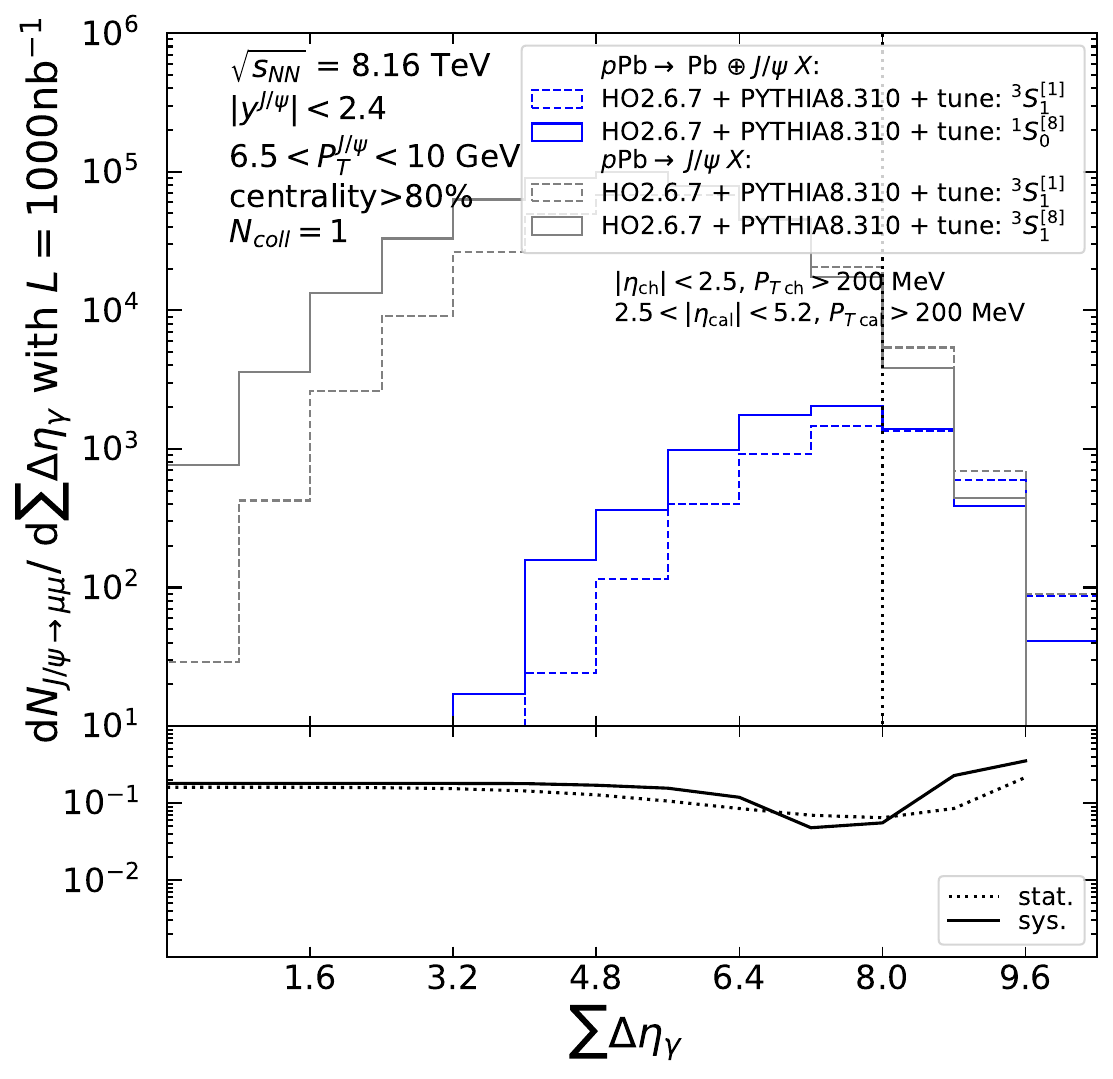}}
      
    \caption{ Comparison between (a) $\Delta\eta_\gamma$ and (b) $\sum\Delta\eta_\gamma$ for $J/\psi$ in the CMS low-\pT acceptance using the singlet (dashed) and octet (solid) tunes of direct photoproduction (blue) and hadroproduction (grey) in the peripheral limit with $N_\text{coll}=1$. The lower panel shows the relative statistical (dotted) and systematic (solid) uncertainties as a function of the cut value on $\Delta \eta_\gamma$ or $\sum\Delta\eta_\gamma$. 
    The dotted vertical line indicates the cut value that minimises the statistical uncertainty.  }
 \label{fig:CMS-rapgap-Ncoll1}
\end{figure*}

\section[]{ Neutron-emission probability}\label{appendix:NeutronEmissions}

We use methods described in \cite{Baltz:2002pp} (also in \cite{Broz:2019kpl}) to compute the probability for inclusive photoproduction of $J/\psi$ to be accompanied by $Y$ neutron emissions in $p$Pb collisions
. Here, we assume that a neutron emission from the lead ion is the result of photoabsorption from a photon emitted by the proton. We also assume that these photoabsorption processes are factorisable. The cross section of such a process can be expressed as 
the convolution of three \textit{probabilities}: $\mathcal{P}_{J/\psi}$, $\mathcal{P}_{Yn}$, and $\exp{-\mathcal{P}_H}$, over a surface in impact parameter space, $b$:
\begin{align}
\sigma(p\text{Pb}\rightarrow \text{Pb}(Xn) & \jpsi X) \nonumber= \\ &\int d^2b \mathcal{P}_{J/\psi}  (b) \mathcal{P}_{Yn} (b) \exp{-\mathcal{P}_H(b)},
\label{eq:AA}
\end{align}
The first of these is the probability to produce a $J/\psi$ meson off a nucleus having emitted a photon from the other and therefore contains two contributions:
\begin{equation}
\begin{split}
\mathcal{P}_{\jpsi}(b) =& 
\int dk \frac{d^3n_\text{Pb}(b,k)}{dkd^2b}\sigma_{\gamma p\rightarrow J/\psi X}(k) \\+& \int dk \frac{d^3n_{p}(b,k)}{dkd^2b}\sigma_{\gamma \text{Pb}\rightarrow J/\psi X}(k),
\end{split}
\label{eq:pvb}
\end{equation}
\begin{equation}
\frac{d^3n(b,k)}{dkd^2b} = \frac{Z^2 \alpha k}{\pi^2 \gamma_L^2} k^2\left(\frac{k b}{\gamma_L}\right),
\label{eq:d3n}
\end{equation}
where $\gamma_L$ is the boost factor between the source and target frames and $\alpha=1/137$ is the electromagnetic coupling constant.

{The second probability, $\mathcal{P}_{Yn}(b)$, is that of the Pb ion emitting a number of neutrons, $Y$, as a result of a photonuclear excitation:  we consider $Y\in\{0,1,X\}$, where $X$ is one or more neutrons.}
This probability to emit at least one neutron can be written as
\begin{equation}
\begin{split}
\mathcal{P}_{Xn}(b) =& 1- \mathcal{P}_{0n}(b),\\
\end{split}
\end{equation}
where
\begin{equation}
\begin{split}
\mathcal{P}_{0n}(b) =&  \exp(-\mathcal{P}_{Xn}^1(b))\\
\end{split}
\end{equation}
is the probability to emit zero neutrons. It is expressed in terms of
\begin{equation}
P^1_{Xn}(b) = \int^{k_\text{max}}_{k_\text{min}} dk \frac{d^3n_p(b,k)}{dkd^2b}\sigma_{\gamma \text{Pb}\rightarrow \text{Pb}^\prime+Xn}(k),
\label{eq:P1xn}
\end{equation}
which is the mean number of photonuclear excitations resulting in a final state with at least one neutron as a function of $b$ and where, following~\cite{Baltz:1996as}, $\sigma_{\gamma \text{Pb}\rightarrow \text{Pb}^\prime+Xn}(k)$ is extracted from data~\cite{Veyssiere:1970ztg,Lepretre:1981tf,Carlos:1984lvc,Armstrong:1971ns,Armstrong:1972sa,Michalowski:1977eg,Caldwell:1973bu}. We also consider a final state with one neutron, which is described by the proability \cite{Broz:2019kpl}:
\begin{equation}
\begin{split}
\mathcal{P}_{1n}(b) =& \mathcal{P}_{1n}^1 (b) \exp(-\mathcal{P}_{Xn}^1(b)).
\end{split}
\end{equation}
The quantity $\mathcal{P}_{1n}^1 (b)$ is approximated by imposing appropriate integration limits in \ce{eq:P1xn} (one could of course use precise data for $\gamma \text{Pb}\rightarrow \text{Pb}^\prime+1n$). The integration limits are chosen such that $\gamma \text{Pb}\rightarrow \text{Pb}^\prime+1n$ is the dominant process: the lower limit of integration is the neutron separation energy, which is 7.4~MeV for Pb~\cite{Baltz:1996as}, and the upper limit is 17~MeV, which is approximated from data~\cite{Veyssiere:1970ztg}.

\begin{table}[h!]
\caption{Inclusive \jpsi photoproduction cross section, 
its fractional contribution to the total cross section, and {the median impact parameter,} $\text{med}(b)${,} in $p$Pb and PbPb collision systems with different requirements on forward neutron emissions. }
\begin{center}
\begin{tabular}{ |c|ccc| } 
\hline
 & $\sigma$ & \% of total &$\text{med}(b)$  \\
\hline
\multicolumn{4}{|c|}{$p$ Pb @ LHC ($\sqrt{s_{NN}}=8.16$~TeV, $\gamma_{\text{CM}} = 4350$)}\\
\multicolumn{4}{|c|}{$p$ Pb $\rightarrow$ Pb+$X$n $\oplus$ $J/\psi$ $X$} \\

\hline
total & 55~$\mu$b & 100\% & 41~fm \\
$0n$ & 55~$\mu$b & 99.99\% & 41~fm \\
$1n$ & 3~nb & 0.005\% & 11~fm \\
$Xn$ &  7~nb & 0.01\% & 11~fm \\

\hline

\multicolumn{4}{|c|}{PbPb @ LHC ($\sqrt{s_{NN}}=5.12$~TeV, $\gamma_{\text{CM}} = 2730$)}\\
\multicolumn{4}{|c|}{PbPb $\rightarrow$ Pb+$X$n $\oplus$ $J/\psi$ $X$} \\

\hline
total & 12~mb & 100\% & 64~fm \\
$0n$ & 10~mb & 82\% & 92~fm \\
$1n$ & 0.6~mb & 5\% & 23~fm \\
$Xn$ &  2~mb & 18\% & 21~fm \\
 \hline
\end{tabular}
\end{center}
\label{tb:sigmaresults}
\end{table}

The final quantity to compute is $\mathcal{P}_{H}(b)$, which is the mean number of projectile nucleons that interact hadronically at least once. It follows that $\exp{-\mathcal{P}_{H}(b)}$ is the probability for no additional strong interactions. We use the hard-sphere approximation to simplify this expression:     \[
  \exp{-\mathcal{P}_{H}(b)}= 
\begin{cases}
    0,& \text{if }  b< R_1 + R_2,\\
    1, &\text{if }b> R_1 + R_2.
\end{cases}
\]  

The resulting cross sections and median impact parameters, $\text{med}(b)$, for the inclusive photoproduction of $J/\psi$ in $p$Pb collisions at 8.16~TeV and PbPb collisions at 5.12~TeV appear in Table \ref{tb:sigmaresults}. As can be seen, neutron emissions are suppressed by $\mathcal{O}(10^{-4})$ in $p$Pb collisions and by $\mathcal{O}(0.2)$ in PbPb collisions.

\section[]{Additional $z$ and $W_{\gamma p}$ reconstruction plots}\label{appendix:reconstructionCMS}

\begin{figure*}
\centering
\subfloat[]{\includegraphics[width=0.5\textwidth]{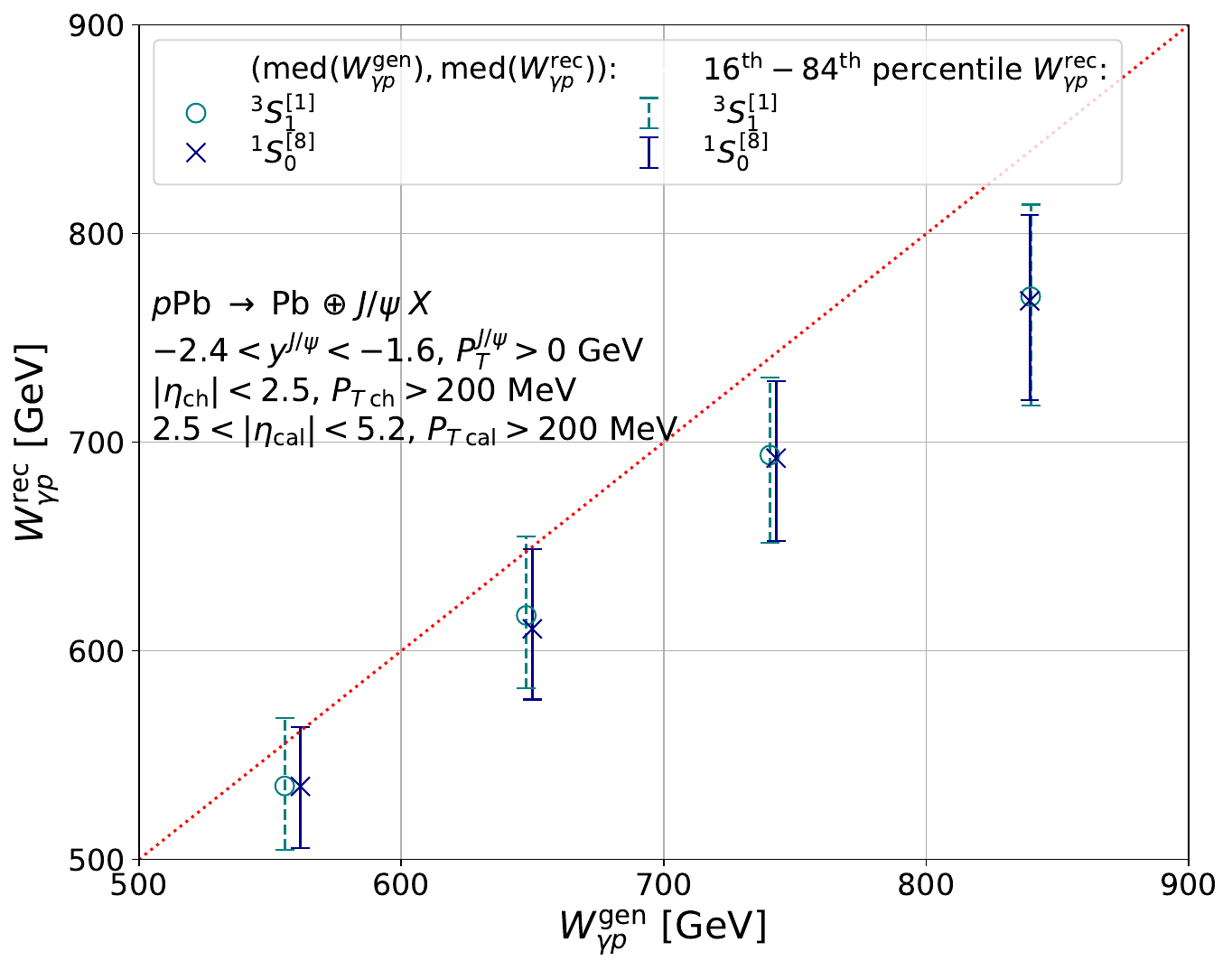}}
\subfloat[]{\includegraphics[width=0.5\textwidth]{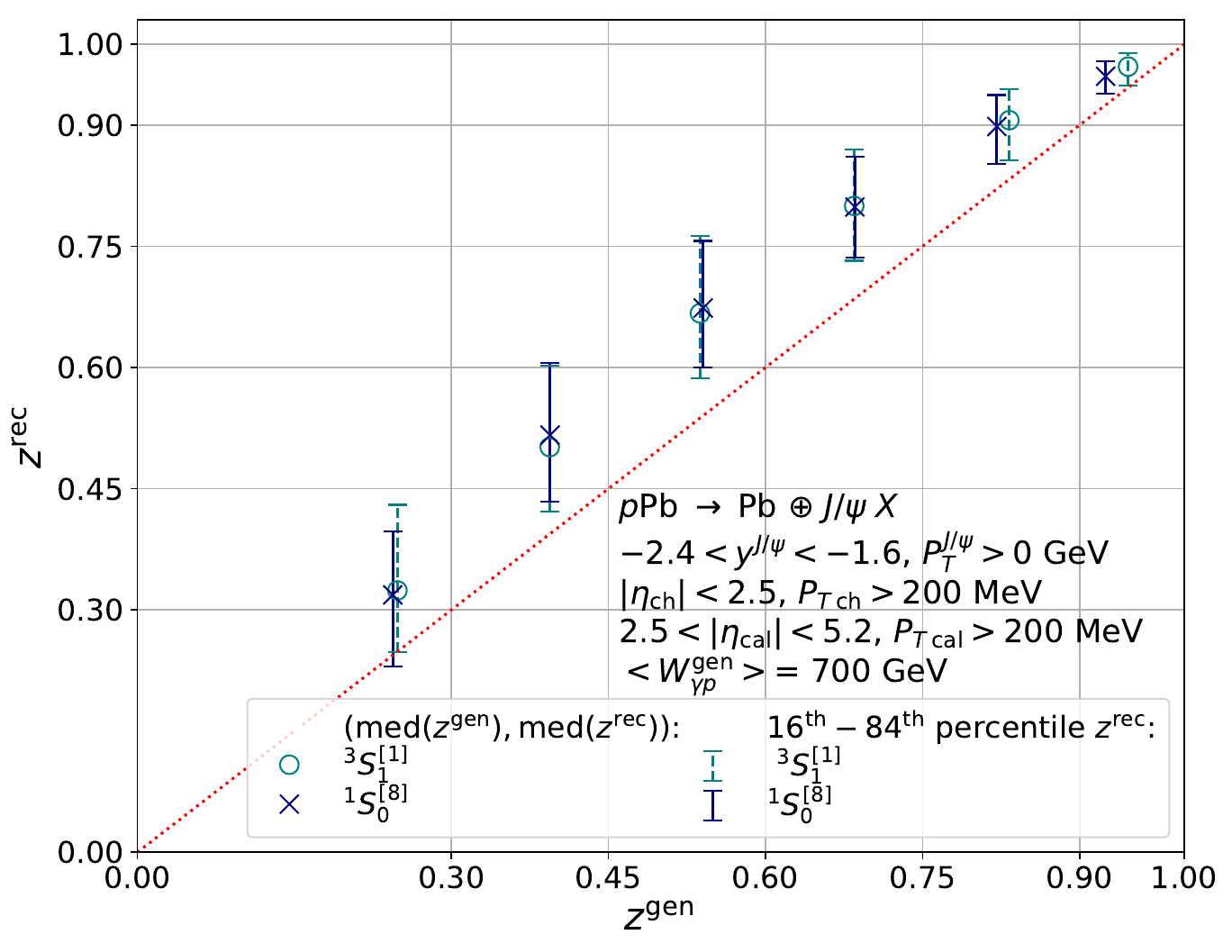}}\\
\subfloat[]{\includegraphics[width=0.5\textwidth]{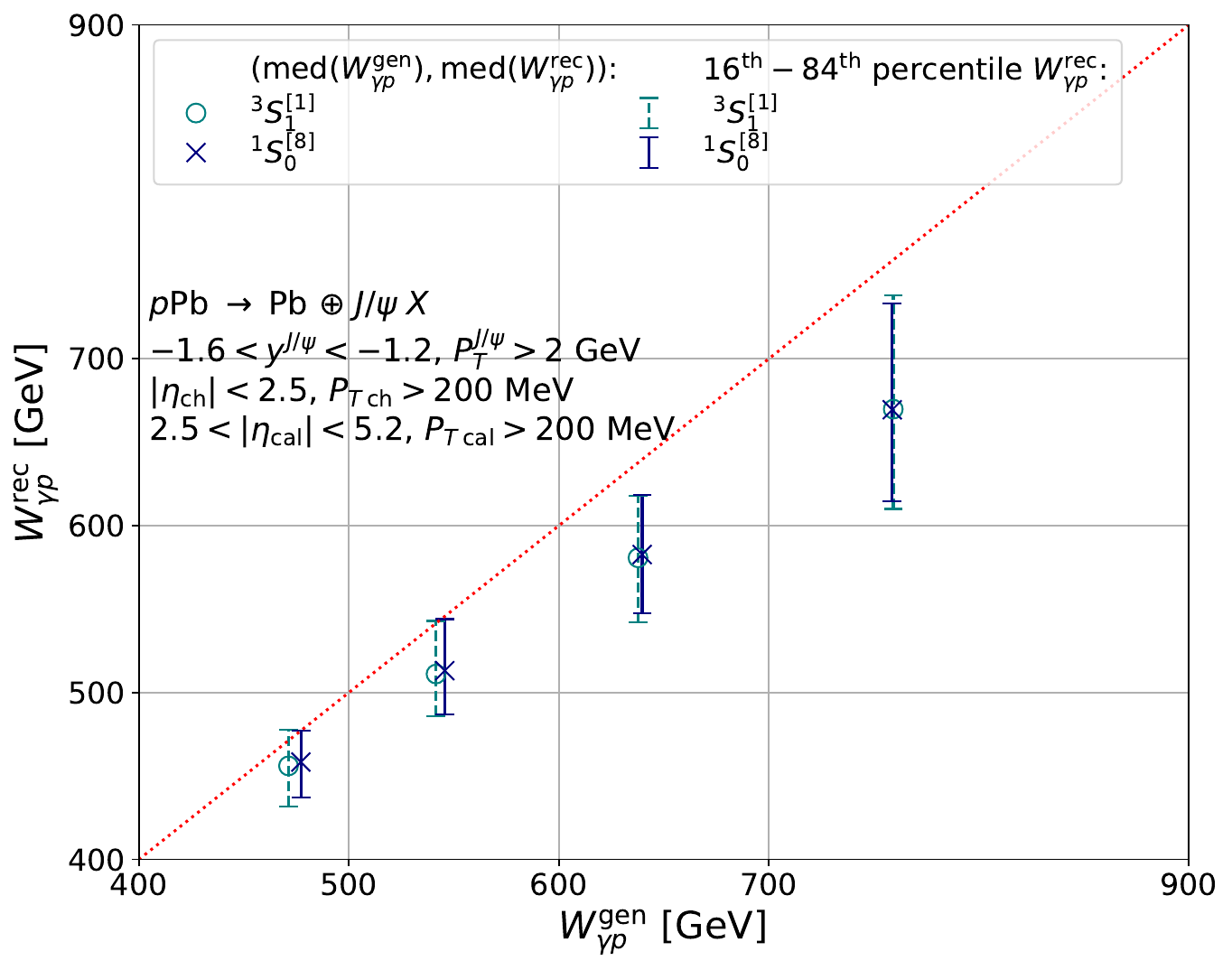}}
\subfloat[]{\includegraphics[width=0.5\textwidth]{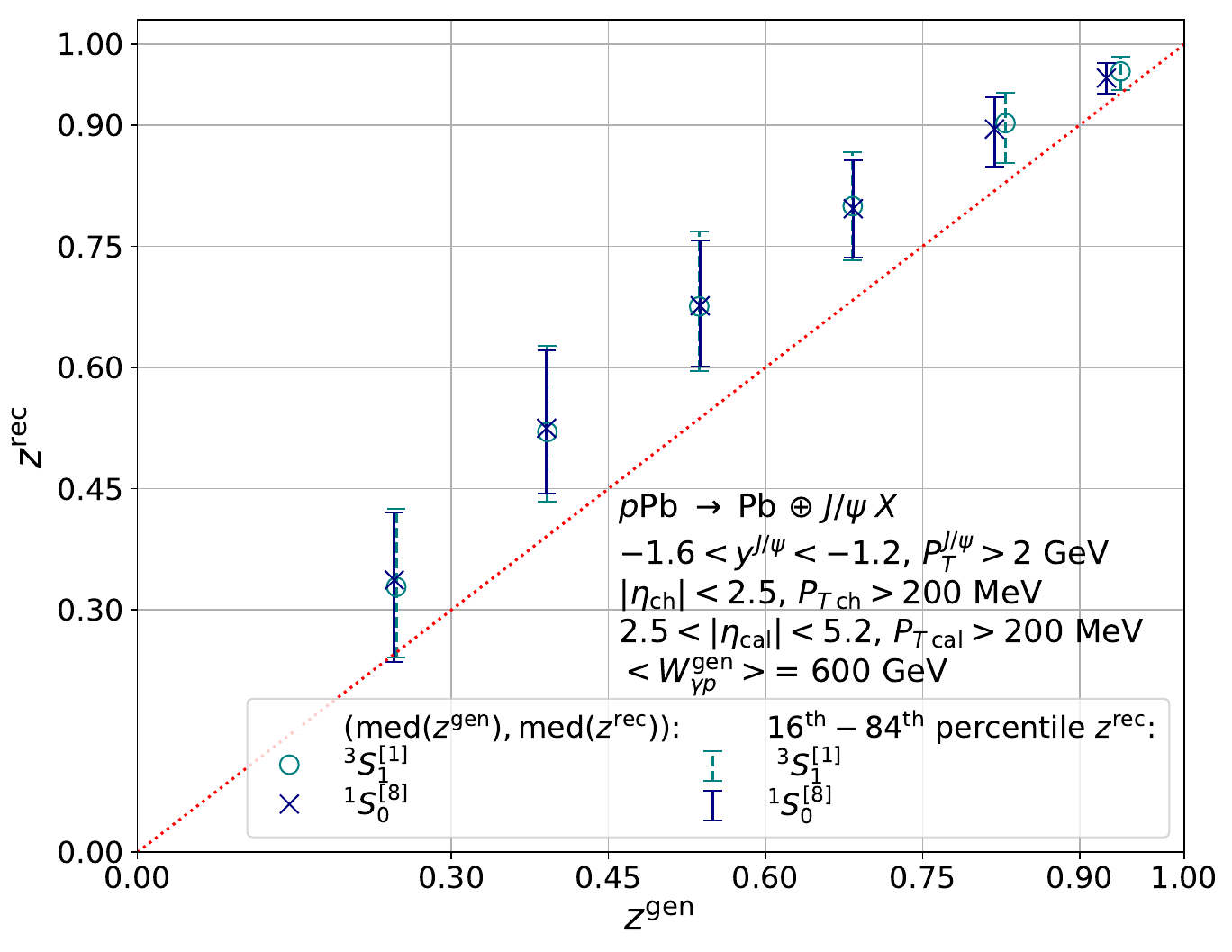}}\\
\subfloat[]{\includegraphics[width=0.5\textwidth]{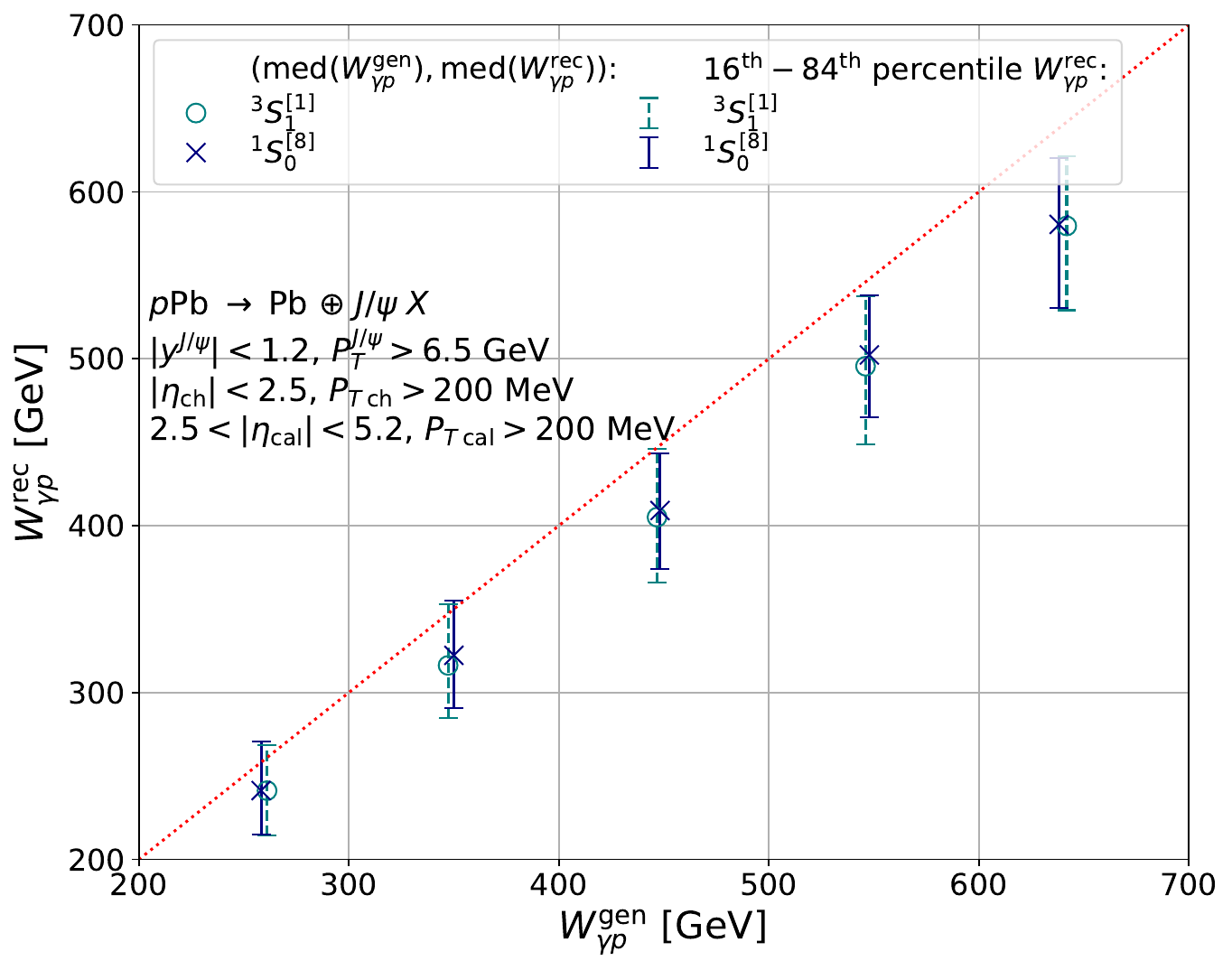}}
\subfloat[]{\includegraphics[width=0.5\textwidth]{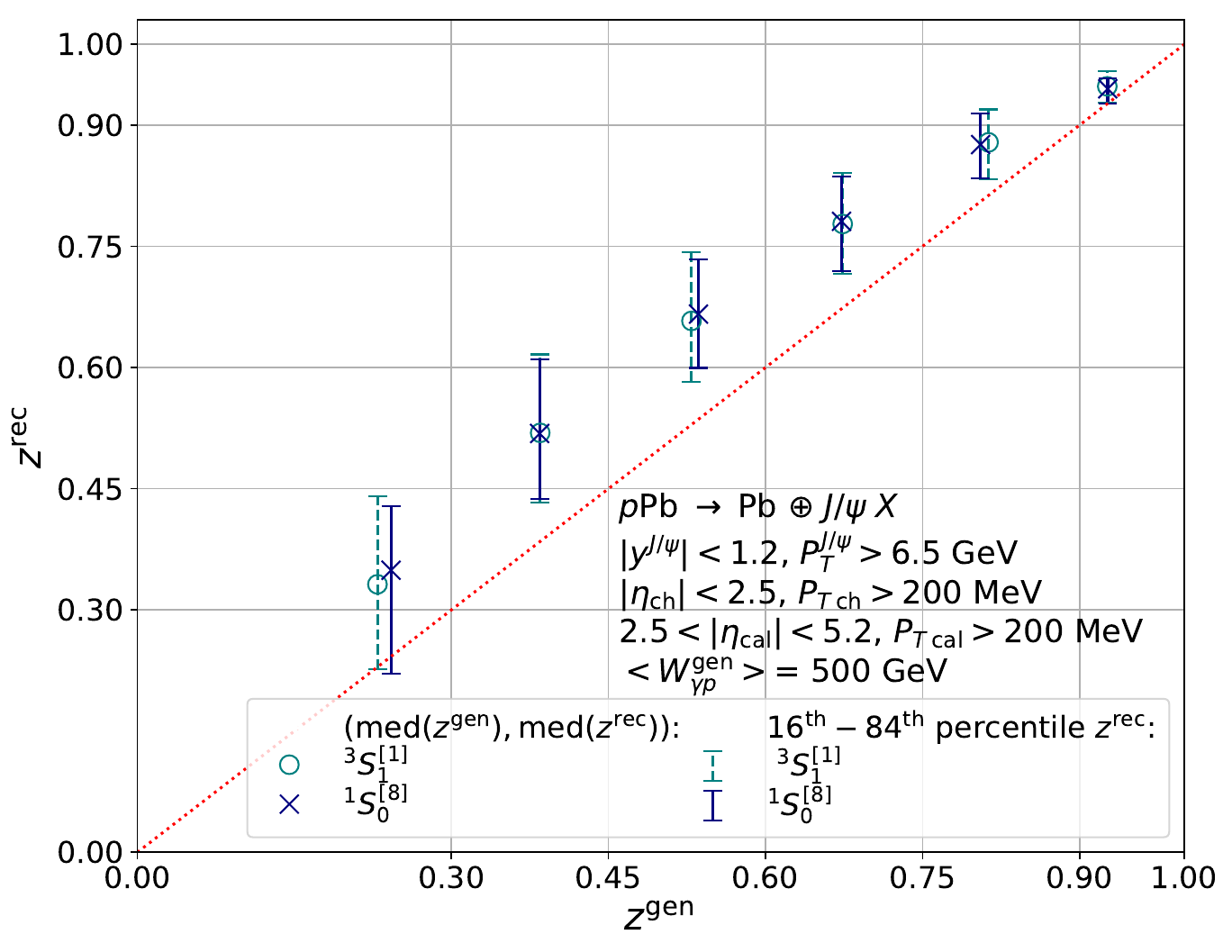}}\\
\caption{The median reconstructed (rec) values as a function of the median generated (gen) values of $W_{\gamma p}$ (a,c,e) and $z$ (b,d,f), using the tuned $^3S_1^{[1]}$ (teal circle) and $^1S_0^{[8]}$ (navy blue cross), for \jpsi reconstructed within the CMS acceptance in the region: (a,b) $-2.4<y^{\jpsi}<-1.6$, (c,d) $-1.6<y^{\jpsi}<-1.2$, and  (e,f) $|y^{\jpsi}|<1.2$. The lower and upper bounds on the error bars indicate the $16^{\rm th}$ and $84^{\rm th}$ percentile on the reconstructed values and the grid lines indicate the chosen binning.}

     \label{fig:recoCMSatyp1}
\end{figure*}

\begin{figure*}
\centering
\subfloat[]{\includegraphics[width=0.5\textwidth]{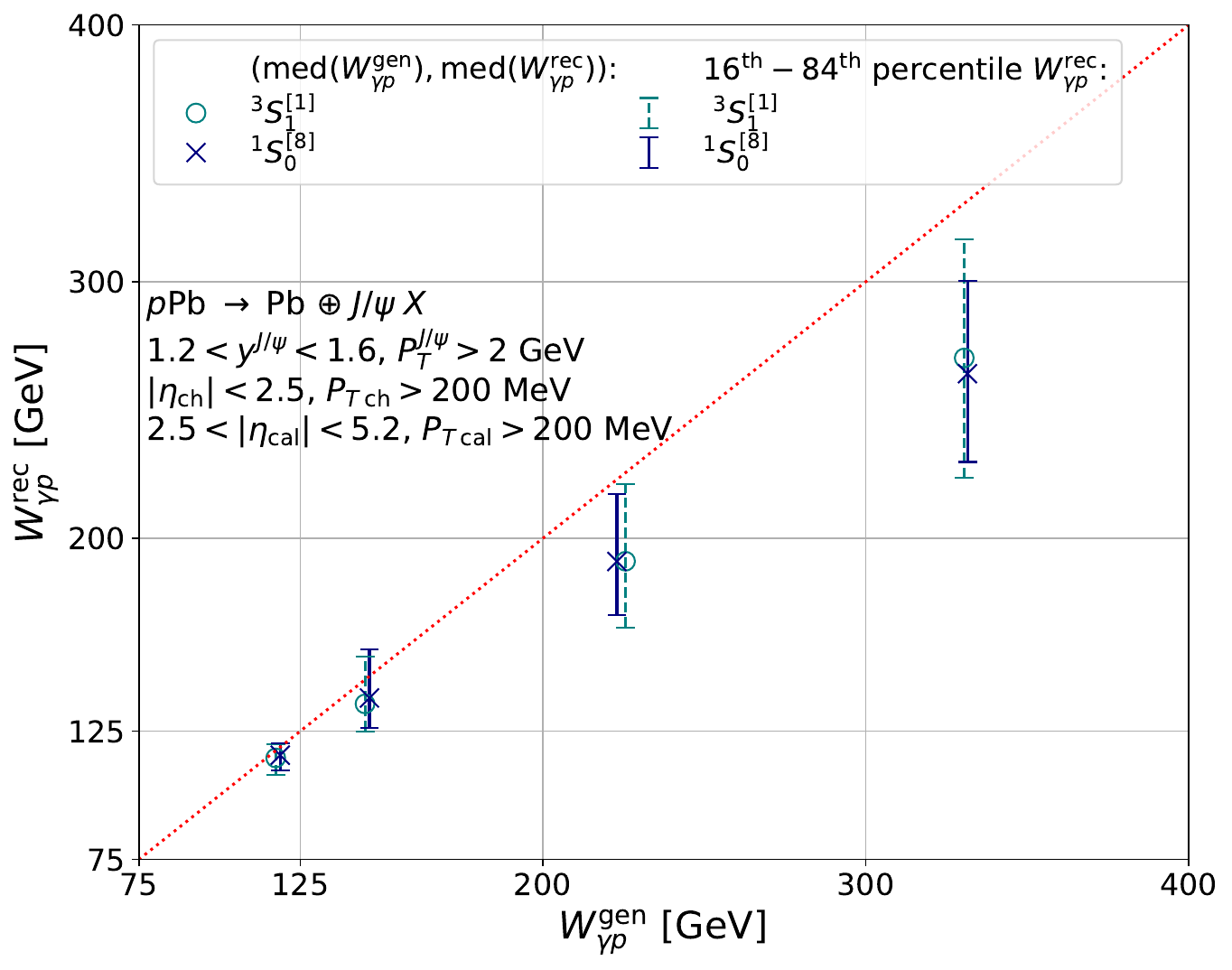}}
\subfloat[]{\includegraphics[width=0.5\textwidth]{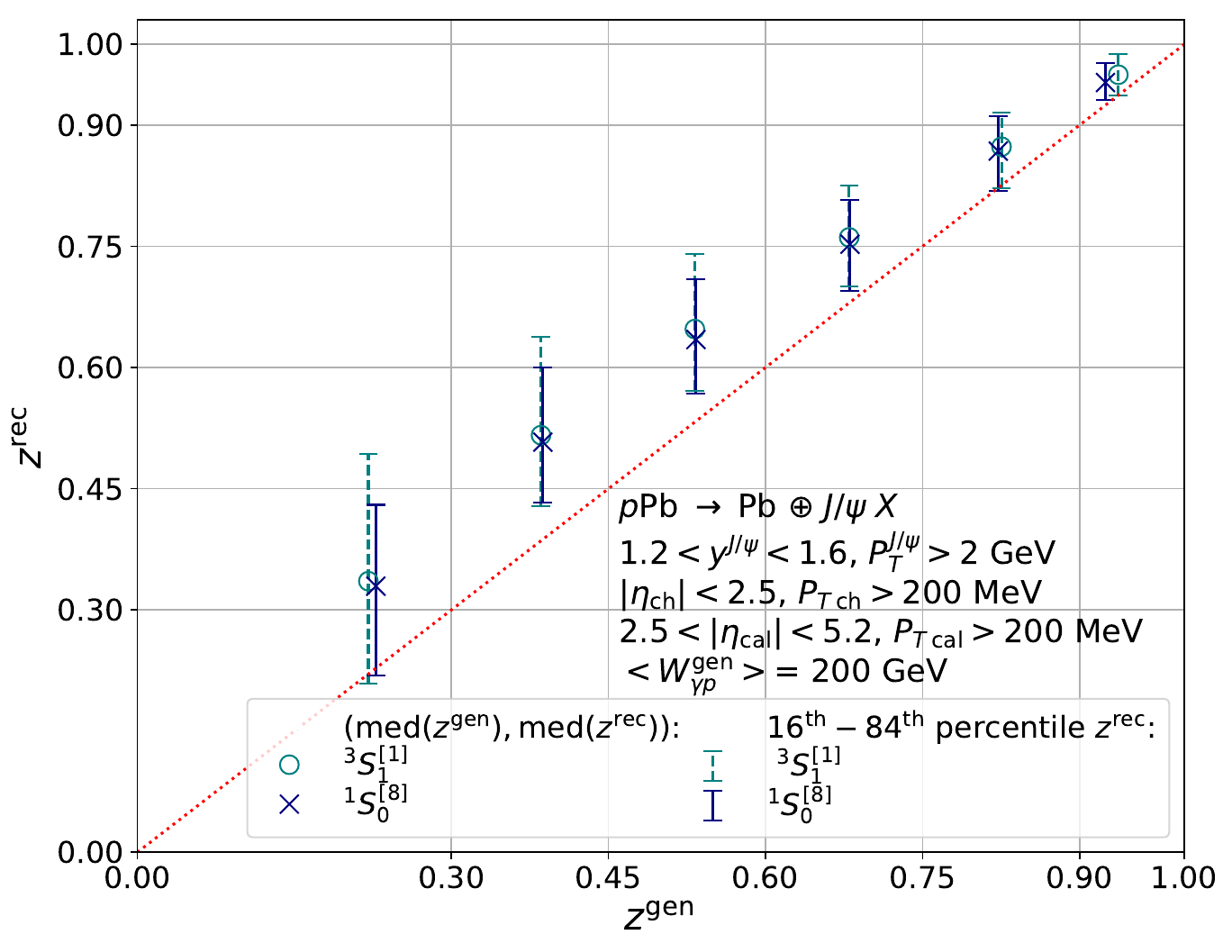}}\\
\subfloat[]{\includegraphics[width=0.5\textwidth]{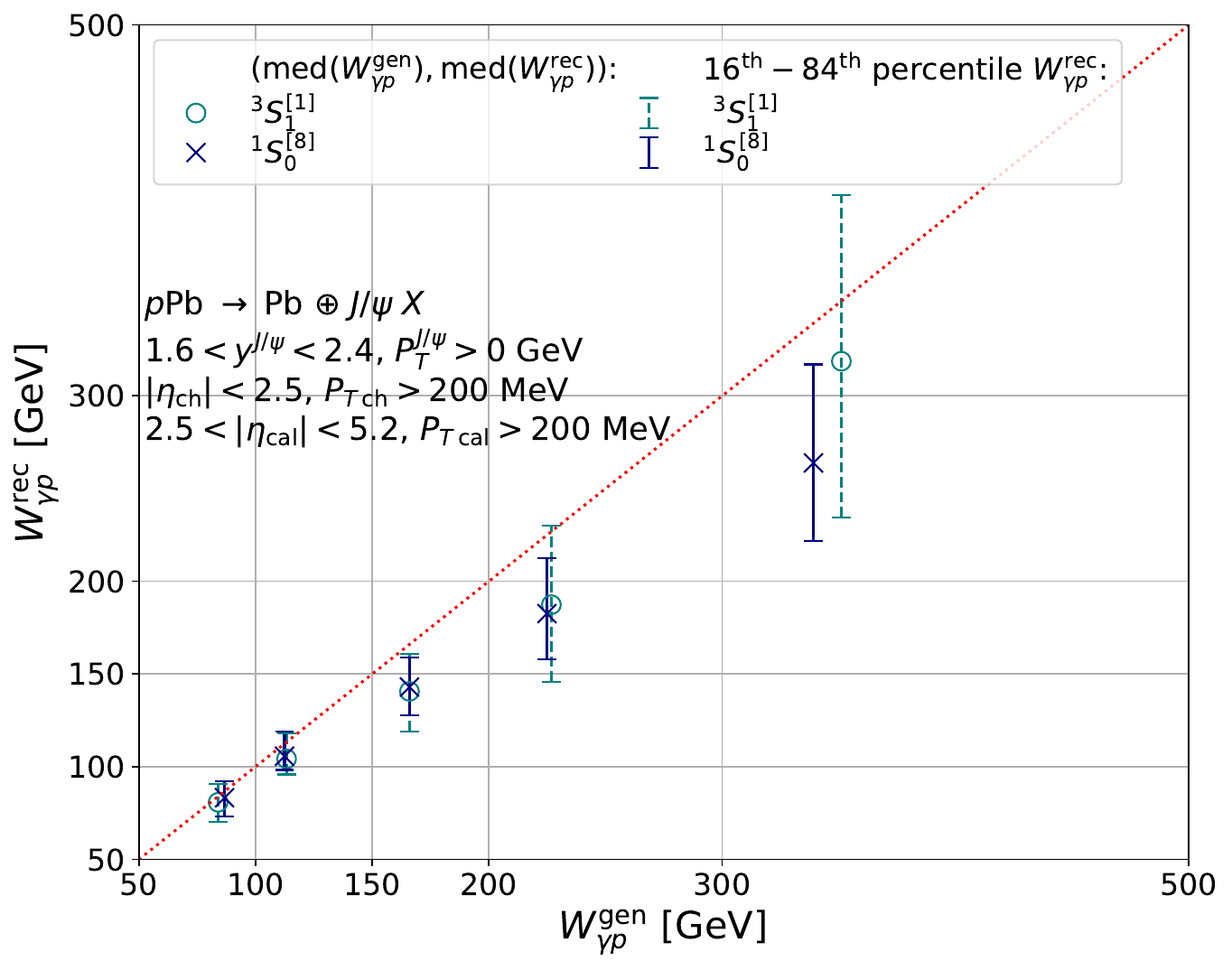}}
\subfloat[]{\includegraphics[width=0.5\textwidth]{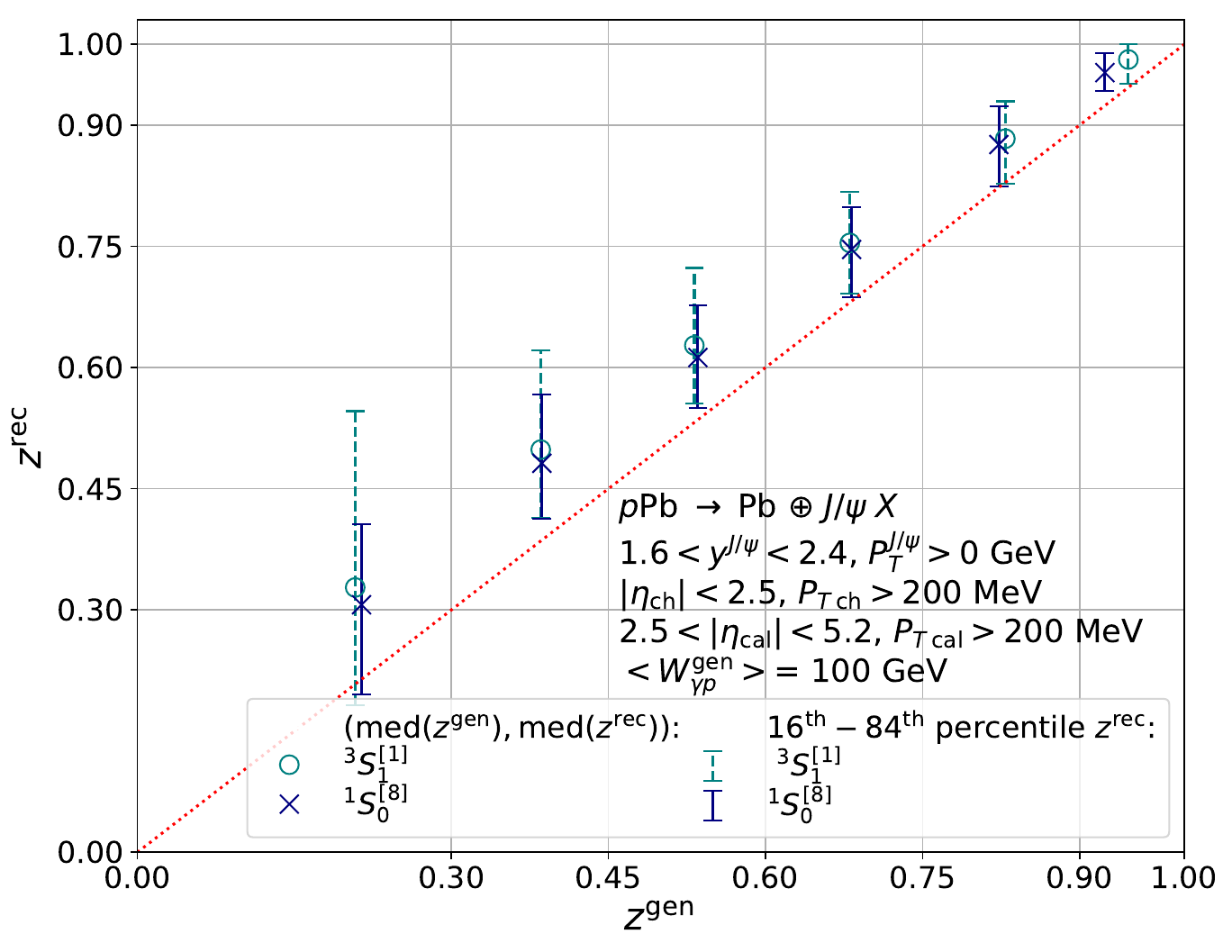}}\\

\caption{The median reconstructed (rec) values as a function of the median generated (gen) values of $W_{\gamma p}$ (a,c) and $z$ (b,d), using the tuned $^3S_1^{[1]}$ (teal circle) and $^1S_0^{[8]}$ (navy blue cross), for \jpsi reconstructed within the CMS acceptance in the region (a,b) $1.2<y^{\jpsi}<1.6$ and (c,d) $1.6<y^{\jpsi}<2.4$. The lower and upper bounds on the error bars indicate the $16^{\rm th}$ and $84^{\rm th}$ percentile on the reconstructed values and the grid lines indicate the chosen binning.}

          \label{fig:recoCMSatyp2}
\end{figure*}

\begin{figure*}
\centering
\subfloat[]{\includegraphics[width=0.5\textwidth]{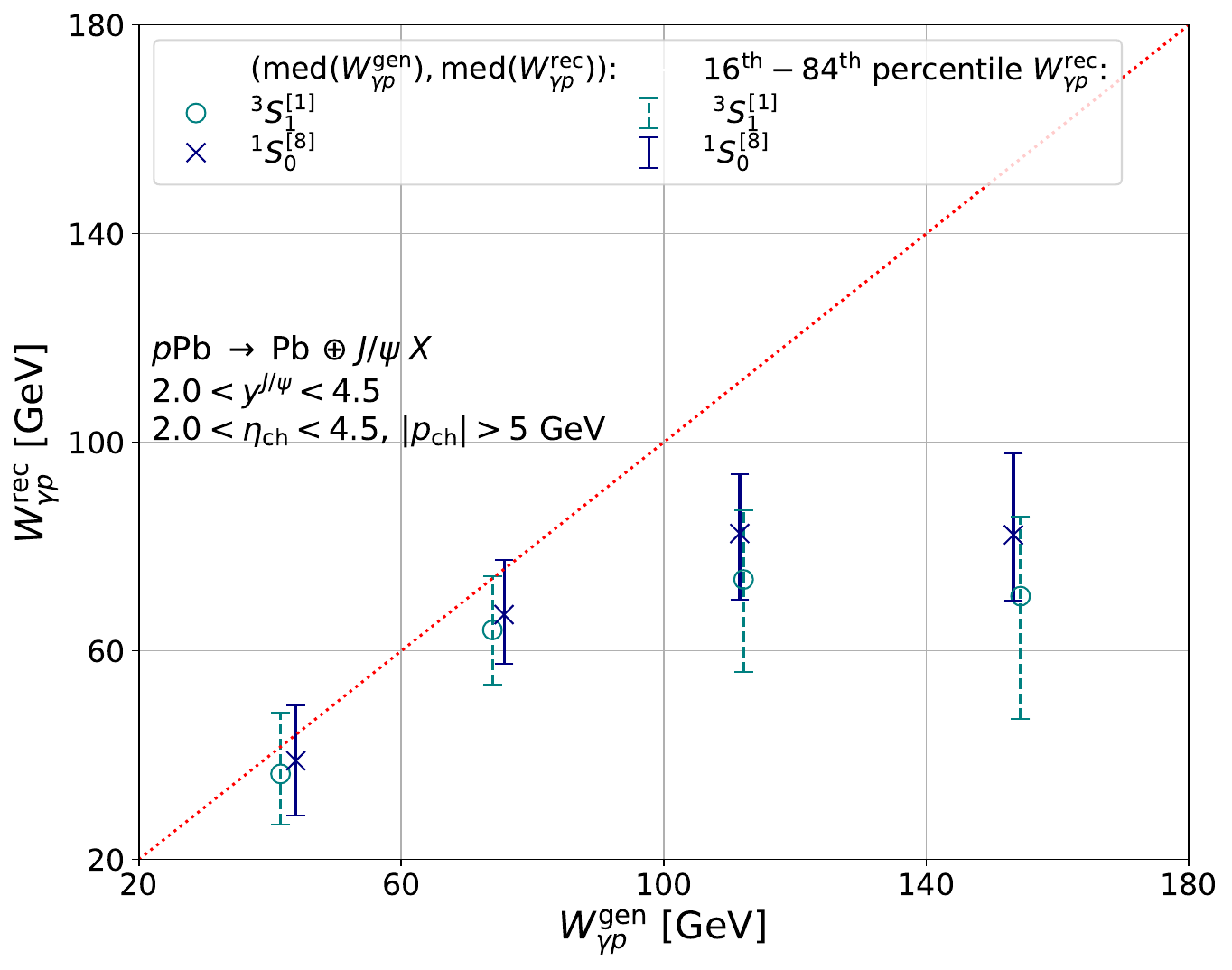}}
\subfloat[]{\includegraphics[width=0.5\textwidth]{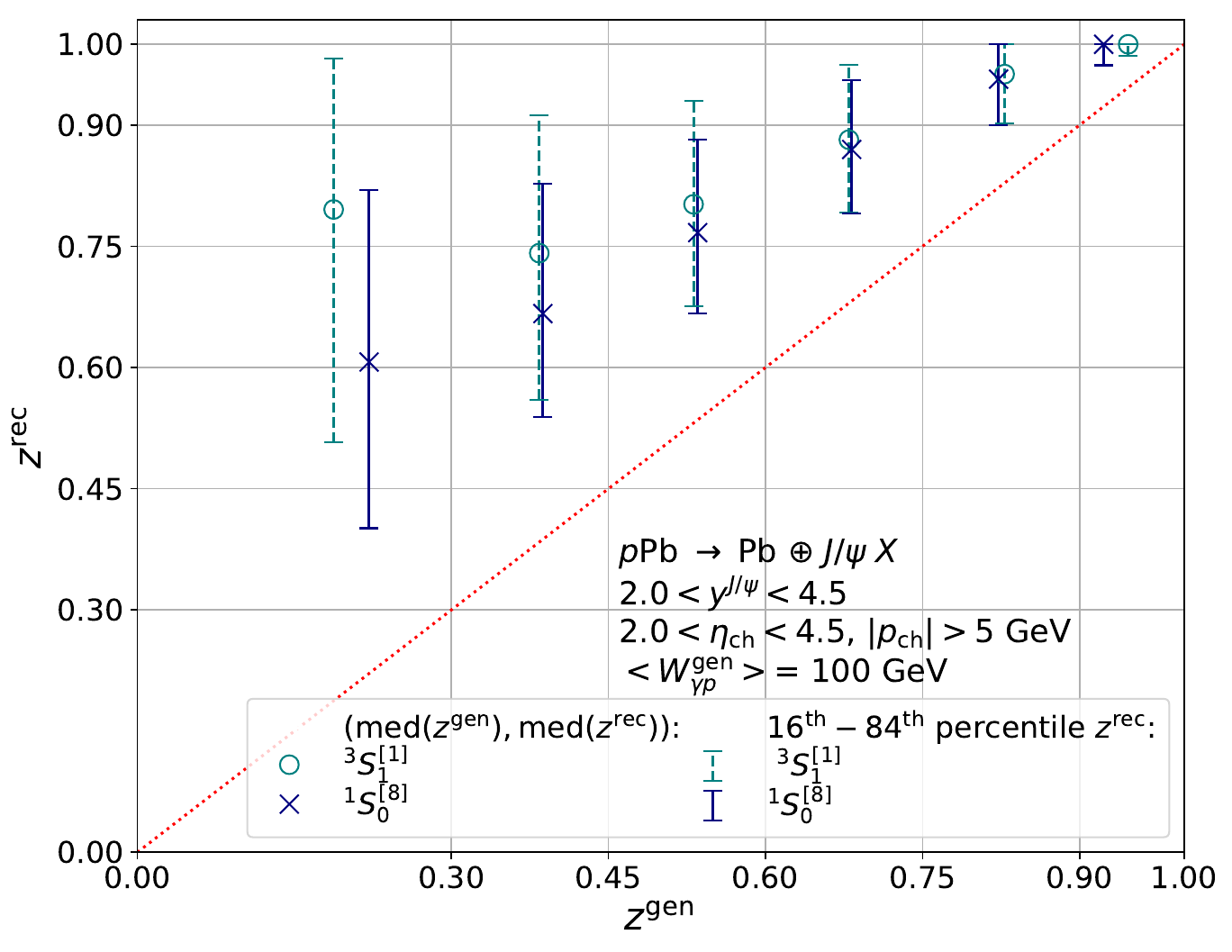}}\\
\subfloat[]{\includegraphics[width=0.5\textwidth]{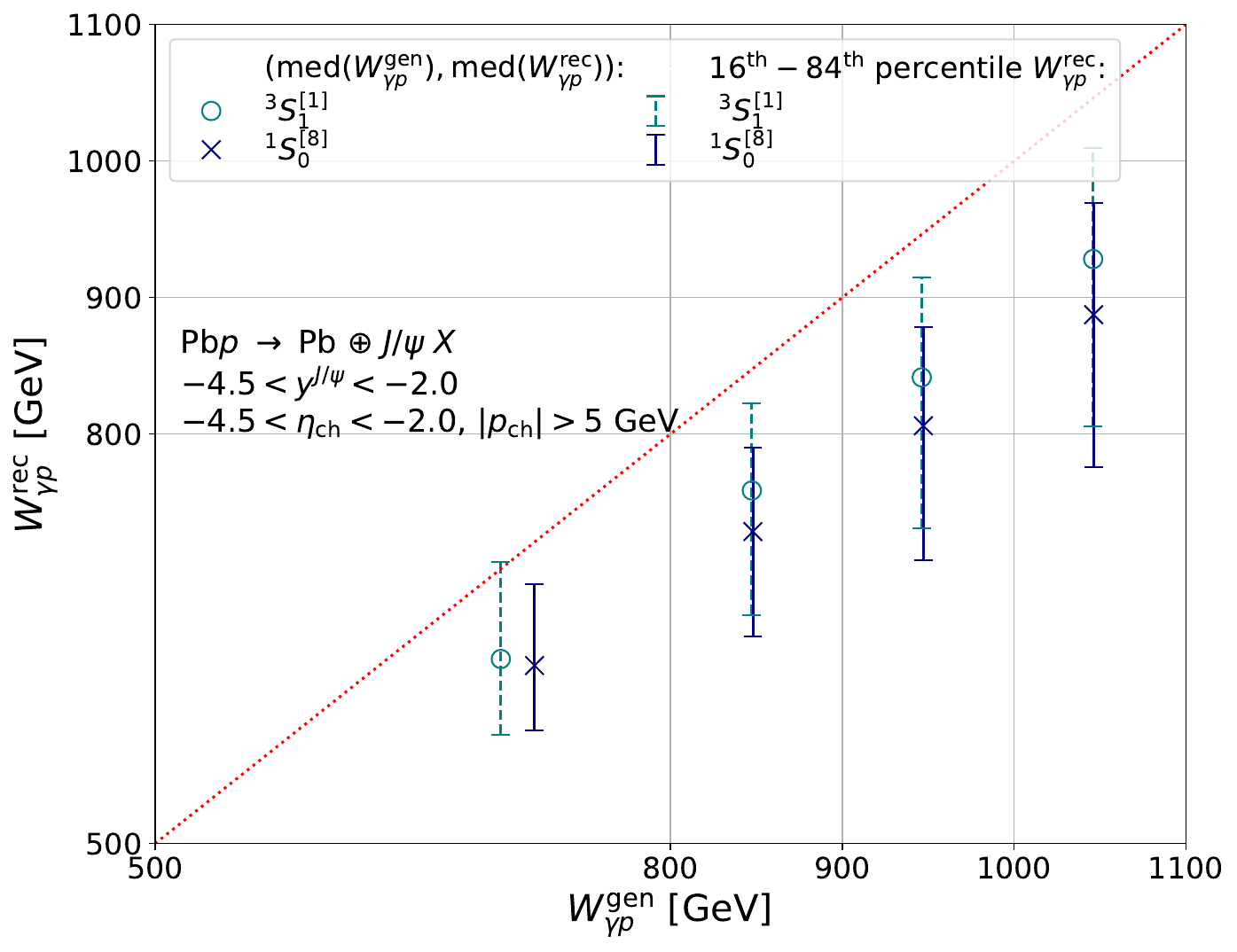}}
\subfloat[]{\includegraphics[width=0.5\textwidth]{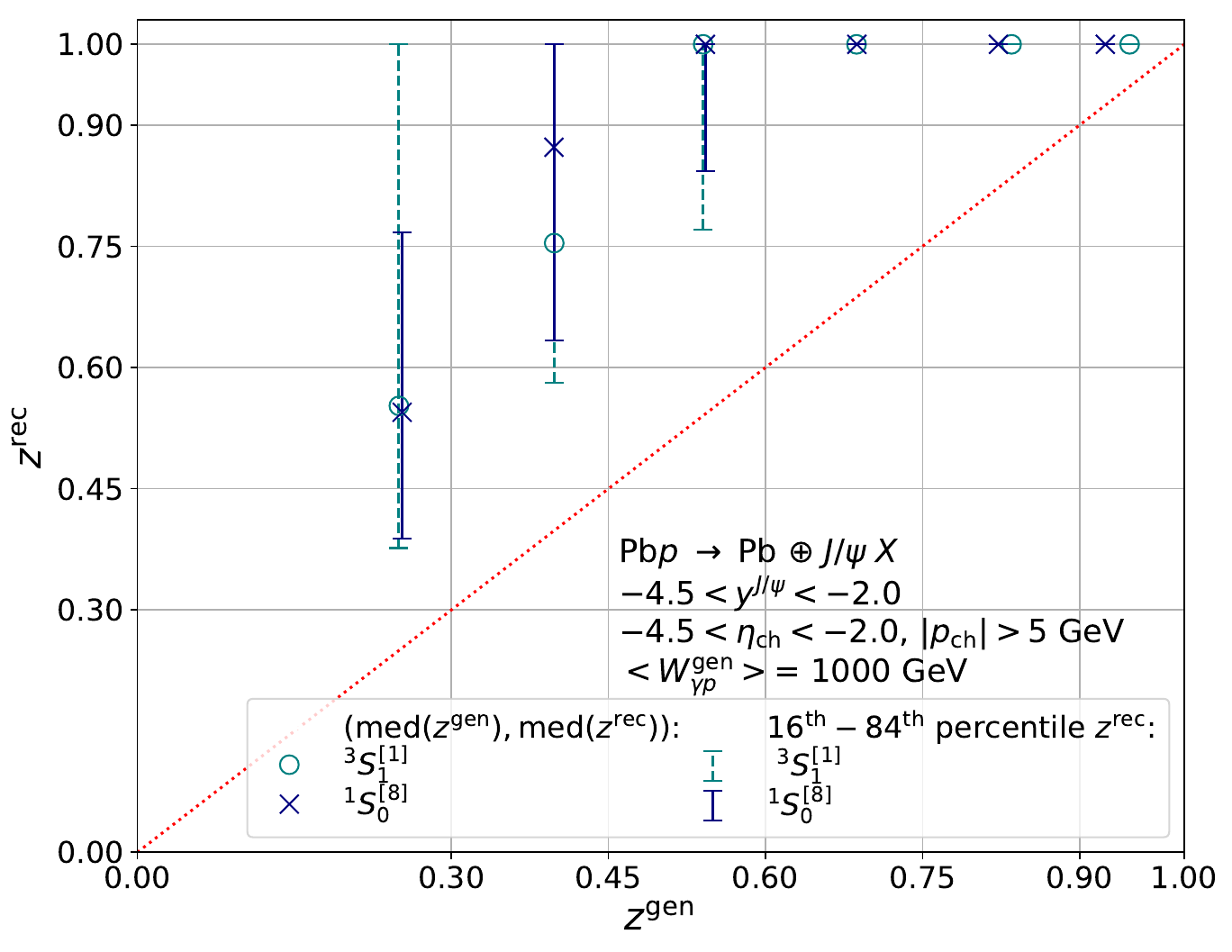}}\\
\caption{The median reconstructed (rec) values as a function of the median generated (gen) values of $W_{\gamma p}$ (a,c) and $z$ (b,d), using the tuned $^3S_1^{[1]}$ (teal circle) and $^1S_0^{[8]}$ (navy blue cross), for \jpsi reconstructed within the LHCb acceptance in the (a,b) $p$Pb and (c,d) Pb$p$ beam configurations. The lower and upper bounds on the error bars indicate the $16^{\rm th}$ and $84^{\rm th}$ percentile on the reconstructed values and the grid lines indicate the chosen binning.}

\label{fig:recoLHCb}
\end{figure*}

Supplementary to Fig.~\ref{fig:recoCMStyp}, the reconstruction capability of the CMS detector in bins of $y^{J/\psi}$ is shown in Figs.~\ref{fig:recoCMSatyp1} and \ref{fig:recoCMSatyp2}, as well as for the LHCb detector in both beam configurations in Figure~\ref{fig:recoLHCb}. In these figures the medians of the reconstructed (rec) and generated (gen) values of $W_{\gamma p}$ and $z$, $\med{W_{\gamma p}^\text{gen,rec}}$ and $\med{z^\text{gen,rec}}$, are plotted using the tuned $^3S_1^{[1]}$ (teal circle) and $^1S_0^{[8]}$ (navy blue cross) simulation samples. The reconstruction bias per bin is the distance between the red, dotted line, where $W_{\gamma p}^\text{rec}=W_{\gamma p}^\text{gen}$ and $z^\text{rec}=z^\text{gen}$, and the teal cross or navy blue circle. The variance of the reconstructed values per bin can be estimated as the difference between the $16^{\rm th}$ and $84^{\rm th}$ percentile values, as represented by the vertical error bars, and the model dependence of the reconstruction can be seen as the difference between values obtained using the $^3S_1^{[1]}$ and $^1S_0^{[8]}$ tunes. The superior reconstruction in CMS with respect to LHCb can be seen as the decreased bias and variance in Figs.~\ref{fig:recoCMSatyp1} and \ref{fig:recoCMSatyp2} with respect to Fig.~\ref{fig:recoLHCb}.

\bibliographystyle{elsarticle-num}

\bibliography{bib}
\end{document}